\documentclass[twocolumn]{WileyMSP-template}
\usepackage{graphicx}
\usepackage{rotating}
\usepackage{amsmath}
\usepackage{amssymb}
\usepackage{bm}
\usepackage{enumerate}
\usepackage{xfrac}
\usepackage[normalem]{ulem}
\usepackage[usenames,dvipsnames]{xcolor}
\usepackage{txfonts}
\usepackage[mathscr]{euscript}
\usepackage[pdftex,
	pdflang={en-US},
    colorlinks=true,
    citecolor = blue,
    urlcolor = blue,
    pdfauthor={},
	pdftitle={Witnessing Entanglement and Quantum Correlations in Condensed Matter: A Review}]{hyperref}
\usepackage{verbatim}
\usepackage{ulem}
\usepackage[version=4]{mhchem}
\usepackage{siunitx}
\usepackage{pdflscape}

\usepackage{mcite}
\usepackage{overcite}
\newcommand{\scite}[1]{\textsuperscript{\protect\cite{#1}}}

\usepackage{mathtools}

\DeclarePairedDelimiter\floor{\lfloor}{\rfloor}

\usepackage[stretch=10,shrink=10]{microtype}

\begin{document}

\title{Witnessing Entang\-lement and Quantum Correlations in Con\-d\-ensed Matter: A Review}
\maketitle

\author{Pontus Laurell*}
\author{Allen Scheie}
\author{Elbio Dagotto}
\author{D. Alan Tennant*}

\begin{affiliations}
Dr. Pontus Laurell, Prof. Elbio Dagotto, Prof. D. Alan Tennant\\
Department of Physics and Astronomy, University of Tennessee, Knoxville, Tennessee 37996, USA\\
E-mail: \url{plaurell@utk.edu}, \url{dtennant@utk.edu}

Dr. Allen Scheie\\
MPA-Q, Los Alamos National Laboratory, Los Alamos, NM 87545, USA

Prof. Elbio Dagotto\\
Materials Science and Technology Division, Oak Ridge National Laboratory, Oak Ridge, Tennessee 37831, USA

Prof. D. Alan Tennant\\
Department of Materials Science and Engineering, University of Tennessee, Knoxville, Tennessee 37996, USA\\
\end{affiliations}

\begin{abstract}
The detection and certification of entanglement and quantum correlations in materials is of fundamental and far-reaching importance, and has seen significant recent progress. It impacts both our understanding of the basic science of quantum many-body phenomena as well as the identification of systems suitable for novel technologies. Frameworks suitable to condensed matter that connect measurements to entanglement and coherence have been developed in the context of quantum information theory. These take the form of entanglement witnesses and quantum correlation measures. 

The underlying theory of these quantities, their relation to condensed matter experimental techniques, and their application to real materials are comprehensively reviewed. In addition, their usage in {\it e.g.} protocols, the relative advantages and disadvantages of witnesses and measures, and future prospects in, e.g., correlated electrons, entanglement dynamics, and entangled spectroscopic probes, are presented. Consideration is given to the interdisciplinary nature of this emerging research and substantial ongoing progress by providing an accessible and practical treatment from fundamentals to application. Particular emphasis is placed on quantities accessible to collective measurements, including by susceptibility and spectroscopic techniques. This includes the magnetic susceptibility witness, one-tangle, concurrence and two-tangle, two-site quantum discord, and quantum coherence measures such as the quantum Fisher information.
\end{abstract}

\date{\today}

\section{Introduction}
Given a device or material sample, how can we detect, certify, and quantify its quantum entanglement and coherence, or, more generally, ``quantumness''? This is a central question for applications\scite{Guehne2009, BESreport2016, NSFreport2018, Friis_2018, Yu2022a, Wang_2022}, because these are precisely the properties that enable quantum systems and technologies to outperform their classical counterparts. It is also important in fundamental quantum many-body physics, as methods for entanglement and coherence detection methods can access previously unavailable quantitative information about their wave functions and enrich our understanding of quantum states of matter.

As with many questions of such a fundamental nature, the proposed answers are only partial and still developing. This is largely because quantum many-body states have very rich structures, allowing for a plethora of ways in which their degrees of freedom may be entangled or---more generally---quantum correlated\scite{RevModPhys.80.517, RevModPhys.81.865, Laflorencie2016, Chiara2018, zeng2019quantum, bayat2022entanglement}. This fact, along with experimental considerations, means that different classes of many-body systems call for different diagnostic approaches. In this review, we focus on achievements in, and methods for, detecting quantum correlations in \emph{quantum materials}\scite{Tokura_2017, keimer2017physics, Cava2021, Iyengar2023}. A broader perspective on different experimental platforms for many-body physics and a pedagogical introduction to the classification of quantum correlations was provided in a recent review\scite{Frerot2023}. Some readers may also be interested in a shorter outlook article\scite{Cruz2023} that briefly reviews experimental entanglement measurements in low-dimensional metal complexes. 

\begin{figure}[t]
    \includegraphics[width=0.98\columnwidth]{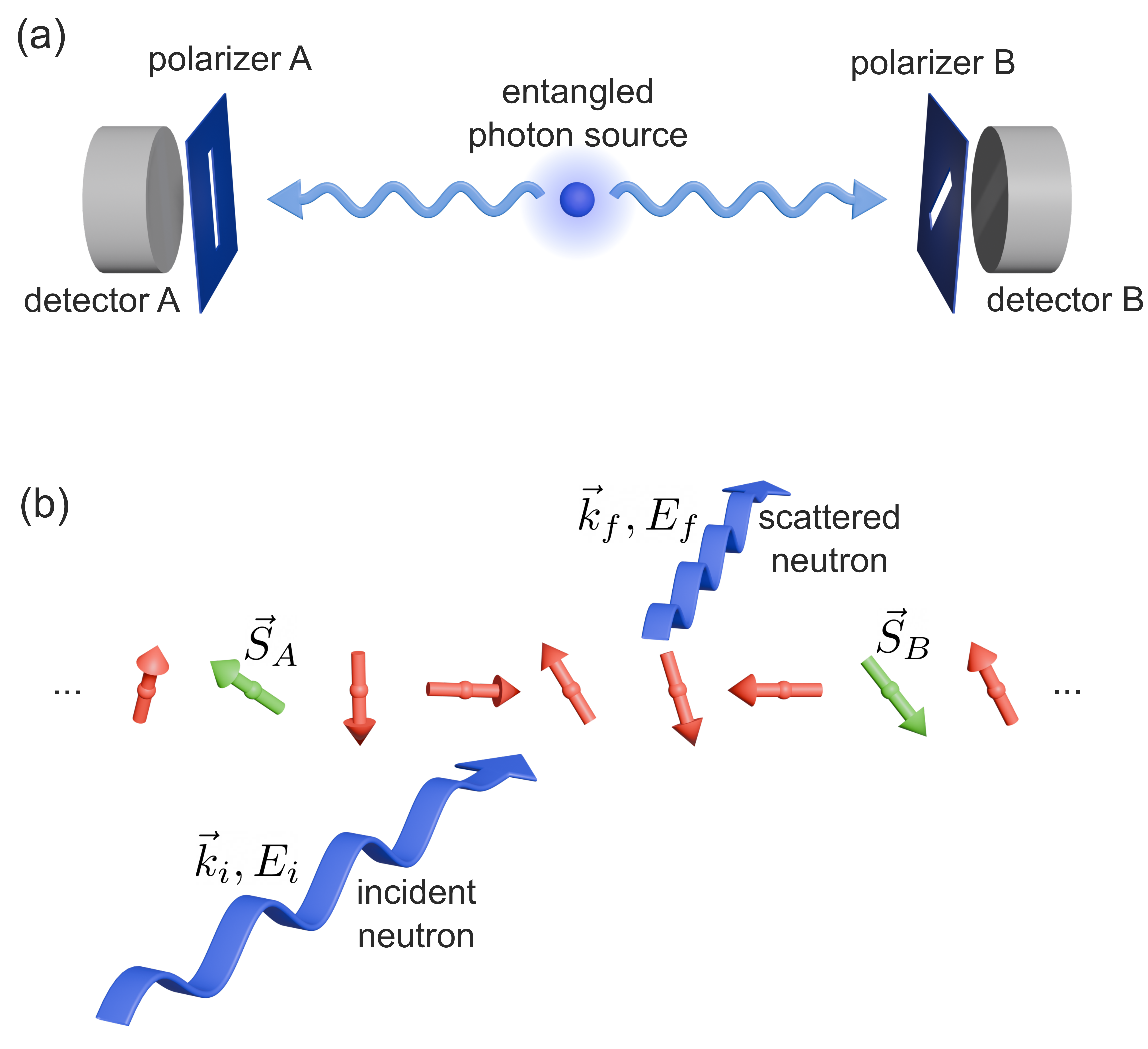}
    \caption{Connections between Bell's inequality measurements and scattering experiments. (a) In an optical Bell's type experiment\scite{PhysRevLett.28.938} the entanglement between photons is witnessed by making a series of measurements of the polarizations of photon pairs. A combination of correlations such as the CHSH witness\scite{PhysRevLett.23.880} are then applied to determine if the photons are indeed entangled. (b) In condensed matter systems entanglement can likewise be witnessed using a combination of correlation functions between components.
    Here we consider the example of two spins $\mathbf{S}_{A/B}$ (which may be spatially separated) in a chain.
    The correlation between their spin components can be measured using, for example, neutron scattering.
    Incident neutrons in a prepared polarization spin state $\alpha$ $(=x,y,z)$ scatter from the spins in a material transferring energy $\hbar\omega=E_i-E_f$ and momentum $\textbf{Q}=\textbf{k}_i-\textbf{k}_f$ to the material where the scattered neutron state is $\beta$. Fourier transforming the scattering cross section $S(\textbf{Q},\omega)$ for different polarizations allows the real space and time correlations between spins $\langle S^{\vphantom{y}x}_{A} S^{\vphantom{y}x}_{B}\rangle, \langle S^{\vphantom{y}x}_{A} S^{y}_{B}\rangle, \langle S^{\vphantom{y}z}_{A} S^{\vphantom{y}z}_{B}\rangle $ {\it etc.} to be reconstructed. Various entanglement witnesses can then be applied. Other experimental techniques can witness entanglement in a similar manner both in and out of equilibrium.}
    \label{fig:schematic}
\end{figure}

Quantum materials are condensed matter systems built up from electronic or spin degrees of freedom. Their physics is typically driven by local electronic interactions originating in the Pauli exclusion principle and Coulomb force, yet allows for a wealth of emergent physical states at meso- and macroscopic scales. Some of the most intriguing and sought-after quantum phases of matter are known to be highly entangled; see, for example, quantum critical states\scite{PhysRevLett.90.227902, PhysRevD.96.126007} and topologically ordered states\scite{RevModPhys.89.041004}, including fractional quantum Hall states\scite{PhysRevLett.96.110404, PhysRevLett.96.110405} and quantum spin liquids\scite{RevModPhys.89.025003, Savary2017, Broholm2020}. Notably, topological order enables emergent fractional quasiparticles, which may be of use for fault-tolerant topological quantum computing technologies\scite{Kitaev2003, RevModPhys.80.1083}. However, their unambiguous experimental demonstration has proven to be very challenging, which calls for reconsideration of our experimental probes and the information we extract from them. Although there currently is no known way to experimentally measure the topological entanglement in such states, it is a worthy goal to pursue, and perhaps one that may be informed by the recent progress reviewed here on probing the more local entanglement in strongly correlated electron systems.

In contrast to other many-body platforms---such as networks of superconducting qubits\scite{Krasnok2024}, nitrogen-vacancy (NV) centers\scite{Doherty2013, Pezzagna2021}, or cold atoms in optical lattices\scite{RevModPhys.82.2313, Schaefer2020}---quantum materials are more closely packed and typically allow only for collective measurements. Indeed, addressing individual degrees of freedom of quantum materials is extremely difficult given the microscopic scales and geometry, while interferometry and quantum state tomography are rarely possible because of the large numbers of constituent particles. 
\scite{Nakamura_2020, Kundu_2023, PhysRevX.13.041012}
Instead, we rely largely on transport experiments, thermodynamic measurements and solid-state spectroscopy techniques. The latter includes linear-response techniques that are highly sensitive to correlations between local degrees of freedom, such as inelastic neutron scattering, x-ray scattering, etc. The most promising way of using these probes to extract information about the entanglement of materials is through \emph{entanglement witnesses}. These are observables that serve as order parameters for specific classes of entangled states. A particularly important class of witnesses are related to quantum coherence measures. These include the quantum Fisher information\scite{Hauke2016}, which has recently attracted significant experimental attention. 
Witnesses associated with collective measurements are useful not only for investigating the suitability of materials for future applications; since condensed matter systems can be easier to hold at thermal equilibrium than other many-body platforms, such witnesses also offer a promising approach for studying fundamental properties of entanglement and quantum correlations in thermal states.

This review is outlined as follows. We first conceptually introduce the idea of entanglement witnesses, and their role in different platforms for many-body physics in Section~\ref{sec:conceptualbackground}. We then proceed with the mathematically heavier Sec. \ref{sec:theory}, in which we present a variety of experimentally relevant witnesses, and provide important derivations. In Sec. \ref{sec:applications} we survey experimental applications of witnesses to detect and quantify entanglement in the solid state. In Sec. \ref{sec:broaderperspective} we provide a broader and forward-looking perspective on entanglement detection in quantum materials. In the more specialized Sec.~\ref{sec:techdevelopments} we focus on technical developments and challenges for entanglement detection in quantum materials using scattering experiments. The review ends with a brief Conclusion section, Sec. \ref{sec:conclusion}. Appendix \ref{sec:qfi:linearresponse} provides a brief introduction to linear response theory, setting a consistent notation.

\section{Conceptual Background} \label{sec:conceptualbackground}

It was realized early on that quantum mechanics had a subtle unexpected consequence known as entanglement \scite{PhysRev.47.777, PhysRev.48.696, Schroedinger1935}. In brief, this is a phenomenon of a group of quantum degrees of freedom, in which the quantum state of an individual degree of freedom can depend on the state of the others, even across vast distances. 
It was not {\it a priori} clear if this effect was due to the probabilistic nature of quantum mechanics, or if it would be better explained using a hidden-variable theory (i.e. a deterministic extension of quantum mechanics with additional variables). 
Decades later, Bell showed that entanglement has experimental consequences in the form of inequalities constructed from correlations of observables\scite{PhysicsPhysiqueFizika.1.195, Bell_2004}, now known as Bell inequalities. Violations of these inequalities allow ruling out classical behavior and local hidden-variable theories. 
One key insight is that information about entanglement is embedded in two-point correlation functions, which remain a crucial component in entanglement detection. Indeed, the 2022 Nobel prize in physics was awarded to Aspect, Clauser and Zeilinger ``for experiments with entangled photons, establishing the violation of Bell inequalities and pioneering quantum information science''. 

Although many such Bell inequalities have been der\-ived\scite{Guehne2009, RevModPhys.86.419}, a common formulation is the CHSH inequality\scite{PhysRevLett.23.880}. If we assume particles $A$ and $B$ (Bell envisioned electron spins, but experiments often measure photon polarization) and the pairwise spin/photon polarization is measured at different orientations ($A_0$, $A_1$, $B_0$, $B_1$; see Fig.~\ref{fig:schematic}(a)) in four separate sub-experiments, there is an inequality that is always satisfied \emph{classically} (i.e. when no entanglement and/or no hidden variables exist): 
\begin{align}
    \langle A_0 B_0 \rangle + \langle A_0 B_1\rangle + \langle A_1 B_0 \rangle - \langle A_1 B_1\rangle &\leq 2.
\end{align}
In the presence of quantum entanglement (in this case non-locality as the particles are separated by a distance), this inequality can be violated, and the upper bound for a quantum theory\scite{Cirelson1980} is
\begin{align}
    \langle A_0 B_0 \rangle + \langle A_0 B_1\rangle + \langle A_1 B_0 \rangle - \langle A_1 B_1\rangle &\leq 2\sqrt{2}.
\end{align}
Violations of the classical inequality have at this point been measured to very high precision in loophole-free ways\cite{Aspect_1999, RevModPhys.71.S288, Hensen_2015, PhysRevLett.115.250401, PhysRevLett.115.250402, PhysRevLett.119.010402, PhysRevLett.121.080404, Storz_2023}, and this is a well-understood landmark in quantum science.

\begin{figure}[t]
    \centering
    \includegraphics[width=89mm]{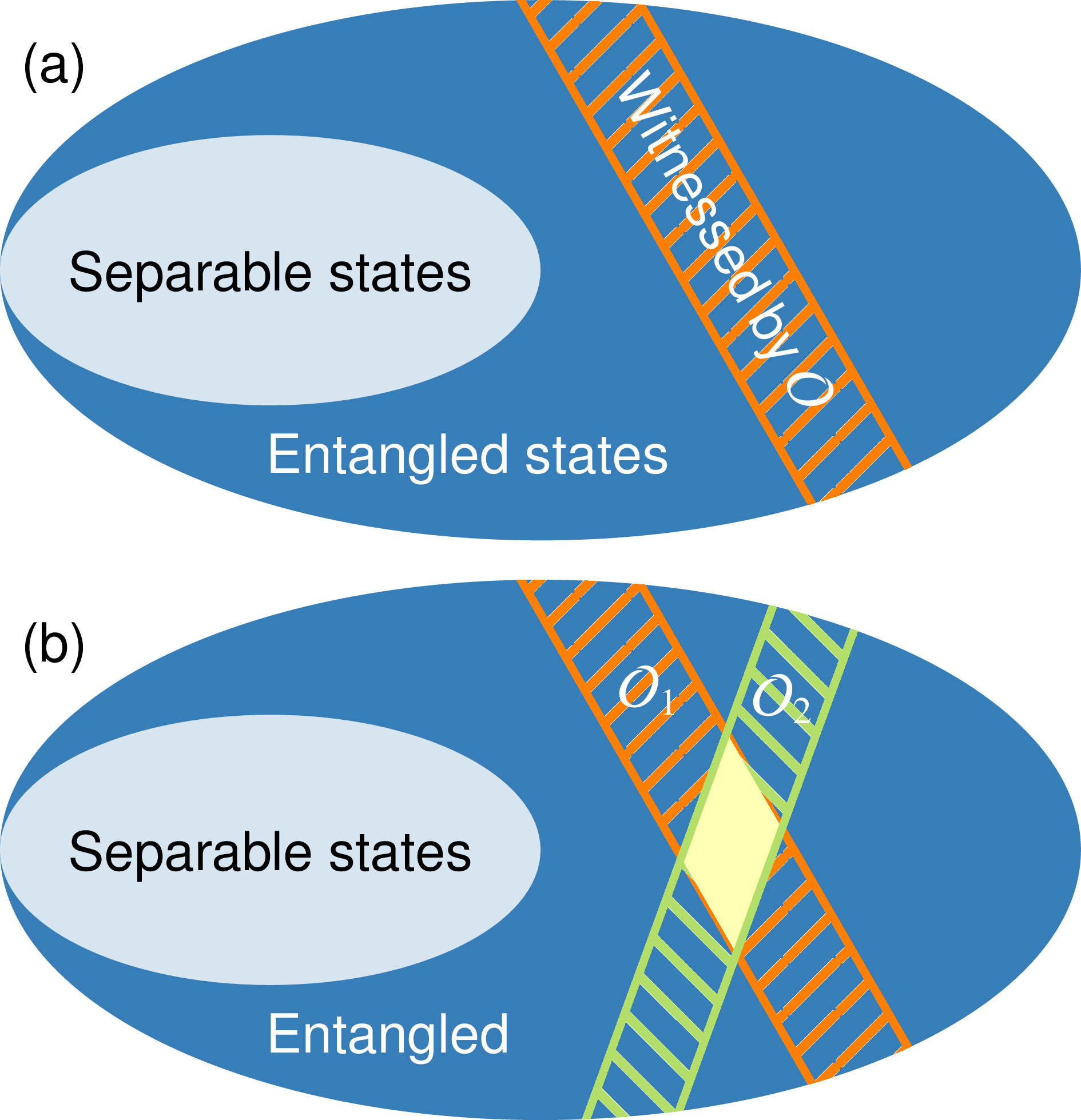}
    \caption{Entanglement witnesses. (a) Schematic picture of a generic Hilbert space. There exists a set of separable states (i.e. states that are not entangled) and a---typically larger---set of entangled states. It is possible to construct observables $\mathcal{O}$ whose (experimental) measurement allows identifying if the system is in a subset of entangled states or not. The subset is represented by the orange hatched area. These observables are called entanglement witnesses. Figure inspired by \scite{Guehne2009}. 
    (b) Here, two witnesses $\mathcal{O}_1$ and $\mathcal{O}_2$ can witness entanglement in the subsets represented by the orange and green hatched areas, respectively. If both witnesses indicate entanglement, the state of the system is constrained to the overlapping region represented by the yellow shaded area. In general, by using multiple distinct witnesses, it is possible to infer additional information and, in principle, triangulate where in the Hilbert space the state lives. 
    }
    \label{fig:witness}
\end{figure}

However, Bell-type inequalities have important limitations when applied to many-body systems. Constructions only exist for a subset of multipartite entangled states, 
and tend to require an exponentially growing number measurements \scite{Guehne2009}. The measurements are also often sensitive to noise \scite{Guehne2009}. A closely related approach is the use of \emph{entanglement witnesses} \scite{Guehne2009, Horodecki1997, Terhal2000, PhysRevA.62.052310, PhysRevA.72.012321, Chruscinski2014}. These quantities are generally functionals of the density matrix $\rho$ encoding a quantum state that allow for distinguishing a subset of entangled states from separable states; see Fig.~\ref{fig:witness} for an illustration. Associated with the functional is an inequality, which, if satisfied, indicates that the state of the system falls within that subset of entangled states. 
If the inequality is not satisfied, the system may be in a separable state or potentially another subset of entangled states.

They key advantage of witnesses is that they enable inferring knowledge about the entanglement of a system from partial information about its state. This is extremely useful in the context of macroscopic materials, which involve Avogadro's numbers of degrees of freedom. The challenge presenting itself is finding appropriate quantum mechanical expectation values accessible to experimental probes. It is particularly natural to consider spectroscopic experiments, which directly measure two-point correlation functions and allow immediate analogies with Bell's inequality experiments; see Fig.~\ref{fig:schematic}(b). However, a wider range of observables have been proposed as witnesses in the solid state, including magnetization \cite{PhysRevA.69.022304, PhysRevLett.93.167203}, spatial correlations between two spins \cite{PhysRevA.69.022304, PhysRevLett.93.167203, PhysRevA.72.032309, PhysRevA.73.012110, PhysRevA.74.022322}, magnetic susceptibility \cite{Wie_niak_2005, PhysRevA.73.012110}, heat capacity \cite{PhysRevB.78.064108, Singh2013}, static structure factors \cite{PhysRevLett.106.020401, Kwek2011, PhysRevB.107.054422}, and dynamic susceptibilities \cite{Hauke2016}. Combinations of witnesses---as in Fig.~\ref{fig:witness}(b)---may be used in protocols for identifying states of interest.

This review focuses on a set of quantities that have been applied to experimental 
condensed matter systems 
namely (i) the magnetic susceptibility entanglement witness $\chi_\mathrm{EW}$ \cite{Wie_niak_2005, PhysRevA.73.012110}, (ii) the one-tangle entanglement witness $\tau_1$ \cite{PhysRevA.61.052306, PhysRevA.69.022304, PhysRevLett.93.167203}, (iii) the concurrence ($C$) and two-tangle ($\tau_2$) entanglement witnesses \cite{PhysRevA.61.052306, PhysRevA.69.022304, PhysRevLett.93.167203}, (iv) the two-site quantum discord measure of quantum correlations \cite{PhysRevLett.88.017901, PhysRevA.80.022108}, and (v) the quantum Fisher information entanglement witness \cite{Hauke2016}. These are summarized in Table~\ref{tab:witnesses}, and will be discussed in detail later.

\begin{table*}[tb]
  \centering
  \caption{Table of entanglement and quantum correlation witnesses. These are described in depth in Section \ref{sec:theory}, and their application to materials outlined in Section \ref{sec:applications}. The physical interpretation of different witnesses may be compared with Fig.~\ref{fig:types}. Relevant experimental techniques (Sec. \ref{sec:applications}) are listed.
  $\chi_m$ denotes magnetic susceptibility, $C_m$ denotes magnetic heat capacity, and $S(\mathbf{Q})$ denotes the static spin structure factor, which may be probed using neutrons. Spectroscopy refers to energy-resolved dynamic susceptibility probes using, for example, neutrons, photons, muons or electrons.
  Studies of materials where the witnesses have been applied are given in Table \ref{tab:Materials}.   
  }
  \begin{tabular}{cccc}
    {\bf Witness} & {\bf Key Formulae} & {\bf Physical Interpretation} & {\bf Experimental Probes} \\
    \hline
    Susceptibility ($\chi_\mathrm{EW}$) & Eq. \eqref{eq:suspectibility:witness} & generalized spin-squeezing & $\chi_m$  \\
    One-tangle ($\tau_1$) & Eq. \eqref{eq:one-tangle} & one-site entanglement & diffraction  \\
    Two-tangle ($\tau_2$) & Eq. \eqref{eq:two-tangle}& total pairwise entanglement & $S(\textbf{Q})$  \\
    Concurrence ($C$) & Eqs. (\ref{eq:concur},\ref{eq:concurrence:dimer}) & pairwise entanglement & $S(\textbf{Q})$, $C_v$, $\chi_m$  \\
    Quantum discord [$Q(\rho)$] & Eqs. (\ref{eq:discord:definition},\ref{eq:discord:XYZ},\ref{eq:discord:Heisenberg}) & pairwise quantum correlations& $S(\textbf{Q})$, $C_v$, $\chi_m$ \\
    Quantum Fisher Information (QFI) & Eqs. (\ref{eq:QFI:mixed}, \ref{eq:QFI:Hauke}) & multi-partite entanglement& spectroscopy \\
    Quantum variance (QV)   & Eq. \eqref{eq:generalquantumcoherencemeasure} & multi-partite entanglement& spectroscopy \\
    Skew information (SI)    & Eqs. (\ref{eq:generalquantumcoherencemeasure}, \ref{eq:skewfilter}) & multi-partite entanglement& spectroscopy \\
    Spatial quantum correlations & Eq. \eqref{eq:spatialquantumcorrelation} & quantum fluctuations & spectroscopy \\
    \hline
  \end{tabular}
  \label{tab:witnesses}
\end{table*}

\begin{figure}[t]
    \includegraphics[width=\columnwidth]{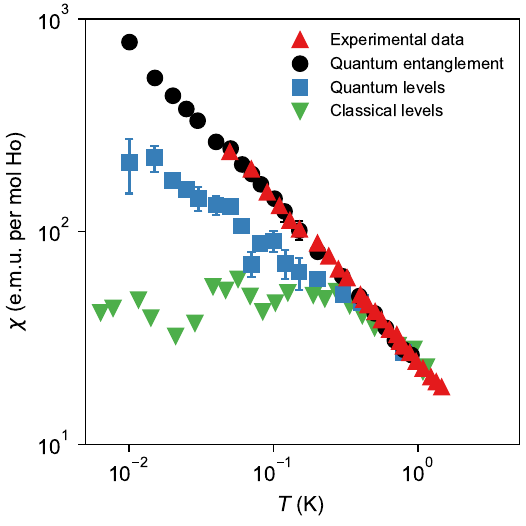}
    \caption{Magnetic susceptibility $\chi$ as a function of the temperature of the dilute, dipolar Ising magnet \ce{LiHo_{0.045}Y_{0.0955}F4}. To theoretically reproduce the experimentally measured susceptibility data at low temperatures it is necessary to use a fully quantum mechanical modeling. 
    Adapted from Figure 1 of Ref.\scite{Ghosh2003}. See also Ref.~\scite{Vedral2003Nature} for a contemporary perspective.
    }
    \label{fig:dipoleising}
\end{figure}

Note that witnesses is not the only path to certifying the presence of entanglement in materials. Entanglement can also be inferred by conventional data analysis methods for systems that can be modeled accurately by theory, as in Fig.~\ref{fig:dipoleising} \scite{Ghosh2003}, and in systems that host specific entangled states with clear signatures, such as from fractionalized excitations \scite{PhysRevLett.70.4003, Christensen2007, Mourigal2013, Piazza2015}. However, due to the reliance on specific models, 
such approaches to demonstrating entanglement are inherently much more limited in scope than witnesses. In general, conventional analysis and entanglement witness analysis synergize and can be combined to provide additional insight.

We next briefly introduce important quantum material classes where entanglement witness and quantum correlation functions have already been applied, or are likely to be applied in the near future. Quantum magnets, correlated electron systems, and quantum fluids provide a wide range of important and exotic quantum phases of matter.

Quantum magnets comprise arrays of coupled localized spin degrees of freedom that realize cooperative quantum behavior \cite{Tennant2019}. They are found in a wide array of solid state materials including low-dimensional materials \cite{Vasiliev2018} and those with frustrated lattices, including triangular, kagome, and pyrochlore motifs, enabling, e.g., quantum spin liquids \cite{Savary2017, Broholm2020}. 
External stimuli (e.g., pressure, magnetic fields) 
can be used to manipulate spin states in these materials, potentially driving systems through entanglement transitions and quantum phase transitions \cite{Sachdev_2011}, which are associated with very high degrees of entanglement \cite{PhysRevLett.90.227902, PhysRevD.96.126007}. Quantum magnets are actively being pursued as the basis of future quantum technologies including sensors, quantum magnonics\cite{Yuan2022}, topological quantum devices, and information storage and processing. They also often represent the simplest realizations of model systems, and can be emulated in quantum simulators and computers. 

(Strongly) correlated electron materials involve interactions between charge, spin, and orbital degrees of freedom. This important class of materials includes unconventional superconductors, multiferroics, spintronic materials, heavy fermion materials\cite{Paschen_2020}, and many other types of systems. Because of their complexity, more complicated entanglement patterns may be expected than in quantum magnets, including entanglement between spin and charge sectors, or between different orbitals. Overall, such materials are often challenging to model quantitatively, and much remains unknown about the structure of quantum correlations within them. Of particular interest in these materials is the study of entanglement near quantum phase transitions and out-of-equilibrium.

The quantum liquids helium-3 and helium-4 remain in liquid states down to temperatures where their de Broglie wavelengths become of order the atomic spacings and take on quantum properties such as superfluidity \cite{volovik}. These cover a remarkable range of quantum phases including topological states as a function of mixture, pressure, dimensional reduction/confinement, and temperature. Entanglement is an important aspect of these states where {\it e.g.} superfluid droplets show area law entanglement \cite{Herdman2017}, and confined helium in one-dimensional pores form highly entangled Luttinger liquids \cite{delmaestro_2011}. The remarkable properties of these states are seen in thermodynamic and transport properties as well as from neutron scattering techniques \cite{DelMaestro2022,Dmowski2017}.

\begin{figure*}[tb]
    \includegraphics[width=\textwidth]{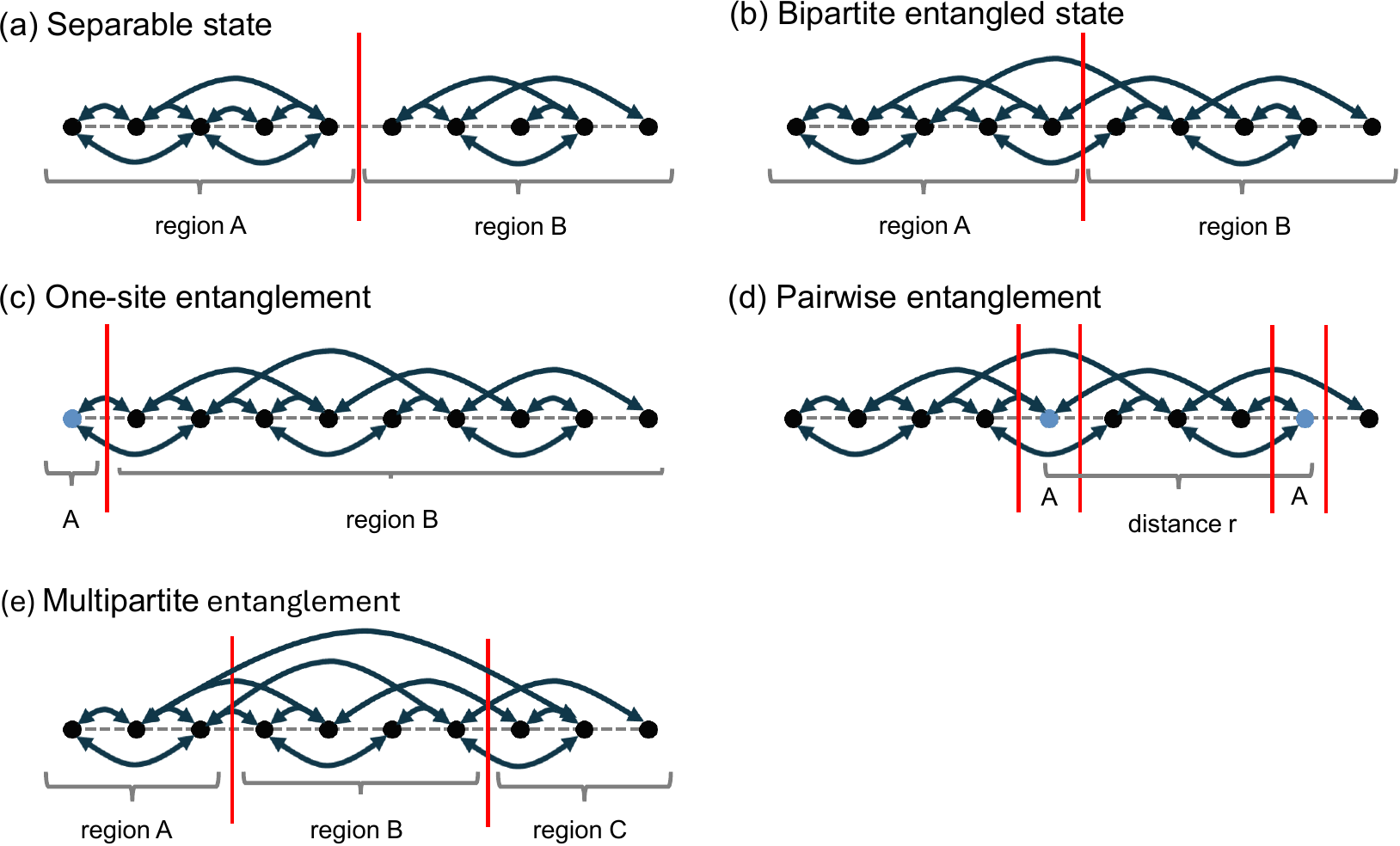} 
    \caption{Entangled states. The arrows represent entangling correlations between sites, and red vertical lines represent partitions of the system into different regions. For simplicity of notation, in the following we assume the quantum state of the full system $|\Psi\rangle$ is a pure state. However, these notions generalize to mixed states as described in the text. 
    (a) A separable state, i.e. non-entangled state, for which $|\Psi\rangle=|\psi_A\rangle \otimes |\psi_B \rangle$. 
    (b) A typical bipartite entangled state, for which $|\Psi\rangle=|\psi_A\rangle \otimes |\psi_B \rangle$.
    (c) The one-site reduced density matrix is obtained by keeping a single site in region A and tracing out all degrees of freedom in the rest of the system. In spin-$1/2$ systems, the entanglement between the single site and the system can be witnessed through the one-tangle.
    (d) The two-site reduced density matrix is obtained by isolating two sites in region A --- which does not need to be contiguous --- and tracing out degrees of freedom in B, which contains all sites not in A. In spin-$1/2$ systems, the pairwise entanglement between two sites separated by a distance $r$ can be witnessed through the concurrence. A related measure is the two-tangle, which is a witness of the total pairwise entanglement.
    (e) For multipartite entangled states it is possible to introduce additional partitions. Here, a state with at least tripartite entanglement is shown, for which $|\Psi\rangle \neq |\psi_A\rangle \otimes |\psi_B\rangle \otimes |\psi_C$. The strength of multipartite entanglement can be characterized by the entanglement depth, which can be witnessed using the quantum Fisher information and related measures.
    }
    \label{fig:types}
\end{figure*}

\section{Theory of entanglement and witnesses}\label{sec:theory}
The goal of this section is to provide a mathematical introduction to entanglement and to provide derivations for some experimentally applicable entanglement witnesses, all in a notation tailored to condensed matter physicists. By collecting the derivations in one place instead of tens of papers we hope to make the material more accessible. 
To keep our focus and scope, we will use certain results from quantum information and quantum metrology without proof, but state clearly when we do so.

\subsection{Bipartite and multipartite entanglement}
The most commonly studied form of entanglement is bipartite entanglement; see Fig.~\ref{fig:types}(a,b). Here, bipartite refers not to a bipartite lattice, but to a splitting of a quantum system's Hilbert space $\mathcal{H}$ into two partitions, $A$ and $B$. How one splits $\mathcal{H}$ --- the choice of the so-called entanglement cut --- is, in principle, arbitrary and depends on which degrees of freedom one is interested in. Here we envision a cut in real space, but note that, e.g., momentum-space and inter-sublattice cuts have also attracted interest. 
The state of the system can be described by a density matrix $\rho$. Tracing over the degrees of freedom in $B$ yields the reduced density matrix of $A$, $\rho_A = \mathrm{Tr}_B \left[ \rho\right]$, and vice versa for $\rho_B$. The reduced density matrix contains all information about the subsystem, and can thus be used for calculating expectation values $\langle O_A\rangle = \mathrm{Tr} \left[ \rho_A \mathcal{O}_A\right]$, where $\mathcal{O}_A$ is an operator with support on $A$. The von Neumann entropy, 
\begin{align}
    S_\mathrm{vN} = -\mathrm{Tr} \left[ \rho_A \log \rho_A \right],
    \label{eq:vonNeumann}
\end{align}
can be viewed as an extension of the Gibbs entropy in statistical mechanics, or the Shannon entropy in classical information theory \scite{kardar2007statistical, Witten_2020}. In fact, when expressed in the diagonal basis with $\lambda_j$ the $j$th eigenvalue of $\rho$, $S_\mathrm{vN}=-\sum_j \lambda_j \log \lambda_j$, is equivalent to the Shannon entropy. $S_\mathrm{vN}=0$ only for pure states, i.e. states that can be written $\rho=|\psi\rangle\langle \psi|$. The maximal value of $S_\mathrm{vN}$ is $\log(N)$, where $N$ is the dimension of the Hilbert space, which is reached for a maximally entangled state. (Conventionally, quantum information theory uses base-$2$ logarithms, while condensed matter theory tends to use the natural logarithm ($\ln$).)

To make things more concrete, let us first consider the simple example of a two-site system in a singlet ground state. The wave function is
\begin{align}
    \left| \psi \right\rangle   &=  \frac{1}{\sqrt{2}} \left( \left| \uparrow\right\rangle_A \left| \downarrow \right\rangle_B - \left| \downarrow \right\rangle_A \left| \uparrow\right\rangle_B \right),
\end{align}
where $\left| \psi\right\rangle_{A/B}$ is the wave function of the degree of freedom labeled $A$ or $B$. The basis for each site is $\{\left|\uparrow\right\rangle\, \left|\downarrow\right\rangle\}$, consisting of one-site eigenstates with well-defined spin projections onto a suitable quantization axis ($\hat{z}$). The density matrix is given by $\rho=\left| \psi \right\rangle\left\langle \psi \right|$. The reduced density matrix of $A$ becomes
\begin{align}
    \rho_A  &=   \mathrm{Tr}_B\left[ \rho \right]  = \left\langle \uparrow\middle|_B\, \rho \middle| \uparrow\right\rangle_B + \left\langle \downarrow\middle|_B\, \rho \middle| \downarrow\right\rangle_B\\
            &= \frac{1}{2} \left[ \left| \uparrow\middle\rangle_A \middle\langle \uparrow\right|_A + \left| \downarrow\middle\rangle_A \middle\langle \downarrow\right|_A \right] = \begin{pmatrix}
                \frac{1}{2} &   0\\
                0           &   \frac{1}{2}
            \end{pmatrix},
\end{align}
which has the form of a mixed state, and
the entanglement entropy becomes $S_{vN}=\log 2$ (the logarithm of a diagonal matrix $A$ with nonzero elements $A_{11},A_{22},\dots$ being $\mathrm{diag}\left( 
\log A_{11}, \log A_{22}, \dots\right)$), i.e. we have a maximally entangled state. In contrast, the spin-polarized state $\left|\uparrow\middle\rangle_A\middle|\uparrow\right\rangle_B$ gives $S_\mathrm{vN}=0$. In both cases, the composite state of the two-site system is pure, but the reduced state is only pure in the polarized case. The mixedness of the reduced density matrix in the singlet case is at the heart of entanglement.

We next turn to the general case of bipartite entanglement, closely following Ref.~\cite{Guehne2009}, as we also did in Ref.~\cite{PhysRevB.103.224434}. A state is considered entangled if it is not separable.  
Consider an arbitrary state, which may be described with the density matrix $\rho=\sum_i p_i |\phi_i\rangle\langle \phi_i|$, where $0\leq p_i\leq 1$ is the probability of the pure state $|\phi_i\rangle$. We say that $\rho$ is separable if it can be expressed $\rho=\sum_i p_i\, \rho_i^A \otimes \rho_i^B$, where $\rho^A_i$ and $\rho^B_i$ are constructed from states in $A$ and $B$, respectively. A special case is the product states, for which $\rho=\rho^A \otimes \rho^B$. It is worth noting that identifying whether a given $\rho$ is separable or not, known as the separability problem, has been shown to be $NP$ hard in general \cite{Gurvits2003, Guehne2009}. (This means that, while the runtime for verifying a solution grows as a polynomial of the size of the Hilbert space, the runtime for any algorithm obtaining the solution is expected to grow exponentially.)

An important generalization of the von Neumann entanglement entropy, Eq. \eqref{eq:vonNeumann}, to a mixed state $\rho$ is known as the entanglement of formation\cite{PhysRevA.54.3824, PhysRevLett.80.2245}. It quantifies the average amount of entanglement that is necessary to generate the state. For a system partitioned into parts $A$ and $B$, it is defined 
\begin{equation}
    E_\mathrm{form}\left( \rho\right)   =  \inf \left\{ \sum_i p_i S\left( \rho_{A,i}\right) \right\}, \label{eq:formation}
\end{equation}
where the infimum is taken over all possible decompositions of $\rho$ into pure states, i.e. all ensembles of states $\left| \psi_i\right\rangle$ and probabilities $i$ such that $\rho=\sum_i p_i \left| \psi_i\right\rangle\left\langle \psi_i\right|$. For each of the pure states $\rho_i=\left| \psi_i\right\rangle\left\langle \psi_i\right|$ one performs a partition, obtaining the reduced density matrix $\rho_{A,i}$ and calculates the von Neumann entropy. $E_\mathrm{form}$ is zero only for separable states. We note that quantities of this type, defined using an optimization operation, are common in quantum information theory. While such constructions may appear far removed from experimental measurements, the entanglement of formation for pairwise entanglement can be witnessed using the concurrence; see Sec.~\ref{sec:concurrence-two-tangle}.

The above definitions for separable and bipartite entangled states 
generalize to the case of multipartite entanglement \cite{Guehne2009, Hyllus2012}; see Fig.~\ref{fig:types}(e). A state is said to be fully separable if it can be written $\rho=\sum_i p_i \rho_i^{(1)} \otimes \dots \rho_i^{(N)}$, where $N$ is the number of degrees of freedom in $\mathcal{H}$, for example the number of sites or particles. States that cannot be expressed this way possess some degree of entanglement, which can be quantified in the language of $m$-separability (or producibility). A pure state is said to be $m$ separable ($m-\mathrm{sep}$) if it can be expressed as 
\begin{align}
    \left|\phi_{m-\mathrm{sep}}\right\rangle  &=\otimes_{l=1}^M|\phi_l\rangle,  \label{eq:mseparable:purestate}
\end{align} where $|\phi_l\rangle$ is a state of $N_l\leq m$ particles and $\sum_{l=1}^M N_l = N$. $|\phi_{m-\mathrm{sep}}\rangle$ has $m$-partite entanglement if it is $m$ separable but not $(m-1)$ separable. Similarly, a mixed state has $m$-partite entanglement if it can be written 
\begin{align}
    \rho_{m-\mathrm{sep}}   &=\sum_l p_l \left| \phi_{m-\mathrm{sep}} \middle\rangle\middle\langle \phi_{m-\mathrm{sep}} \right|.   \label{eq:mseparable:mixedrho}
\end{align}

\subsection{Magnetic susceptibility}
In an important paper from 2005, Wie\'sniak et al.\cite{Wie_niak_2005} dem\-onstrated that the magnetic susceptibility, a thermodynamic quantity, provides sufficient information to witness entanglement in certain spin systems. This approach has since been used in the analysis of a substantial number of materials. The construction is as follows. Consider a system of $N$ spin-$S$ spins held at thermal equilibrium with temperature $T$. The zero-field isothermal magnetic susceptibility along each of the orthogonal $a=x,y,z$ spin-space directions is given by
\begin{align}
    \chi_a  &=  \frac{g^2 \mu_B^2}{k_BT} \Delta^2 \left( M_\mathrm{tot}^a \right) ,\label{eq:susceptibility:variance}
\end{align}
where $M_\mathrm{tot}^a=\sum_{i=1}^N S_i^a$ is the total magnetization along $\hat{a}$, $g$ is the Land\'e $g$-factor, $\mu_B$ the Bohr magneton, $k_B$ the Boltzmann constant, and $\Delta^2 \left( M \right) = \left\langle M^2\right\rangle-\left\langle M\right\rangle^2$ denotes variance.

A key observation is that, in a product state, the variance of the magnetization equals the sum of variances of individual spin operators. (This is analogous to the statement in probability theory that the variance of a sum of \emph{independent} random variables equals the sum of variances of the individual variables.) Assuming the $g$-factor is isotropic, the average $\bar{\chi}=\left( \chi_x + \chi_y + \chi_z\right)/3$ then satisfies
\begin{align}
    \bar{\chi}  &=  \frac{g^2 \mu_B^2}{3k_BT} \left[ \Delta^2 \left( M_\mathrm{tot}^x \right) + \Delta^2 \left( M_\mathrm{tot}^y \right) + \Delta^2 \left( M_\mathrm{tot}^z \right) \right]  \\
                &=  \frac{g^2 \mu_B^2}{3k_BT}    \sum_{i=1}^N    \left[ \Delta^2 \left( S_i^x\right) + \Delta^2 \left( S_i^y\right) + \Delta^2 \left( S_i^z\right)\right]    \label{eq:susceptibility:pure:step}\\
                &=  \frac{g^2 \mu_B^2}{3k_BT}    \sum_{i=1}^N    \left[ \langle \mathbf{S}_i\cdot\mathbf{S}_i\rangle - \langle S_i^x\rangle^2 - \langle S_i^y\rangle^2 - \langle S_i^z\rangle^2 \right]  \\
                &\geq   \frac{g^2 \mu_B^2 N \hbar^2 S}{3k_BT}
\end{align}
since $\langle \mathbf{S}_i\cdot\mathbf{S}_i\rangle=\hbar^2 S(S+1)$ and $\sum_a \langle S_i^a\rangle^2 \leq \hbar^2 S^2$. This generalizes to a generic separable state $\rho$, which can be viewed as a convex mixture (or weighted mean) of $n$ product states, each with probability $p_n\geq 0$ ($\sum_n p_n=1$), such that the density matrix is $\rho=\sum_n p_n \otimes_{i=1}^N \rho_n^i$. (There always exists an ensemble of states for which $\rho$ can be decomposed in this fashion.) The derivation proceeds as before, but due to the convexity, step \eqref{eq:susceptibility:pure:step} is replaced by \cite{PhysRevA.68.032103}
\begin{align}
    \bar{\chi}  & {\geq}  \frac{g^2 \mu_B^2}{3k_BT}    \sum_n p_n \sum_{i=1}^N    \left[ \Delta^2 \left( S_i^x\right)_n + \Delta^2 \left( S_i^y\right)_n + \Delta^2 \left( S_i^z\right)_n\right]  \nonumber  \\
                &\geq \frac{g^2 \mu_B^2 N \hbar^2S}{3k_BT}
\end{align}
where $\Delta^2 \left( S_i^a\right)_n$ is the variance taken in state $\rho_n$. It thus follows that the system must be entangled if the entanglement witness associated with the susceptibility, $\chi_\mathrm{EW}$, satisfies 
\begin{align}
    \chi_\mathrm{EW}\equiv \bar{\chi}  &\leq   \frac{g^2 \mu_B^2 N \hbar^2S}{3k_BT}.  \label{eq:suspectibility:witness}
\end{align}
for the case in point.
A less general, but physically transparent derivation of the same bound was obtained for dimerized magnets \cite{PhysRevA.73.012110}. 
We note that both derivations assume an isotropic $g$-factor, which strongly constrains which classes of magnetic materials this witness is suitable for. We also note that the bound can be understood as a generalized spin squeezing inequality \cite{Guehne2009, PhysRevA.79.042334}.

\subsection{One-tangle}
The following will use a notation matching Ref.~\cite{Fubini2006}. We consider a system of $N$ spin-$1/2$ degrees of freedom 
(or, equivalently, qubits) and single out a single spin at site $j_0$, placing it in region $A$; see Fig.~\ref{fig:types}(c). Region $B$ describes the rest of the system, such that the many-body Hilbert space is given by $\mathcal{A}\otimes\mathcal{B}$. The basis for the one-site Hilbert space is again $\{\left|\uparrow\right\rangle\, \left|\downarrow\right\}$. A pure state of the system can be written
\begin{align}
	\left| \psi \right\rangle	&=	\sum_{\nu\in\mathcal{A}} \left| \nu \right\rangle \otimes \sum_{\Gamma\in\mathcal{B}} c_{\nu\Gamma} \left| \Gamma\right\rangle.
\end{align}
The probability for our spin to be in the state $|\nu\rangle$ is $p_\nu = \sum_\Gamma \left| c_{\nu\Gamma}\right|^2$. The normalization condition $\left\langle \psi \middle| \psi \right\rangle=1$ correctly implies $\sum_\nu p_\nu=1$. The pure state $|\psi\rangle$ corresponds to the density matrix
\begin{align}
	\rho	&=	\left| \psi\right\rangle\left\langle \psi\right| \\
			&=	\sum_{\nu_1\in\mathcal{A}} \left| \nu_1 \right\rangle \otimes \sum_{\Gamma_1\in\mathcal{B}} c_{\nu_1\Gamma_1} \left| \Gamma_1\right\rangle \sum_{\nu_2\in\mathcal{A}} \left\langle \nu_2 \right|\otimes  \sum_{\Gamma_2\in\mathcal{B}} c^\ast_{\nu_2\Gamma_2} \left\langle \Gamma_2\right\rangle.
\end{align}
The one-site reduced density matrix $\rho^{(1)}$ is obtained by tracing out all degrees of freedom in $\mathcal{B}$, i.e.
\begin{align}
	\rho^{(1)}	&=	\sum_{\Gamma\in\mathcal{B}}	\left\langle \Gamma \middle| \psi \middle\rangle \middle\langle \psi \middle| \Gamma \right\rangle =	\sum_{\nu_1,\nu_2\in\mathcal{A}}	\left| \nu_1 \middle\rangle\middle\langle \nu_2 \right|	\sum_{\Gamma\in\mathcal{B}} c_{\nu_1\Gamma} c^\ast_{\nu_2\Gamma}.	\label{eq:rhoonesitereduced}
\end{align}
We may now use that a generic $2\times 2$ matrix $M$ can be decomposed in terms of the identity and Pauli matrices through
\begin{align}
    M   &=  \sum_\mu a_\mu \sigma^\mu,\quad \mu \in \{ 0, x, y, z\},
\end{align}
where $\sigma^0=I$ and $a_\mu=\mathrm{Tr}\left[ \sigma^\mu M \right]/2$. $a_\mu\in\mathbb{R}$ if $M$ is Hermitian, otherwise $a_\mu \in \mathbb{C}$ in general. Since $\rho^{(1)}$ contains all information about the subsystem $\mathcal{A}$, these traces can be interpreted as measurements $\langle \sigma_{j_0}^\mu \rangle$. It follows that 
\begin{align}
	\rho^{(1)}	&=		\begin{pmatrix}
		\frac{1}{2} + \left\langle S_{j_0}^z\right\rangle & \left\langle S_{j_0}^x\right\rangle - i \left\langle S_{j_0}^y\right\rangle \\
		\left\langle S_{j_0}^x\right\rangle + i \left\langle S_{j_0}^y\right\rangle & 	\frac{1}{2} - \left\langle S_{j_0}^z\right\rangle
	\end{pmatrix}\\
							&=	\frac{\sigma^0}{2} + \sum_a \langle S_{j_0}^a\rangle \sigma^\mu,  \quad a\in \{ x,y,z\}
\end{align}
where $S_j^a = \sigma_j^a/2$ is a spin-$1/2$ operator.

The single-site von Neumann entanglement entropy is given by $S_\mathrm{vN}^{(1)}	=	-\mathrm{Tr}\left[ \rho^{(1)} \ln \rho^{(1)} \right].$ By using the Taylor series $\ln \left( 1+x\right)	=	x	-\frac{x^2}{2} + \frac{x^3}{3} + \dots$ and retaining only the first (linear) term, we obtain
\begin{align}
	S_\mathrm{vN}^{(1)}	&\approx	-\mathrm{Tr}\left[ \rho^{(1)} \left( \rho^{(1)}-1\right) \right] = 1-\mathrm{Tr} \left[ \left(\rho^{(1)}\right)^2\right],
\end{align}
which is proportional to the \emph{linear entropy} \cite{PhysRevA.67.022110}
\begin{align}
	S_L	[\rho^{(1)}]	&=	\frac{d}{d-1} \left\{ 1 - \mathrm{Tr} \left[ \left( \rho^{(1)} \right)^2 \right]\right\},
\end{align}
where $d$ is the dimension of $\rho$ (here, $d=2$). This quantity is a measure of mixedness in quantum states, here normalized to lie in the range $[0,1]$, with $0$ for completely pure states, and $1$ for completely mixed states. (Conversely, $\gamma=\mathrm{Tr}\left[\rho^2\right]$ is known as the purity.) The advantage of the linear entropy over the regular entanglement entropy is that it can be computed without diagonalizing the reduced density matrix. 
The one-tangle \cite{PhysRevA.61.052306, PhysRevLett.96.220503} is now given by
\begin{align}
	\tau_1	&=	S_L \left[ \rho^{(1)}\right] =	4\mathrm{det}\, \rho^{(1)}	= 1-4\sum_\mu \langle S_{j_0}^\mu \rangle^2.
 \label{eq:one-tangle}
\end{align}
One should keep in mind that $j_0$ is a single site. If it is fully classically ordered, e.g. $\langle S_{j_0}^z\rangle=1/2$, $\tau_1$ vanishes, indicating a pure state. If it is entirely quantum disordered, i.e. $\langle S_{j_0}^\mu\rangle=0$, $\tau_1=1$, indicating that our single site is maximally mixed with the rest of the system (assuming nonzero interactions). 
In translation-invariant systems, the expectation values at site $j_0$ can be replaced with appropriate averages throughout the system (taking ordering vectors into account). Then $\tau_1$ can be experimentally accessed, e.g., through measurements of Bragg peaks. 

It is also important to remember that we assumed the state $|\psi\rangle$ of the entire system to be a pure state. Strictly speaking, this construction is thus only valid at $T=0$, although it is qualitatively useful also at low temperatures. It is possible to construct a one-tangle also for mixed states \cite{PhysRevLett.96.220503}, however the resulting expressions involves an optimization over all possible pure state decompositions, that has yet to be turned into a useful entanglement witness.

\subsection{Concurrence and two-tangle}\label{sec:concurrence-two-tangle}
The two-site reduced density matrix is obtained analogously. We choose two spin-$1/2$ degrees of freedom at sites $i$ and $j$ to make up region $A$, and represent the rest of the system by $B$; see Fig.~\ref{fig:types}(d). A variety of bases for the two-site Hilbert space $\mathcal{A}$ have been considered in the literature \cite{Fubini2006, PhysRevB.102.064409}, but we will focus on the so-called standard basis $|\nu\rangle \in \{ |\uparrow\uparrow\rangle,\, |\uparrow\downarrow\rangle,\,|\downarrow\uparrow\rangle,\, |\downarrow\downarrow\rangle\}$. The reduced density matrix $\rho^{(2)}_{ij}$ is again given by Eq.~\eqref{eq:rhoonesitereduced}, but is now a $4\times 4$ matrix. Such matrices can be expressed
\begin{align}
	A	&=	\sum_{\mu}\sum_{\nu}	\frac{1}{4} \left( \sigma^\mu \otimes \sigma^\nu \right) \left\langle \sigma^\mu \sigma^\nu \right\rangle,  \label{eq:representation:fourbyfour}
\end{align}
where $a_{\mu\nu}	=	\frac{1}{4}\mathrm{Tr}\left[ \left( \sigma^\mu \otimes \sigma^\nu \right) A\right]$. Relating the matrix elements to measurements of different operators is straightforward, and the result in the general case can be found in Ref.~\cite{PhysRevA.68.060301}. In the following we will assume that parity (or $\mathbb{Z}_2$) symmetry is present, i.e. that the magnetization along $\hat{z}$ has to stay constant or change in steps of $2$. In other words, the Hamiltonian lacks terms such as $S^z S^+$ and magnetic fields along $\hat{x}$ or $\hat{y}$. Under this assumption one obtains the reduced density matrix \scite{PhysRevA.69.022304}
\begin{align}
	\rho^{(2)}_\mathrm{ij}	&=		\begin{pmatrix}
		a		&	0		& 0		& c \\
		0		&	x		& z		& 0\\
		0		&	z^\ast	& y		& 0\\
		c^\ast	&	0		& 0		& b
	\end{pmatrix}\label{eq:rho2:z2}
\end{align}
where
\begin{align}
	a	&=	\frac{1}{4} + M^z_{ij} + g^{zz}_{ij},\\
	b	&=	\frac{1}{4} - M^z_{ij} + g^{zz}_{ij},\\
	x	&=	\frac{1}{4} + \delta S^z_{ij} - g^{zz}_{ij},\\
	y	&=	\frac{1}{4} - \delta S^z_{ij} - g^{zz}_{ij},\\
	c	&=	g^{xx}_{ij} - g^{yy}_{ij} - i \left( g^{xy}_{ij} + g^{yx}_{ij} \right),\\
	z	&=	g^{xx}_{ij} + g^{yy}_{ij} + i \left( g^{xy}_{ij} - g^{yx}_{ij} \right),
\end{align}
and where we have defined
\begin{align}
	g^{\alpha\beta}_{ij}	&=	\left\langle S_i^\alpha S_j^\beta\right\rangle = \frac{1}{4} \left\langle \sigma_i^\alpha \sigma_j^\beta\right\rangle,\\
	M^z_{ij}	&=	\frac{1}{2} \left( \langle S_i^z \rangle + \langle S_j^z \rangle \right)	=	\frac{1}{4} \left( \langle \sigma_i^z \rangle + \langle \sigma_j^z \rangle \right),\\
	\delta S^z_{ij}	&=	\frac{1}{2} \left( \langle S_i^z \rangle - \langle S_j^z \rangle \right)	=	\frac{1}{4} \left( \langle \sigma_i^z \rangle - \langle \sigma_j^z \rangle \right).
\end{align}
The presence of additional symmetries can constrain Eq.~\eqref{eq:rho2:z2} further. In particular, translational invariance enforces $\delta S^z_{ij} =0$ and thus $x=y$. If the Hamiltonian is real, for example if there is no Dzyaloshinski-Moriya interaction, $g^{xy}_{ij}=g^{yx}_{ij}$, and $c,z\in \mathbb{R}$. In the presence of $U(1)$ symmetry, as in the XXZ model, $g^{xx}_{ij}=g^{yy}_{ij}$ and $c=0$.

Wootters \scite{PhysRevLett.80.2245}, extending work by Hill and Wootters \scite{PhysRevLett.78.5022}, proved that the entanglement of formation, Eq. \eqref{eq:formation}, of a two-site reduced density matrix in any mixed state satisfies
\begin{align}
    E_\mathrm{form} \left( \rho\right)  &=  h \left( \frac{1+\sqrt{1-C^2\left( \rho\right)}}{2} \right) \label{eq:entform:concurrence}
\end{align}
where $h(x)=-x\log x - \left( 1-x\right) \log \left( 1-x\right)$. $C$ is the concurrence, which is defined as
\begin{align}
    C\left( \rho \right)    &=  \max \left\{ 0, 2\lambda_\mathrm{max} - \mathrm{Tr} \left[ R\right] \right\},
    \label{eq:concurrenceformula}
\end{align}
where $\lambda_\mathrm{max}$ is the largest eigenvalue of $R=\sqrt{ \rho^{(2)}_{ij} \widetilde{\rho^{(2)}_{ij}}}$, and
\begin{align}
	\widetilde{\rho^{(2)}_{ij}}	&=	\left( \sigma^y \otimes \sigma^y \right) \left( \rho^{(2)}_{ij}\right)^\ast \left( \sigma^y \otimes \sigma^y \right)
\end{align}
is the time-reversed copy of $\rho^{(2)}_{ij}$. (This is just the action of the standard time-reversal operator for two spin-$1/2$ degrees of freedom \scite{Sakurai1994}.)

Both ${\rho^{(2)}_{ij}}$ and $\widetilde{\rho^{(2)}_{ij}}$ are positive semi-definite Hermitian matrices (their eigenvalues represent probabilities), but $\rho^{(2)}_{ij} \widetilde{\rho^{(2)}_{ij}}$ is generally non-Hermitian. We may use that the matrix square root $\sqrt{{\rho^{(2)}_{ij}}}$ is also a positive semi-definite Hermitian matrix, write $\rho^{(2)}_{ij} \widetilde{\rho^{(2)}_{ij}}=\sqrt{\rho^{(2)}_{ij}} \sqrt{\rho^{(2)}_{ij}} \widetilde{\rho^{(2)}_{ij}}$ and exploit that $\sqrt{\rho^{(2)}_{ij}} \sqrt{\rho^{(2)}_{ij}} \widetilde{\rho^{(2)}_{ij}}$ has the same eigenvalues as $\sqrt{\rho^{(2)}_{ij}} \widetilde{\rho^{(2)}_{ij}} \sqrt{\rho^{(2)}_{ij}}$, which is Hermitian. Then, for all nonzero vectors $v$ we have
\begin{align}
	v^\dagger	\sqrt{\rho^{(2)}_{ij}}\widetilde{\rho^{(2)}_{ij}}\sqrt{\rho^{(2)}_{ij}} v	&=	\left( \sqrt{\rho^{(2)}_{ij}} v\right)^\dagger \widetilde{\rho^{(2)}_{ij}} \left(\sqrt{\rho^{(2)}_{ij}} v\right)	\geq 0
\end{align}
guaranteeing that the eigenvalues of $R$ are non-negative. Alternatively, the eigenvalues of $R$ can be obtained by taking square roots of the eigenvalues of $R^2=\rho^{(2)}_{ij} \widetilde{\rho^{(2)}_{ij}}$.

Here we find that the eigenvalues of $R^2$ are (in no particular order)
\begin{align}
	\lambda_1	&=	\left| \sqrt{ab}-|c|\right|,  \label{eq:lambda1}\\
	\lambda_2	&=	\left| \sqrt{ab}+|c|\right| = \sqrt{ab}+|c|,\\
	\lambda_3	&=	\left| \sqrt{xy}-|z|\right|,  \label{eq:lambda3}\\
	\lambda_4	&=	\left| \sqrt{xy}+|z|\right| = \sqrt{xy} + |z|.
\end{align}
We note that it is easy to miss the absolute value signs on $c$ and $z$ if these quantities are not formally treated as complex throughout. $\lambda_2\geq \lambda_1$ and $\lambda_4\geq \lambda_3$, so we have two candidates for $\lambda_\mathrm{max}$ to consider for Eq. \eqref{eq:concurrenceformula}. The constraint that the eigenvalues of $\rho^{(2)}_{ij}$ are non-negative implies $\sqrt{ab}	\geq |c|,$ and $\sqrt{xy}	\geq |z|,$ and the concurrence becomes
\begin{align}
    C	&=	2\max \left\{ 0, |c| - \sqrt{xy}, |z|-\sqrt{ab} \right\}.
    \label{eq:concur}
\end{align}
We now expand this result out for some useful cases, following Ref.~\cite{PhysRevA.69.022304}. 
In the general parity-symmetric case,
\begin{align}
	C	&=	2\max \left\{ 0, \sqrt{\left( g^{xx}_{ij} - g^{yy}_{ij} \right)^2 + \left( g^{xy}_{ij}+ g^{yx}_{ij} \right)^2}  \right.\nonumber\\
		&\left. - \sqrt{\left(\frac{1}{4} - g^{zz}_{ij}\right)^2 - \left(\delta S^z_{ij}\right)^2}, \sqrt{\left( g^{xx}_{ij} + g^{yy}_{ij} \right)^2 + \left( g^{xy}_{ij} - g^{yx}_{ij} \right)^2} \right.\nonumber\\
		&\left. -\sqrt{\left( \frac{1}{4} + g^{zz}_{ij}\right)^2 -\left(M^z_{ij}\right)^2 }\right\}.
\end{align}
Requiring translational invariance and a real Hamiltonian, this simplifies to
\begin{align}
	C	&=	2\max \left\{ 0, \left| g^{xx}_{ij} - g^{yy}_{ij} \right| - \frac{1}{4} + g^{zz}_{ij},  \right.\nonumber\\
	&\left. \left| g^{xx}_{ij} + g^{yy}_{ij} \right| -\sqrt{\left( \frac{1}{4} + g^{zz}_{ij}\right)^2 -\left(M^z_{ij}\right)^2 }\right\},
\end{align}
where $g^{\alpha\beta}_{ij}$ only depends on the distance between $i$ and $j$. 
In the case of the isotropic $S=1/2$ Heisenberg spin model, this simplifies further to
\begin{align}
	C	&=	2\max \left\{ 0, 2\left| g^{zz}_{ij}\right|- \sqrt{\left( \frac{1}{4} + g^{zz}_{ij}\right)^2 -\left(M^z_{ij}\right)^2 }\right\}.
\end{align}
In the absence of order,
\begin{align}
	C	&=	2\max \left\{ 0,\, 2\left| g^{zz}_{ij}\right|- \frac{1}{4} - g^{zz}_{ij}\right\}.
\end{align}
If $g^{zz}_{ij}<0$, 
\begin{align}
	C	&=	2\max \left\{ 0,\, -3 g^{zz}_{ij}- \frac{1}{4}\right\},
\end{align}
which means the concurrence for antiferromagnetic Heis\-enberg dimers (see Sec.~\ref{sec:applications:dimers}) is given by
\begin{align}
	C	&=	2\max \left\{ 0, -\langle \mathbf{S}_0 \cdot \mathbf{S}_{d_1}\rangle -\frac{1}{4} \right\},  \label{eq:concurrence:dimer}
\end{align}
where $\langle \mathbf{S}_0 \cdot \mathbf{S}_{d_1}\rangle$ is the intradimer spin-spin correlation function.

As Equation~\eqref{eq:entform:concurrence} shows, the entanglement of formation increases monotonically with $0\leq C\leq 1$. Thus the concurrence is a proper entanglement measure that can be experimentally obtained through the measurement of appropriate spin correlation functions and order parameters. 
Alternatively, for sufficiently symmetric Hamiltonians, the concurrence can be expressed in terms of other thermodynamic quantities, such as magnetic susceptibility \cite{PhysRevB.77.104402} and the internal energy \cite{PhysRevA.63.052302, Wang2002}. 
Finally, one can define the two-tangle as the square of all concurrences, 
\begin{equation}
    \tau_2  =   \sum_{i\neq j}  C^2_{ij}.
    \label{eq:two-tangle}
\end{equation}
(In the case of one-dimensional translation-invariant systems this is commonly written $\tau_2=\sum_{r>0} C_r^2$ where $r=\left| j-i\right|$ is the separation between the sites.) 
It was conjectured 
\cite{PhysRevA.61.052306} 
and later proved 
\cite{PhysRevLett.96.220503} that $\tau_2 \leq \tau_1$. This inequality reflects a limit to how strong pairwise entanglement can be in a system, a trade-off known as monogamy of entanglement \cite{PhysRevA.61.052306, Terhal_2004}. Interestingly, strongly correlated ground states in condensed matter systems can be linked to monogamy \cite{PhysRevLett.96.220503, Scheie2023}. 
The ratio $\tau_2/\tau_1$, sometimes called the entanglement ratio, has been used as an estimate for the fraction of the total entanglement that is stored in pairwise entanglement at $T=0$ \cite{PhysRevLett.93.167203, PhysRevA.74.022322}. 
Due to monogamy, the concurrence generically decays quickly with distance and number of neighbors \cite{RevModPhys.80.517, Syljuaasen_2004, PhysRevA.69.062314, PhysRevLett.93.167203, PhysRevLett.94.147208, Fubini2006, PhysRevA.74.022322, Baroni_2007}. It is thus expected to be most useful for systems where the entanglement is encoded in strong short-range correlations, such as dimerized magnets and spin clusters. In contrast, as discussed in Ref.~\cite{PhysRevB.103.224434} concurrence is much less powerful in quantum spin liquids, and has been found to vanish for all pairs of spin sites in the exactly solvable Kitaev spin liquid \cite{PhysRevLett.98.247201}.

\subsection{Two-site quantum discord}
The certification of genuine quantum correlations that go beyond entanglement is also of interest. The type of such quantum correlations that has received the most attention is known as the (bipartite) quantum discord \cite{PhysRevLett.88.017901, LHenderson_2001}. We review the general formulation of this quantity in Sec.~\ref{sec:discord:general} and then specialize to two-site quantum discord (two-site QD) in Sec.~\ref{sec:discord:twosite}. Analytical expressions relating the quantum discord of XYZ spin systems in the absence of order are given in Sec.~\ref{sec:discord:heisenberg}. The important special case of isotropic Heisenberg exchange is addressed explicitly.

\subsubsection{General formulation}\label{sec:discord:general}
To formulate the quantum discord, one considers two \emph{classically} equivalent ways of writing the mutual information between two probability distributions. The quantum discord is then defined as the difference between their quantum generalizations; see Refs.~\cite{RevModPhys.84.1655, Chiara2018} for excellent reviews. Nonzero discord thus indicates the inequivalence of two expressions that would be equal in a classical state, and, accordingly, implies the presence of nonclassical correlations. We note that separable (i.e. non-entangled) states can have finite discord, and that it has been shown that typical states in an arbitrary Hilbert space have nonzero quantum discord \cite{PhysRevA.81.052318}.

Here we will proceed directly to a suitable definition, following Refs.~\cite{PhysRevA.77.042303, PhysRevB.78.224413, PhysRevA.80.022108, PhysRevA.81.042105, PhysRevA.82.069902}. 
We consider a bipartition of a system described by density matrix $\rho$ into two subsystems described by reduced density matrices $\rho_A=\mathrm{Tr}_B \left[ \rho\right]$ and $\rho_B=\mathrm{Tr}_A \left[ \rho\right]$, respectively. The quantum mutual information is defined
\begin{align}
	I	\left( \rho\right)	&=	S\left( \rho_A\right) + S\left( \rho_B\right)	-S\left( \rho\right)	\label{eq:qmi:definition}\\
							&=	S\left( \rho_A\right) - S\left( \rho \middle| \rho_B\right)	\label{eq:qmi:one}
\end{align}
where $S\left( \rho\right)= -\mathrm{Tr} \left[ \rho^A \log \rho^A \right]$ is the von Neumann entropy of $\rho^A$, and where $S\left( \rho \middle| \rho_B\right)	=	S\left( \rho \right) -S\left(\rho_B\right)$ 
is a quantum generalization of the conditional entropy. The quantum mutual information $I\left( \rho\right)$ can be viewed as a measure of the total correlations between subsystems $A$ and $B$, and is always non-negative. (Note also that $I\left( \rho\right)$ is \emph{not} an entanglement measure; a mixture of separable states is not entangled but can have nonzero quantum mutual information.)

Alternatively, the conditional entropy can be generalized using measurement operations performed only on $B$. Let $\{B_k\}$ be a set of one-dimensional projection operators (each $B_k$ projecting onto a single outcome $k$). If measurement outcome $k$ is obtained, the state becomes
\begin{align}
	\rho_k	&=	\frac{1}{p_k} \left( I \otimes B_k \right) \rho \left( I \otimes B_k\right)  \label{eq:discord:measurement:rhok}
\end{align}
where $p_k = \mathrm{Tr}\left[ \left( I \otimes B_k \right) \rho \left(I \otimes B_k \right)\right]$, and $I$ is the identity operator acting only on $A$. Then the conditional entropy can be expressed
\begin{align}
	S \left( \rho \middle| \left\{ B_k \right\} \right)	&=	\sum_k p_k S\left( \rho_k \right),  \label{eq:discord:conditionalentropy}
\end{align}
in which case the quantum mutual information can be defined using the alternative expression
\begin{align}
	J	\left( \rho\middle| \{ B_k\}\right)	&= S\left( \rho_A\right) - S \left( \rho \middle| \{ B_k \}\right).	\label{eq:qmi:two}
\end{align}
Any difference between Eqs.~\eqref{eq:qmi:one} and \eqref{eq:qmi:two} has to be due to quantum effects on the correlation between $A$ and $B$!

To make the definition independent of the specific choice of measurement $\{ B_k\}$ one defines
\begin{align}
	C	\left( \rho\right)	&=	\sup_{\{ B_k \}} J \left( \rho\middle| \{ B_k\}\right),	\label{eq:discord:classicalcorr}
\end{align}
which is a measure of the classical correlation between $A$ and $B$ \cite{PhysRevLett.88.017901, LHenderson_2001}. The supremum is taken over all sets possible choices of $\{ B_k\}$, making the procedure and measure general. The quantum discord is then defined
\begin{align}
	Q\left( \rho\right)		&=	I \left( \rho\right) - C\left( \rho\right),    \label{eq:discord:definition}
\end{align}
and becomes a measure of genuine quantum correlations. The discord is bounded from above by the entanglement entropy of the measured subsystem, $Q\left( \rho\right)	\leq S\left( \rho_B\right)$ \cite{PhysRevA.84.042124}.

\subsubsection{Two-site formulation}\label{sec:discord:twosite}

In general, the optimization operation in Eq.~\eqref{eq:discord:classicalcorr} is complicated and often needs to be carried out numerically. However, analytical progress can be made by restricting the discussion to quantum discord between only two sites in a many-body system, and by considering sufficiently symmetric Hamiltonians. We first obtain the two-site reduced density matrix by tracing out other parts of the system, and assume a real-valued Hamiltonian, in which case we obtain from Eq.~\eqref{eq:rho2:z2}
\begin{align}
\rho^{(2)}_\mathrm{ij}	&=		\begin{pmatrix}
a		&	0		& 0		& c \\
0		&	x		& z		& 0\\
0		&	z		& y		& 0\\
c		&	0		& 0		& b
\end{pmatrix}\label{eq:rho2:z2:real}
\end{align}
Now let us decompose this two-site system into parts $A$ and $B$ that each describe a single site. In other words, $\rho^A$ and $\rho^B$ will be one-site reduced density matrices, and $\rho=\rho^{(2)}_\mathrm{ij}$. It is convenient to rewrite $\rho_{ij}^{(2)}$ using the representation Eq.~\eqref{eq:representation:fourbyfour}. One obtains \cite{PhysRevA.80.022108}
\begin{align}
	\rho^{(2)}_\mathrm{ij}	&=	\frac{1}{4} \left[ I\otimes I + \sum_{i=1}^3 c_i \sigma^i \otimes \sigma^i + c_4 I\otimes \sigma^3 + c_5 \sigma_3 \otimes I \right],	\label{eq:rho:fordiscord}
\end{align}
where
\begin{align}
	c_1	&=	2z+2c,\\
	c_2	&=	2z-2c,\\
	c_3	&=	a+b-x-y,\\
	c_4	&=	a-b-x+y,\\
	c_5	&=	a-b+x-y.
\end{align}
(The coefficient of $I\otimes I$ satisfies $c_0=a+b+x+y=1$.)
The eigenvalues are then
\begin{align}
	\lambda_1	&=	\frac{1}{4} \left[ 1+c_3 + \sqrt{\left( c_1-c_2\right)^2 + \left(c_4+c_5\right)^2}\right],	\label{eq:eig1:rho:sarandyform}\\
	\lambda_2	&=	\frac{1}{4} \left[ 1+c_3 - \sqrt{\left( c_1-c_2\right)^2 + \left(c_4+c_5\right)^2}\right],	\label{eq:eig2:rho:sarandyform}	\\
	\lambda_3	&=	\frac{1}{4} \left[ 1-c_3 + \sqrt{\left( c_1+c_2\right)^2 + \left(c_4-c_5\right)^2}\right],	\label{eq:eig3:rho:sarandyform}	\\
	\lambda_4	&=	\frac{1}{4} \left[ 1-c_3 - \sqrt{\left( c_1+c_2\right)^2 + \left(c_4-c_5\right)^2}\right].	\label{eq:eig4:rho:sarandyform}
\end{align}
We also obtain
\begin{align}
	\rho_A	&=	\mathrm{Tr}_B \left[ \rho \right] = \left\langle \uparrow \Biggm|_B\, \rho_{ij}^{(2)} \middle| \uparrow\right\rangle_B + \left\langle \downarrow \Biggm|_B\, \rho_{ij}^{(2)} \middle| \downarrow\right\rangle_B\\
			&=	\frac{1}{2} \begin{pmatrix}	1 + c_5	&	0\\
			0	&	1-c_5\end{pmatrix},\\
	\rho_B	&=	\frac{1}{2} \begin{pmatrix}	1 + c_4	&	0\\
	0	&	1-c_4\end{pmatrix},
\end{align}
which have the eigenvalues
\begin{align}
	r_1^A	&=	\frac{1+c_5}{2}, \quad
	r_2^A	=	\frac{1-c_5}{2},\\
	r_1^B	&=	\frac{1+c_4}{2},\quad
	r_2^B	=	\frac{1-c_4}{2},
\end{align}
leading to the one-site entanglement entropies
\begin{align}
	S\left( \rho_A\right)	&=	- \left( r_1^A \log r_1^A + r_2^A \log r_2^A \right),\label{eq:discord:onesiteentropy:A}\\
	S\left( \rho_B\right)	&=	- \left( r_1^B \log r_1^B + r_2^B \log r_2^B \right).
\end{align}
We also have that $S\left( \rho_{ij}^{(2)}\right) = -\sum_{i=1}^4 \lambda_i \log \lambda_i$, with $\lambda_i$ as given in Eqs.~	\eqref{eq:eig1:rho:sarandyform}--\eqref{eq:eig4:rho:sarandyform}. These entropies are sufficient to calculate the quantum mutual information \eqref{eq:qmi:definition}.

Turning to the classical correlations, we next need to parametrize our measurements $\{ B_k\}$. Let
\begin{align}
	\Pi_k	&=	\left| k\middle\rangle\middle\langle k\right|,	\quad	k\in \{0,1\}
\end{align}
be a projector acting on subsystem $B$ along the basis element $\left| k\right\rangle$. Then any von Neumann measurement operator can be expressed as
\begin{align}
	B_k	&=	V\Pi_k V^\dagger,
\end{align}
where $V$ is a general SU(2) matrix,
\begin{align}
	V	&=	tI	+	i\vec{y} \cdot \vec{\sigma},
\end{align}
and where $t\in \mathbb{R}$, $\vec{y}=\left( y_1,y_2,y_3\right) \in \mathbb{R}^3$ and $t^2+\vec{y}\cdot \vec{y}=1$. (This implies $t\in \left[ -1,+1\right]$, $y_i\in \left[ -1,+1\right]$, $i\in\{1,2,3\}$.) As stated before, after a measurement on $B$ the state $\rho_{ij}^{(2)}$ changes into an ensemble $\{ \rho_k, p_k\}$; see Eq.~\eqref{eq:discord:measurement:rhok}. It is convenient to write
\begin{align}
	p_k	\rho_k	&=	\left( I\otimes B_k\right) \rho_{ij}^{(2)} \left( I\otimes B_k\right)	\\
				&=	\left[ I \otimes \left( V\Pi_kV^\dagger\right)\right] \rho_{ij}^{(2)} \left[ I \otimes \left( V\Pi_k V^\dagger\right)\right].
\end{align}
This product can be straightforwardly evaluated using the mixed-product property of the Kronecker product for square matrices $A,B,C,D,$
\begin{align}
	\left( A\otimes B\right) \left( C\otimes D\right)	&=	AC\otimes BD,
\end{align}
and the useful relations
\begin{align}
	V^\dagger \sigma^1 V	&=	\left( t^2 +y_1^2 -y_2^2 - y_3^2\right) \sigma^1 + 2\left( ty_3 +y_1y_2\right) \sigma^2\nonumber\\
    &+ 2\left( -ty_2 +y_1y_3\right) \sigma^3,	\label{eq:discordidentity:sigmax}\\
	V^\dagger \sigma^2 V	&=	2\left( -ty_3 +y_1y_2\right) \sigma^1 + \left( t^2 +y_2^2 -y_1^2 - y_3^2\right) \sigma^2 \nonumber\\
	&+ 2\left( ty_1 +y_2y_3\right) \sigma^3,\\
	V^\dagger \sigma^3 V	&=	2\left( ty_2 +y_1y_3\right) \sigma^1 + 2\left( -ty_1 +y_2y_3\right) \sigma^2  \nonumber\\
	&+ \left( t^2 +y_3^2 -y_1^2 - y_2^2\right) \sigma^3,
\end{align}
and $\Pi_0	\sigma^3 \Pi_0	=	\Pi_0,$	$\Pi_1	\sigma^3 \Pi_1	=	-\Pi_1,$	$\Pi_k \sigma^1 \Pi_k = \Pi_k \sigma^2 \Pi_k = 0 \quad \forall k,$ as well as $V^\dagger I V=I$ and $\Pi_k I \Pi_k = \Pi_k$. We will also define
\begin{align}
	z_1	&=	2 \left( -ty_2 + y_1 y_3\right),\\
	z_2	&=	2\left( ty_1 + y_2 y_3\right),\\
	z_3	&=	t^2 + y_3^2 - y_1^2 - y_2^2.
\end{align}
One obtains
\begin{align}
	p_0 \rho_0	&=	\frac{1}{4} \left( I + c_1 z_1 \sigma^1 + c_2 z_2 \sigma^2 + c_3 z_3 \sigma^3 \right.\nonumber\\
				&\left.  + c_4 z_3 I + c_5 \sigma^3 \right) \otimes V\Pi_0 V^\dagger,	\\
	p_1 \rho_1	&=	\frac{1}{4} \left( I - c_1 z_1 \sigma^1 - c_2 z_2 \sigma^2 - c_3 z_3 \sigma^3 \right.\nonumber\\
	&\left.  - c_4 z_3 I + c_5 \sigma^3 \right) \otimes V\Pi_1 V^\dagger.
\end{align}
Using $\mathrm{Tr} \left[ A\otimes B\right]	=	\mathrm{Tr} \left[ A\right] \mathrm{Tr}\left[ B\right],$ and $\mathrm{Tr}\left[ V\Pi_0 V^\dagger \right] = \mathrm{Tr}\left[ V\Pi_1 V^\dagger\right]	=	t^2 +y_1^2 + y_2^2 + y_3^2 = 1,$
one finds
\begin{align}
	p_0	&=	\mathrm{Tr}\left[ p_0 \rho_0\right] = \frac{1}{2} \left( 1+ c_4 z_3 \right),\\
	p_1 &=	\mathrm{Tr}\left[ p_1 \rho_1\right] = \frac{1}{2} \left( 1- c_4 z_3 \right),
\end{align}
as the Pauli matrices are traceless and $\mathrm{Tr}\left[ I \right] = 2$. After collecting terms, the density matrices can be expressed
\begin{align}
\rho_0	&=	\frac{1}{2} \left\{ I + \frac{1}{1+ c_4 z_3} \left[ c_1 z_1 \sigma^1 + c_2 z_2 \sigma^2 + \left( c_3 z_3 + c_5 \right)\sigma^3 \right]\right\}\nonumber\\
& \otimes \left( V\Pi_0 V^\dagger\right),\\
\rho_1	&=	\frac{1}{2} \left\{ I + \frac{1}{1- c_4 z_3} \left[ -c_1 z_1 \sigma^1 - c_2 z_2 \sigma^2 - \left( c_3 z_3 - c_5 \right)\sigma^3 \right]\right\}\nonumber\\
& \otimes \left( V\Pi_1 V^\dagger\right).
\end{align}

\subsubsection{Expressions for XYZ and Heisenberg spin systems}\label{sec:discord:heisenberg}

We now specialize to the case $c_4=c_5=0$ relevant to spin systems in the absence of order (i.e. $M_{ij}^z=\delta S_{ij}^z=0$). Then we have $p_0=p_1=1/2$ and obtain that both $\rho_0$ and $\rho_1$ have eigenvalues $\left\{ 0,\, 0,\, \frac{1+\theta}{2},\, \frac{1-\theta}{2}\right\}$ with $\theta=\sqrt{\left| c_1 z_1\right|^2 + \left| c_2 z_2 \right|^2 + \left| c_3 z_3\right|^2}$. We get $S\left( \rho_0\right) = S\left( \rho_1\right)$ where
\begin{align}
	S\left( \rho_0\right) = &=	-\frac{1-\theta}{2} \log \frac{1-\theta}{2} - \frac{1+\theta}{2} \log \frac{1+\theta}{2},
\end{align}
and the conditional entropy with respect to $\{ B_k\}$, Eq.~\eqref{eq:discord:conditionalentropy}, becomes
\begin{align}
    S\left( \rho_{ij}^{(2)} \middle| \{ B_k \} \right)	
    &= -\frac{1-\theta}{2} \log \frac{1-\theta}{2} \nonumber\\
    &- \frac{1+\theta}{2} \log \frac{1+\theta}{2}.
\end{align}
It is convenient to work with binary base logarithms, for which the single-site entropy \eqref{eq:discord:onesiteentropy:A} simplifies to $S\left( \rho_A\right) = \log_2 2 = 1$ in units of bits. The quantum mutual information, calculated using the alternative definition Eq.~\eqref{eq:qmi:two}, is 
\begin{align}
	J	\left( \rho_{ij}^{(2)} \middle| \{ B_k\}\right)	
	&= \frac{1-\theta}{2} \log_2 \left( 1-\theta\right) \nonumber\\
        &+ \frac{1+\theta}{2} \log_2 \left( 1+\theta\right).
\end{align}

Now define
\begin{align}
	c_{ij}^\mathrm{max}	&=	\max \left\{ \left| c_1 \right|, \left| c_2 \right|, \left| c_3 \right| \right\}.
\end{align}
Then
\begin{align}
	0	\leq	\theta	&=	\sqrt{\left| c_1 z_1\right|^2 + \left| c_2 z_2 \right|^2 + \left| c_3 z_3\right|^2}	\nonumber\\
		&\leq \sqrt{\left(c_{ij}^\mathrm{max}\right)^2 \left( \left| z_1\right|^2 + \left| z_2\right|^2 + \left| z_3\right|^2\right)} = c_{ij}^\mathrm{max},
\end{align}
since $z_1^2+z_2^2+z_3^2=1$. This implies that the optimization to be done to evaluate the classical correlations simply corresponds to
\begin{align}
	\sup_{\{ B_k\}}\theta	&=	\sup_V \theta	= c_{ij}^\mathrm{max}.
\end{align}
In other words, the classical correlations become [Eq. \eqref{eq:discord:classicalcorr}]
\begin{align}
	C\left( \rho_{ij}^{(2)} \right)	
    &=	\frac{1-c_{ij}^\mathrm{max}}{2} \log_2 \left( 1-c_{ij}^\mathrm{max}\right) \nonumber\\
    &+ \frac{1+c_{ij}^\mathrm{max}}{2} \log_2 \left( 1+c_{ij}^\mathrm{max}\right).
\end{align}

Under our assumption of $c_4=c_5=0$, $c_1	=	4g^{xx}_{ij},$ $c_2	=	4g^{yy}_{ij},$ and $c_3	= 4g^{zz}_{ij},$ such that the quantum discord  \eqref{eq:discord:definition} is entirely expressible in terms of two-site spin-spin correlations. Explicitly, for XYZ spin systems,
\begin{align}
&Q^\mathrm{XYZ}_{ij} \left( \rho_{ij}^{(2)}\right)	=	I\left( \rho_{ij}^{(2)}\right) - C\left( \rho_{ij}^{(2)} \right)	\\
&=	\left( \frac{1}{4} - g^{xx}_{ij} - g^{yy}_{ij} - g^{zz}_{ij} \right) \log_2 \left( 1 - 4g^{xx}_{ij} - 4g^{yy}_{ij} - 4g^{zz}_{ij}\right) \nonumber\\
&+ \left( \frac{1}{4} - g^{xx}_{ij} + g^{yy}_{ij} + g^{zz}_{ij}\right) \log_2 \left( 1 - 4g^{xx}_{ij} + 4g^{yy}_{ij} + 4g^{zz}_{ij} \right) \nonumber\\
&+ \left( \frac{1}{4} + g^{xx}_{ij} - g^{yy}_{ij} + g^{zz}_{ij}\right) \log_2 \left( 1 + 4g^{xx}_{ij} - 4g^{yy}_{ij} + 4g^{zz}_{ij}\right)\nonumber\\
&+ \left( \frac{1}{4} + g^{xx}_{ij} + g^{yy}_{ij} - g^{zz}_{ij}\right) \log_2 \left( 1 + 4g^{xx}_{ij} + 4g^{yy}_{ij} - 4g^{zz}_{ij} \right)\nonumber\\
& - \frac{1-c_{ij}^\mathrm{max}}{2} \log_2 \left( 1 - c_{ij}^\mathrm{max}\right) \nonumber\\
& - \frac{1+c_{ij}^\mathrm{max}}{2} \log_2 \left( 1+ c_{ij}^\mathrm{max} \right),
\label{eq:discord:XYZ}
\end{align}
with $c_{ij}^\mathrm{max} = \max \left\{ \left| g_{ij}^{xx} \right|\, \left| g_{ij}^{yy} \right|\, \left| g_{ij}^{zz} \right| \right\}$.

The case of spin systems with isotropic Heisenberg interactions, including spin chains and dimers, is of particular experimental relevance. By symmetry, $g_{ij}^{xx}=g_{ij}^{yy}=g_{ij}^{zz}$. (Note that this holds for arbitrary range interactions.) Defining $G_{ij}=4g^{zz}_{ij},$
\begin{align}
    Q_{ij}^\mathrm{Heis}\left( \rho\right)	&=	\frac{1}{4} \left[ \left( 1- 3G_{ij} \right)\log_2 \left( 1-3G_{ij}\right) \right.\nonumber\\
    &\left.+ 3\left( 1 +G_{ij} \right) \log_2 \left( 1+G_{ij}\right) \right]\nonumber\\
    &- \frac{1}{2} \left[ \left( 1+ |G_{ij}|\right) \log_2 \left( 1+|G_{ij}|\right) \right.\nonumber\\
    &\left.+ \left( 1-|G_{ij}|\right) \log_2 \left( 1-|G_{ij}|\right) \right],
    \label{eq:discord:Heisenberg}
\end{align}
as stated in, for example, Ref.~\cite{10.1063/1.4862469}.

\subsection{Quantum Fisher information}
We introduce the quantum Fisher information in Section \ref{sec:qfi:intro} and derive corresponding entanglement bounds in Section \ref{sec:qfi:bounds}. 
The Hauke et al.~\cite{Hauke2016} relation is proven in Section \ref{sec:qfi:Haukeproof} using linear response theory results derived in Appendix \ref{sec:qfi:linearresponse}. Section \ref{sec:qfi:generalization} describes a generalization to quantum variance and skew information.

\subsubsection{Introduction}\label{sec:qfi:intro}
The quantum Fisher information (QFI) is a witness of multipartite entanglement, and was shown to be measurable using spectroscopic techniques in a seminal paper by Hauke et al.~\cite{Hauke2016}. However, the concept has its origin in quantum metrology, and specifically the quantum theory of phase estimation \cite{helstrom1976quantum, Holevo2011, PhysRevLett.72.3439}. For reviews of this field we recommend Refs.\cite{Pezze2014, Toth_2014}. The QFI can also be viewed as a distance metric on the space of quantum states \cite{PhysRevLett.72.3439} and plays an important role in quantum information geometry \cite{Lambert2023}, topics beyond the scope of this article.

One of the most celebrated quantum metrology results is the quantum Cram\'er-Rao bound for the maximal precision with which one can measure a parameter $\nu$,
\begin{align}
	\left( \Delta \nu \right)^2    &\geq	\frac{1}{MF_\mathcal{Q} \left[ \rho ; \mathcal{O} \right]} ,	\label{eq:quantumcramerrao}
\end{align}
where $\left( \Delta \nu \right)^2$ denotes the variance of $\nu$, $M$ is the number of independent measurements and the quantity $F_\mathcal{Q} \left[ \rho ; \mathcal{O} \right]$ is the QFI in a state $\rho$ and for an operator $\mathcal{O}$ coupling to $\nu$. (A relevant example is the case where $\nu$ represents a magnetic field, and $\mathcal{O}$ a spin operator.) As we will see later, higher QFI corresponds to stronger entanglement, in the sense of larger entanglement depths. Equation~\eqref{eq:quantumcramerrao} thus shows that highly entangled states are required to reach the highest possible levels of measurement precision; a larger entanglement depth enabling higher precision. Conversely, the precision of a measurement can imply the presence of specific entangled states.

The QFI related to an operator $\mathcal{O}$ and a generic state described by the density matrix $\rho$ is given by
\begin{align}
	F_\mathcal{Q} \left[ \rho; \mathcal{O} \right]	&=	2\sum_{\lambda,\lambda'}	\frac{\left( p_\lambda - p_{\lambda'} \right)^2}{p_\lambda + p_{\lambda'}} \left| \left\langle \lambda \middle| \mathcal{O} \middle| \lambda'\right\rangle \right|^2,	\label{eq:QFI:mixed}
\end{align}
where $\left| \lambda\right\rangle$ and $\left| \lambda'\right\rangle$ are eigenstates of $\rho$ with eigenvalues $p_\lambda$ and $p_{\lambda'}$, respectively. This notation emphasizes that the eigenvalues can be interpreted as probabilities. Note that the sum excludes terms with $p_\lambda + p_{\lambda'}=0$. For the case of a pure state the expression simplifies to $F_\mathcal{Q} = 4 \left( \Delta \mathcal{O} \right)^2,$ where $\left( \Delta \mathcal{O} \right)^2=\left\langle \mathcal{O}^2\right\rangle - \left\langle \mathcal{O}\right\rangle^2$ is the variance. 

In the following, we will assume a system of $N$ sites or particles, and that $\mathcal{O}$ is a sum of local, bounded Hermitian operators $\mathcal{O}_j$, i.e.
\begin{align}
	\mathcal{O}&= \sum_{j=1}^N \mathcal{O}_j.
\end{align}
Concretely, in spectroscopic scattering experiments, $\mathcal{O}$ often represents the momentum space Fourier transform of on-site operators. In the context of spin-polarized inelastic neutron scattering, for example, $\mathcal{O}$ can represent $S^\mu_\mathbf{k} = N^{-1/2} \sum_i e^{i\mathbf{k} \cdot \mathbf{r}_i} S_i^\mu$, where $\mathbf{k}$ is the wave vector and $\mathbf{r}_i$ is the position of site $i$. In non-polarized neutron scattering, a linear combination of the components $\mu$ is considered, weighted by appropriate polarization factors. In the context of nonresonant inelastic xray scattering, $\mathcal{O}$ can represent $n_\mathbf{k} = N^{-1/2} \sum_i e^{i\mathbf{k} \cdot {\bf r}_i} n_i$, where $n_i$ is the electron density operator. 
We also assume a thermal state $\rho = \sum_l p_l \left| \lambda \middle\rangle \middle\langle \lambda \right|$, where the $\left| \lambda\right\rangle$ are energy eigenstates with occupation probabilities given by the Boltzmann distribution, 
\begin{align}
    p_\lambda &= \frac{1}{Z} \exp \left( -\frac{E_\lambda}{k_B T}\right),    \label{eq:boltzmann}
\end{align}
where $E_\lambda$ is the energy eigenvalue, $T$ is the temperature, $k_B$ is the Boltzmann constant, and $Z$ the partition function.

Under these assumptions, Hauke et al.\cite{Hauke2016} showed that the QFI density can be expressed
\begin{align}
	f_\mathcal{Q} \left[ \rho; \mathcal{O}; \mathbf{k}\right]	&=	\frac{4}{\pi}	\int_{0}^\infty \mathrm{d}(\hbar \omega) \tanh \left( \frac{\hbar \omega}{2 k_BT }\right) \chi^{\prime\prime} \left( \mathbf{k}, \hbar \omega, T\right),	\label{eq:QFI:Hauke}
\end{align}
where $\chi^{\prime\prime}$ is the imaginary part of the dynamic susceptibility in state $\rho$ associated with the operator $\mathcal{O}$, and $\mathbf{k}$ is a possible wave vector index. (It is often absorbed into the definition of $\mathcal{O},$ but is convenient to keep explicit in the context of scattering experiments.) Here $f_\mathcal{Q}\equiv F_\mathcal{Q}/N$ is called the QFI density, which is the appropriate quantity in Eq.~\eqref{eq:QFI:Hauke} if the dynamic susceptibility is treated as an intensive quantity (i.e. normalized per site, as is common in condensed matter). The importance of this formula is that $f_\mathcal{Q}$ can be obtained using spectroscopic techniques such as inelastic neutron scattering. As long as $\chi^{\prime\prime}$ can be obtained in absolute units \cite{10.1063/1.4818323} and the above assumptions are met, it provides an experimentally accessible way to bound the entanglement in the system. We note that the requirement of normalization to absolute units amounts to a restriction on what experimental methods are practical for determining the QFI.

\subsubsection{Entanglement bounds}\label{sec:qfi:bounds}
Bounds for the QFI in separable and fully entangled states were obtained in Ref.~\cite{PhysRevLett.102.100401}. Bounds for $m$-separable states were first obtained in Refs.~\cite{Hyllus2012, Toth2012}. These bounds rely on the fact that the QFI is convex in its density matrix argument, meaning that for mixed states $\rho=p\rho_1 + \left( 1-p\right) \rho_2$ and $p\in \left[ 0,1\right]$,
\begin{align}
    F_\mathcal{Q} \left[ \rho \right]  &\leq p F_\mathcal{Q} \left[ \rho_1 \right] + \left( 1-p\right) F_\mathcal{Q} \left[ \rho_2 \right].
\end{align}
This property will not be proven here. It is most straightforwardly derived in a more general discussion of the Fisher information \cite{Pezze2014} by making use of the result that the classical Fisher information is convex \cite{Cohen1968}. This property implies that a mixed $m$-separable state [Equation~\eqref{eq:mseparable:mixedrho}] satisfies
\begin{align}
    F_\mathcal{Q} \left[ \rho_{m-\mathrm{sep}}   \right] &\leq \sum_l p_l F_\mathcal{Q} \left[ \left| \phi_{m_l-\mathrm{sep}} \middle\rangle\middle\langle \phi_{m_l-\mathrm{sep}} \right| \right] \\
                                                        &\leq \sum_l p_l 4\left. \left( \Delta \mathcal{O}\right)^2 \right|_{\left| \phi_{m_l-\mathrm{sep}} \right\rangle}
\end{align}
where the variance of $\mathcal{O}$ is evaluated in the state $\left| \phi_{m_l-\mathrm{sep}} \right\rangle$, which is of the form \eqref{eq:mseparable:purestate}. Since $\mathcal{O}$ is linear,
\begin{align}
    4\left. \left( \Delta \mathcal{O}\right)^2 \right|_{\left| \phi_{m_l-\mathrm{sep}} \right\rangle}   &=  \sum_{l=1}^M    \left. \left( \Delta \mathcal{O}\right)^2 \right|_{\left| \phi_{l} \right\rangle}   \leq \sum_{l=1}^M \left[ \lambda_\mathrm{max}^{(l)} - \lambda_\mathrm{min}^{(l)}\right]^2
\end{align}
where $\lambda_\mathrm{max}^{(l)}$ and $\lambda_\mathrm{min}^{(l)}$ are the largest and smallest eigenvalues of the operator acting on the sites involved in $\left| \phi_l\right\rangle$, i.e. $\mathcal{O}^{(l)}=\otimes_{i=1}^{N_l}\mathcal{O}_i$, where $\mathcal{O}_i$ is the operator acting on a single site. (The integers $M$ and $N_l$ come from the definition \eqref{eq:mseparable:mixedrho}.) We are interested in the case when the same operator is measured on all sites, for which $\lambda_\mathrm{max}^{(l)}=N_l \lambda_\mathrm{max}$ and $\lambda_\mathrm{min}^{(l)}=N_l \lambda_\mathrm{min}$, where $\lambda_\mathrm{max}$ and $\lambda_\mathrm{min}$ are the largest and smallest eigenvalues of $\mathcal{O}_i$. This gives the bound
\begin{align}
    \max_{\rho_{m-\mathrm{sep}}} F_\mathcal{Q} \left[ \rho_{m-\mathrm{sep}} \right]   &\leq \left( \lambda_\mathrm{max} - \lambda_\mathrm{min}\right)^2 \max_{\left\{ N_l\right\}} \sum_{l=1}^M N_l^2,
\end{align}
where the optimization on the right hand side is over all possible partitions satisfying $\sum_{l=1}^M N_l = N$. This is achieved by making the $N_l$ as large as possible. In particular, for $m$-separable states that have $N_l\leq m$,
\begin{align}
    \max_{\left\{ N_l\right\}} \sum_{l=1}^M N_l^2   &=  s m^2 + r^2,\quad s=\floor*{\frac{N}{m}},\, r= N-sm,
\end{align}
where $\floor*{x}$ is the ``floor function'' that equals the largest integer smaller than or equal to $x$.

This means that the QFI in an $m$-separable state is bounded by
\begin{align}
    F_\mathcal{Q}   \left[ \rho_{m-\mathrm{sep}} \right]    &\leq \left( sm^2 + r^2 \right) \left( \lambda_\mathrm{max} - \lambda_\mathrm{min}\right)^2.
\end{align}
Conversely, if
\begin{align}
    F_\mathcal{Q}   \left[ \rho \right]    &>  \left( sm^2 + r^2 \right) \left( \lambda_\mathrm{max} - \lambda_\mathrm{min}\right)^2  \label{eq:qfi:bound}
\end{align}
the system must be at least $(m+1)$-partite entangled.  
The special case of a separable state corresponds to $m=1$, for which the QFI satisfies the bound
\begin{align}
    F_\mathcal{Q}   \left[ \rho_\mathrm{sep} \right]    &\leq N \left( \lambda_\mathrm{max} - \lambda_\mathrm{min}\right)^2.
\end{align}
The bound for a maximally entangled state corresponds to $m=N$, for which
\begin{align}
    F_\mathcal{Q}   \left[ \rho_{N-\mathrm{ent}} \right]    &\leq N^2 \left( \lambda_\mathrm{max} - \lambda_\mathrm{min}\right)^2.
\end{align}
This result is known in quantum metrology as the Heisenberg limit.

The bound \eqref{eq:qfi:bound} simplifies if $m$ is a divisor of $N$. For finite systems this can be checked explicitly. For experimental condensed matter systems we can typically assume that $N$ is large and indeterminate, such that all $m\ll N$ are divisors. Under this assumption, $r=0$ and in terms of the QFI density $f_\mathcal{Q}=F_\mathcal{Q}/N$,
\begin{align}
    f_\mathcal{Q}   \left[ \rho \right]    &> m \left( \lambda_\mathrm{max} - \lambda_\mathrm{min}\right)^2\label{eq:qfid:bound:divisor}
\end{align}
indicates the presence of at least $(m+1)$-partite entanglement. 
We note that it is sometimes assumed that the operator $\mathcal{O}_i$ has a unit spectrum in order to simplify this formula to $f_\mathcal{Q}\left[ \rho\right] > m$. However, when applied to a specific experiment, the eigenvalues of the measured operators can have other ranges. The supplemental material of Ref.~\cite{PhysRevLett.127.037201} discusses the application to inelastic neutron scattering on spin$-S$ magnetic systems. For unpolarized scattering, i.e. without spin-polarization resolution, the bound becomes $f_\mathcal{Q}> 12S m ^2$. Alternatively, this can be expressed as $\mathrm{nQFI}>m$ in terms of the normalized QFI \cite{PhysRevB.103.224434} defined as  $\mathrm{nQFI}=f_\mathcal{Q}/(12S^2)$.

\subsubsection{Proof of the Hauke et al. relation}\label{sec:qfi:Haukeproof}
From the definition of the hyperbolic tangent,
\begin{align}
	\tanh (x)	&=	\frac{e^x -e^{-x}}{e^x+e^{-x}},
\end{align}
it is easy to derive the identity
\begin{align}
	\tanh\left( \frac{x-y}{2}\right)	&=	\frac{e^{-y}-e^{-x}} {e^{-y}+e^{-x}}.
\end{align}
Thus, using Eq.~\eqref{eq:boltzmann},
\begin{align}
	\tanh \left( \frac{E_{\lambda'}-E_\lambda}{2k_BT}\right)	&=	\frac{e^{-\frac{E_\lambda}{k_BT}}-e^{-\frac{E_{\lambda'}}{k_BT}}}{e^{-\frac{E_\lambda}{k_B T}}+e^{-\frac{E_{\lambda'}}{k_BT}}}	\equiv \frac{p_\lambda - p_{\lambda'}}{p_\lambda+p_{\lambda'}}	\label{eq:usefulidentity}
\end{align}
Inserting the K\"allen-Lehmann representation for $\chi''$, Eq. \eqref{eq:chiprimeprime:lehmann} (see Appendix \ref{sec:qfi:linearresponse} for a derivation), into \eqref{eq:QFI:Hauke} yields
\begin{align}
	f_\mathcal{Q}	&=		\frac{4}{\pi}	\sum_{\lambda,\lambda'} \left| \middle\langle \lambda\middle| \mathcal{O}\middle| \lambda\middle\rangle\right|^2 \left( p_\lambda - p_{\lambda'} \right) \nonumber\\
					&\times\int_0^\infty d\left( \hbar\omega\right)\tanh \left( \frac{\hbar\omega}{2k_BT}\right) \pi \delta \left( \hbar\omega - E_{\lambda'} + E_\lambda \right).
\end{align}
Using that $\tanh(x)$ is odd in $x$ and that the delta function is even in its argument, one obtains after relabeling $\lambda\leftrightarrow \lambda'$ in one of the sums,
\begin{align}
	f_\mathcal{Q}	&=	2	\sum_{\lambda,\lambda'} \left| \middle\langle \lambda\middle| \mathcal{O}\middle| \lambda\middle\rangle\right|^2 \left( p_\lambda - p_{\lambda'} \right) \nonumber\\
	&\times\int_{-\infty}^\infty d\left( \hbar\omega\right)\tanh \left( \frac{\hbar\omega}{2k_BT}\right) \delta \left( \hbar\omega - E_{\lambda'} + E_\lambda \right)\\
	&=	2 \sum_{\lambda,\lambda'} \left| \middle\langle \lambda\middle| \mathcal{O}\middle| \lambda\middle\rangle\right|^2 \left( p_\lambda - p_{\lambda'} \right)	\tanh \left( \frac{E_{\lambda'} - E_\lambda}{2k_B T} \right).
\end{align}
Then, using the identity \eqref{eq:usefulidentity}, we recover Equation~\eqref{eq:QFI:mixed}, thus proving the validity of Equation~\eqref{eq:QFI:Hauke}.

\subsubsection{Generalization to quantum variance and skew information}\label{sec:qfi:generalization}
Equation~\eqref{eq:QFI:Hauke} can be generalized to a family of quantum coherence measures\cite{Scheie2023}
\begin{align}
    I \left[ \mathcal{O},\mathbf{k}; h,\rho\right] &=  \frac{1}{\pi}	\int_{0}^\infty \mathrm{d}\left( \omega\hbar\right) \, h \left( \frac{\hbar\omega}{2k_BT}\right) \chi^{\prime\prime} \left( \mathbf{k}, \omega, T\right),    \label{eq:generalquantumcoherencemeasure}
\end{align}
where $h(x)$ is a monotone quantum filter function that satisfies $h(x)\sim x$ for $x\rightarrow 0$ and $h(x)\rightarrow \mathrm{const}$ as $x\rightarrow \infty$ \cite{Petz1996, Gibilisco2009, Frerot2017}. Physically, this represents a high-pass filter for frequencies $\hbar\omega\gg k_B T$. For the QFI $I_\mathrm{QFI}=f_\mathcal{Q}$, $h(x)=4\tanh(x/2)$. 
As described in Ref.~\cite{Scheie2023}, this generalization gives experimental access to additional quantities that have been studied in quantum information theory. In particular, the quantum variance $I_\mathrm{QV}$ \cite{FrerotR2016, Frerot2019} is obtained for the filter function $h(x)=\mathcal{L}(x/2)$, where $\mathcal{L}(x)=\coth x - 1/x$ is the Langevin function, and the Wigner-Yanase-Dyson skew information $I_\alpha$ \cite{Wigner_1963, RevModPhys.50.221} is obtained for
\begin{align}
    h_\alpha \left( x \right)   &=  \frac{\cosh \left( x/2\right) - \cosh \left[ \left( \alpha -1/2\right) x\right]}{\sinh \left( x/2\right)},  \label{eq:skewfilter}
\end{align}
and $0<\alpha<1$ is a parameter introduced by Dyson, with $\alpha=1/2$ originally proposed by Wigner and Yanase. The Fourier transform of Equation \eqref{eq:generalquantumcoherencemeasure} from momentum space into real space defines a family of spatial two-site quantum correlation functions \cite{Scheie2023},
\begin{equation}
C[O_i,O_j;h,\rho] =  \frac{1}{\pi} \int_0^\infty \mathrm{d}\left( \hbar \omega\right)\, h \left( \beta \hbar \omega \right) \chi^{\prime\prime}_{O_i,O_j} \left( \omega \right)~.
\label{eq:spatialquantumcorrelation}
\end{equation}
that will not be covered in detail in this review. Here $i$ and $j$ are site indices, and $\chi^{\prime\prime}_{O_i,O_j} \left( \omega \right)$ is the imaginary part of the two-site dynamical susceptibility.

The inequality chain $I_\mathrm{QV} \leq I_{1/2} \leq f_\mathcal{Q}/4 \leq 2I_{1/2} \leq 3I_\mathrm{QV}$ \cite{luo2004wigner,FrerotR2016} guarantees that the quantum variance and skew information can be used to witness multipartite entanglement, just like the QFI. 
From the experimental perspective, once one has measured $\chi^{\prime\prime}$, 
Equation \eqref{eq:generalquantumcoherencemeasure} is equally straightforward to evaluate for all quantum filter functions. However, theoretical methods may favor certain quantities. For example, the quantum variance and skew information are more accessible than QFI to quantum Monte Carlo methods \cite{FrerotR2016, Frerot2019, Scheie2023}, as they can be computed using alternative expressions that do not involve numerically difficult analytical continuations of the dynamical correlation function from imaginary to real time.

\subsection{Summary}
The preceding pages have introduced a variety of witnesses of entanglement and quantum correlations as well as their derivations and expressions in specific cases. Table \ref{tab:witnesses} provides an overview of the respective witnesses, with references to key equations and applicable experimental techniques for quick reference.

\section{Applications in Condensed Matter}\label{sec:applications}
In this section we review applications of entanglement witnesses to experimental data for condensed matter systems. We group these applications into four rough categories based on the types of systems that have been explored in the literature: systems that realize dimer states, quantum critical spin chains, candidate quantum spin liquid materials, and systems that deviate from the prior three categories. We note that there is an even more extensive literature on theoretical analyses of entanglement witnesses in various spin models that is outside the scope of this article.

\subsection{Dimer systems and spin clusters}\label{sec:applications:dimers}
One of the simplest wave functions that possess entanglement is the singlet state formed by two quantum degrees of freedom, each with internal Hilbert space dimension two, such that its total angular momentum is zero.  This state can be realized in few-body systems, such as two photons with entanglement between their polarizations, or two fermions with entanglement between their spin degree of freedom. In many-body systems, the objects can, for example, be spin-$1/2$ local magnetic moments in a quantum magnet, exciton states in a molecular aggregate \cite{PhysRevA.104.042416}, or qubits in a quantum computer. We will focus on the magnetic case, for which the state is naturally expressed
\begin{equation}
    \left| \psi \right\rangle    =  \frac{1}{\sqrt{2}} \left( \left| \uparrow\right\rangle_1 \left| \downarrow \right\rangle_2 - \left| \downarrow \right\rangle_1 \left| \uparrow\right\rangle_2 \right),
\end{equation}
where $\left| \psi\right\rangle_i$ is the wave function of the $i$th degree of freedom. $\left|\uparrow\right\rangle$ and $\left|\downarrow\right\rangle$ represent eigenstates with well-defined spin projections onto a suitable quantization axis. Crucially, this state cannot be expressed as a sum of product states, i.e. it is non-separable. Indeed, it is one of the four Bell states or EPR pairs, i.e. maximally entangled states of two qubits\scite{Nielsen2010}.

Such spin singlets show up in several quantum magnetic contexts\scite{Vasiliev2018}---including random singlets in strongly disordered systems\cite{Igloi2005, Igloi2018, PhysRevX.8.041040, PhysRevLett.123.087201}, and as coherent superpositions in resonating valence bond (RVB) states\cite{RevModPhys.89.025003, Savary2017}---but the simplest realizations are found in dimerized magnets. It is useful to think of these systems as hosting singlets locked in place at pairs of sites, known as dimers, producing an overall product state with such singlets distributed throughout the lattice. 
Consider the bond alternating Heisenberg chain
\begin{equation}
    H   =  J\sum_j \left( \mathbf{S}_{2j}\cdot \mathbf{S}_{2j+1} + \alpha \mathbf{S}_{2j+1}\cdot\mathbf{S}_{2j+2}\right),   \label{eq:alternatingheisenberg}
\end{equation}
where $J>0$, $\mathbf{S}_j$ is the spin-1/2 operator at site $j$, and $\alpha=J^\prime/J$ is a bond alternation parameter, such that the exchange alternates between $J$ and $J'$. For $\alpha=1$, the usual, uniform Heisenberg antiferromagnetic chain is recovered. For $|\alpha | \ll 1$ 
the Hamiltonian describes a system of weakly coupled dimers. In such cases, one expects strong entanglement between the two sites making up the dimer, and only weak interdimer entanglement.

\begin{figure}
    \includegraphics[width=\columnwidth]{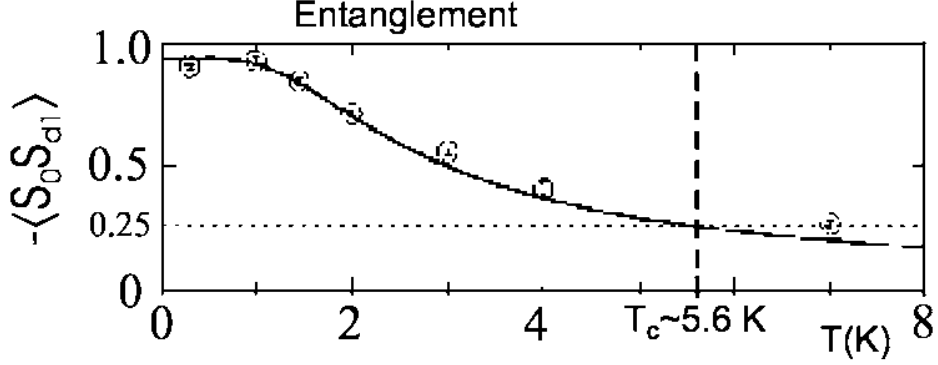}
    \caption{Intradimer correlation $\langle \mathbf{S}_0\cdot\mathbf{S}_d\rangle$ in copper nitrate. Entanglement is witnessed when $\left| \left\langle \mathbf{S}_0\cdot\mathbf{S}_{d_1}\right\rangle\right|>1/4$. \scite{PhysRevB.67.054414, PhysRevB.85.014402} Reproduced with permission \scite{PhysRevA.73.012110} 2006, American Physical Society.}
    \label{fig:dimer:coppernitrate}
\end{figure}

Remarkably, there exists a wide range of materials with weakly coupled magnetic chains, whose magnetic properties can be described by Equation \eqref{eq:alternatingheisenberg} \scite{PhysRevLett.79.745, PhysRevB.59.11384, PhysRevB.60.1197, PhysRevLett.99.087204}. Copper nitrate [\ce{Cu(NO3)2 2.5H2O}, in neutron studies often deuterated to \ce{Cu(NO3)2 2.5D2O}] with $\alpha\approx 0.25$ is considered a model realization of the Hamiltonian, and has received much experimental attention \scite{PhysRev.132.1057, PhysRevB.27.248, PhysRevB.67.054414, PhysRevLett.84.4465, PhysRevB.85.014402, PhysRevB.90.094419}. In 2006, Brukner, Vedral and Zeilinger \scite{PhysRevA.73.012110} reanalyzed previously published inelastic neutron scattering \scite{PhysRevLett.84.4465} and magnetic susceptibility \scite{PhysRev.132.1057} data from an entanglement perspective. Assuming a system at thermal equilibrium described by Equation \eqref{eq:alternatingheisenberg} with spin-isotropic coupling and vanishing magnetic order, and that spin-spin correlations beyond nearest neighbors are negligible compared to the intradimer spin-spin correlation $\left\langle \mathbf{S}_0 \cdot \mathbf{S}_{d_1}\right\rangle$, they derived the following expression for the zero-field magnetic susceptibility (averaged over three orthogonal directions),
\begin{equation}
    \chi    = \frac{g^2 \mu_B^2 N\hbar^2}{k_B T} \left[ \frac{1}{4} + \frac{1}{3} \left \langle \mathbf{S}_0 \cdot \mathbf{S}_{d_1}\right\rangle \right]
\end{equation}
where $g$ is the g factor, and $\mu_B$ is the Bohr magneton. (This follows from Equation \eqref{eq:susceptibility:variance}.) Since $\left|\left\langle \mathbf{S}_{0} \cdot \mathbf{S}_{d_1}\right\rangle \right| \leq \left| \mathbf{S}_{0\vphantom{d_1}} \right| \left| \mathbf{S}_{d_1}\right| \leq 1/4 $ for any separable state, and since the intradimer correlations are antiferromagnetic, one obtains the inequality
\begin{equation}
    \chi    \geq     \frac{1}{6}\frac{g^2 \mu_B^2 N\hbar^2}{k_B T}  .
\end{equation}
Violations of this inequality indicates the system is in a non-separable, i.e. entangled, state, in agreement with Equation \eqref{eq:suspectibility:witness}. Brukner et al. found such violations at temperatures below \SI{5}{K} \scite{PhysRevA.73.012110}, indicating an entangled low-temperature phase. Their complementary analysis of neutron scattering data is shown in Fig.~\ref{fig:dimer:coppernitrate}. By direct inspection, the intradimer spin-spin correlation exceeds the maximal classical value below $T_c\approx$\SI{5.6}{K}. Since this is a spin-isotropic dimer system, the concurrence can be expressed [Equation~\eqref{eq:concurrence:dimer}]
\begin{equation}
    C   = 2\max \left\{ 0, -\left\langle \mathbf{S}_0 \cdot \mathbf{S}_{d_1} \right\rangle - \frac{1}{4} \right\},
\end{equation}
which follows the temperature dependence of $\left\langle \mathbf{S}_0 \cdot \mathbf{S}_{d_1} \right\rangle$. It witnesses the presence of entanglement within the dimers up to $T_c$. The careful reader will note that the intradimer correlation appears to exceeds the maximal value in an isolated spin-1/2 singlet, namely $S(S+1)=0.75$, at low temperatures. However, these values are consistent within the experimental error bars: at \SI{0.3}{K}, $\left\langle \mathbf{S}_0 \cdot \mathbf{S}_{d_1} \right\rangle=0.9(2)$ \scite{PhysRevLett.84.4465}. In general, the extraction of spatial correlation functions is sensitive to normalization and background subtraction procedures.

The initial works by Wie{\'{s}}niak et al. \scite{Wie_niak_2005} and Brukner et al. \scite{PhysRevA.73.012110} inspired many additional experimental characterizations of the entanglement in copper nitrate, including witnesses based on the magnetic susceptibility at finite magnetic fields \scite{Das_2013}, and heat capacity \scite{Singh2013}. In addition, the two-site quantum discord was determined \scite{PhysRevB.84.024418, Singh_2015}. Susceptibility witnesses and concurrence (in some experiments evaluated from heat capacity data) were also applied to several other low-dimensional and molecular dimer magnet compounds, with both $S=1/2$ and higher-$S$ magnetic moments, including \ce{Na2Cu5Si4O14} \scite{PhysRevB.77.104402}, metal carboxylates \scite{PhysRevB.79.054408}, \ce{KNaMSi4O10} (M=Mn, Fe, or Cu) \scite{Soares-Pinto_2009}, \ce{Fe2(\mu2-oxo)-(C3H4N2)6(C2O4)2} \scite{Reis_2012}, nitrosyl iron complexes \scite{10.1063/1.4862469}, \ce{NH4CuPo4\cdot H2O} \scite{doi:10.1063/1.4861732}, and copper acetate \ce{C8H16Cu2O10} \scite{Athira_2023}. Rappoport et al. \scite{PhysRevB.75.054422} witnessed entanglement using the magnetic susceptibility in pyroborate \ce{MgMnB2O5}, hosting a Griffiths phase with $S=5/2$ dimers, and warwickite \ce{MgTiOBO3}, hosting an $S=1/2$ random singlet phase. They found the witness can certify the presence of entanglement in both systems. 
Two-site quantum discord was witnessed in copper acetates and nitrosyl iron complexes \scite{PhysRevB.84.024418, 10.1063/1.4862469}. Concurrence and two-site discord between spatially separated sites (\SIrange{220}{250}{\text{\AA}} apart) were witnessed using magnetization measurements in the chain motifs of \ce{Sr14Cu24O41} (the structure of which features both spin-ladder and chain subsystems) \scite{Sahling2015}. See also Ref.~\cite{PhysRevB.100.235103} for further analysis of the mediators of these quantum correlations.

Finally, entanglement has also been witnessed in the \ce{(Cr7Ni)2} supramolecular dimer system \scite{PhysRevLett.104.037203, Garlatti2017, Garlatti_2019}. This system is made of linked antiferromagnetic rings, where each ring realizes a $S=1/2$ ground state that is robust to applied magnetic fields and weaker inter-ring interactions. They have therefore been proposed as molecular qubits, that could be used as building blocks for quantum computation \scite{PhysRevLett.94.207208, 10.1063/5.0053378, Moreno_Pineda_2021} and simulation \scite{PhysRevLett.107.230502, Moreno_Pineda_2021} platforms. 
The entanglement has been experimentally probed in complexes with two rings, i.e. dimers. This was initially done using the magnetic susceptibility \scite{PhysRevLett.104.037203}, thus probing entanglement in the thermal state. Later, the concurrence in \emph{eigenstates} was probed using inelastic neutron scattering \scite{Garlatti2017} by using the magnetic field to prepare specific, factorized ground states, which is rarely possible in other condensed matter systems.

\subsection{Critical quantum spin chains}

One of the most paradigmatic models in magnetism and many-body quantum physics is the Heisenberg spin chain
\begin{equation}
    H   = J\sum_j \mathbf{S}_j \cdot \mathbf{S}_{j+1},  \label{eq:heisenberg}
\end{equation}
originally introduced by Heisenberg in 1928 \scite{Heisenberg1928}. (Besides its fundamental importance, it also has potential applications in quantum communication \scite{PhysRevLett.91.207901, Bose_2007}.)
We will focus on the antiferromagnetic case, where $J>0$. In the general case, $\mathbf{S}_j$ represents a spin-$S$ operator. The physics of the model turn out to depend crucially on the value of $S$ \scite{Haldane1983, PhysRevLett.50.1153, Affleck1989a, Auerbach1994}: In the case of integer $S$, the excitation spectrum is gapped and the system can host topological edge states, whereas the half-integer case corresponds to a critical system with gapless excitations. The $S=1/2$ chain hosts fractional excitations known as spinons \scite{Faddeev1981}, each carrying spin $1/2$, which should be contrasted with the usual magnon (or spin-wave) excitations which carry spin $1$.

As far as we are aware, entanglement has yet to be experimentally witnessed in systems described by Equation~\eqref{eq:heisenberg} with $S>1/2$. However, theoretical predictions exist for $S=1$ \scite{PhysRevB.99.045117, PhysRevB.108.144414} and $S\geq 5/2$ \scite{PhysRevB.103.224434, PhysRevB.107.059902}. We will thus focus on the quantum spin chain with $S=1/2$ in this section. A natural guess for its ground state is the N\'eel state consisting of alternating up and down spins, $\left| \dots\uparrow\downarrow\uparrow\downarrow\uparrow\downarrow\dots\right\rangle$. However, the N\'eel state is actually not an eigenstate of the Hamiltonian \scite{Auerbach1994}! Instead, the actual ground state is a more complicated state first obtained by Bethe \scite{Bethe1931} and Hulth\'en \cite{Hulthen1938}: a macroscopic singlet with overall spin of zero entangling all sites. Its entanglement content has been characterized theoretically in a large number of ways \scite{RevModPhys.80.517, bayat2022entanglement}. It being a critical state, conformal field theory predicts that the entanglement entropy between a finite subset and the remainder of the system scales logarithmically with the size of the subset \scite{PhysRevLett.90.227902, PhysRevLett.92.096402, Calabrese2004}. It is also associated with substantial multipartite entanglement \scite{PhysRevD.96.126007, PhysRevB.107.054422}, and short-range pairwise entanglement \scite{PhysRevLett.87.017901, Wang2002, PhysRevA.69.062314, PhysRevA.74.022322}.

The presence of entanglement can be inferred from observing signatures of the ground state, such as scattering continua due to the presence of spinons \scite{PhysRevLett.70.4003, PhysRevB.52.13368, Lake2005, Mourigal2013, PhysRevLett.111.137205, Blanc2018, Wu2019, Gao2023}. This is a model-dependent approach that relies on us having a good theoretical understanding of the ground state. It thus cannot be generalized to all systems of interest. However, the very fact that we have a handle on the ground state makes these systems excellent testing grounds for entanglement witnesses, allowing for contrasting between different entanglement measurements. Witnesses based on magnetic susceptibility were applied to \ce{Cu(thiazole)2Cl2}, a copper based polymer system  \scite{Chakraborty2012}, and an organic radical molecular chain \scite{10.1063/1.4824458}, indicating entanglement up to \SI{12}{K} and \SI{28}{K}, respectively.

\begin{figure}
    \includegraphics[width=1.\columnwidth]{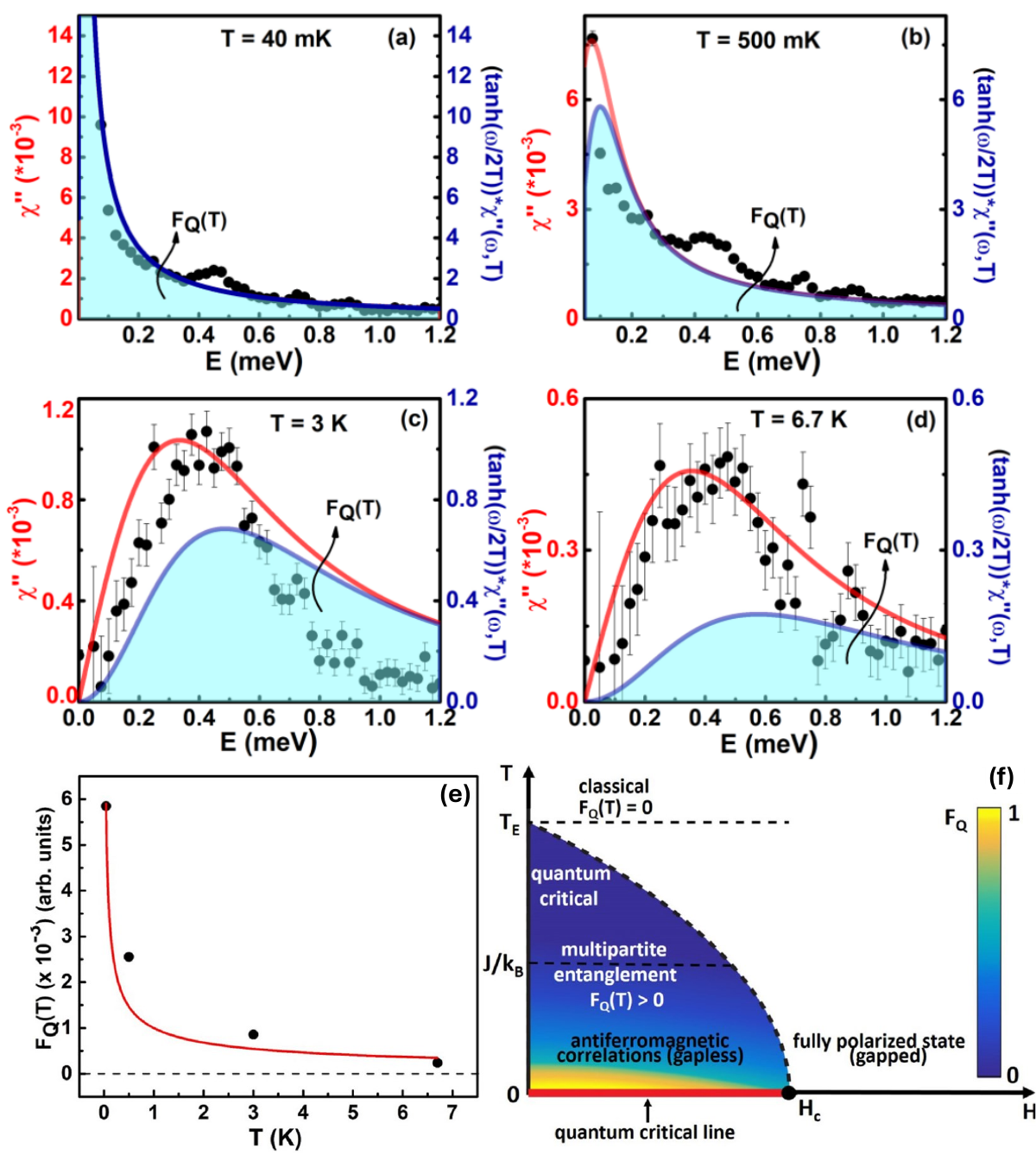}
    \caption{\ce{[Cu(\mu-C2O4)(4-aminopyridine)2(H2O)]_n}. (a)-(d) Inelastic neutron scattering results for a polycrystalline sample at several temperatures. Black circles represent $\chi''\left( \omega\right)$ (left axis) integrated over a range of momentum transfers around the antiferromagnetic momentum. Red solid lines indicate a theoretical fit. Blue solid lines represent the QFI integrand (right axis). The shaded area under this curve determines the QFI.
    (e) Temperature-variation of the QFI. Since $\chi''$ (and thus the QFI) was obtained in arbitrary units, the entanglement depth was not probed directly in this experiment. The temperature scaling is consistent with theoretical expectations for the Heisenberg antiferromagnetic chain (red line). 
    (f) Schematic phase diagram for the Heisenberg antiferromagnetic chain in the presence of an applied magnetic field $H$. 
    Panels from Figures 4 and 5 of Mathew et al.,\scite{PhysRevResearch.2.043329} reproduced under the CC BY 4.0 license\scite{ccby4}. Copyright 2020, G. Mathew et al., published by American Physical Society.}
    \label{fig:mathew}
\end{figure}
A new chapter opened up after the seminal 2016 paper by Hauke et al. \scite{Hauke2016}, showing that multipartite entanglement could be witnessed and the entanglement depth inferred via the quantum Fisher information calculated from dynamical susceptibilities. 
The first experimental work in this direction was a 2020 study by Mathew et al. \cite{PhysRevResearch.2.043329} on a polycrystalline sample of the $S=1/2$ Heisenberg spin chain \ce{[Cu(\mu-C2O4)(4-aminopyridine)2(H2O)]_n}, summarized in Figure~\ref{fig:mathew}. They obtained a temperature scaling of the QFI that is consistent with theoretical expectations for the Heisenberg spin chain, thereby demonstrating the viability of the approach. 
However, since they did not obtain scattering in absolute units, it is unclear what degree of multipartite entanglement was actually witnessed in \ce{[Cu(\mu-C2O4)(4-aminopyridine)2(H2O)]_n}. (The inequality Eq. \eqref{eq:qfid:bound:divisor} that bounds the entanglement depth requires a quantitative determination of the QFI.)

\begin{figure*}[t]
    \includegraphics[width=\textwidth]{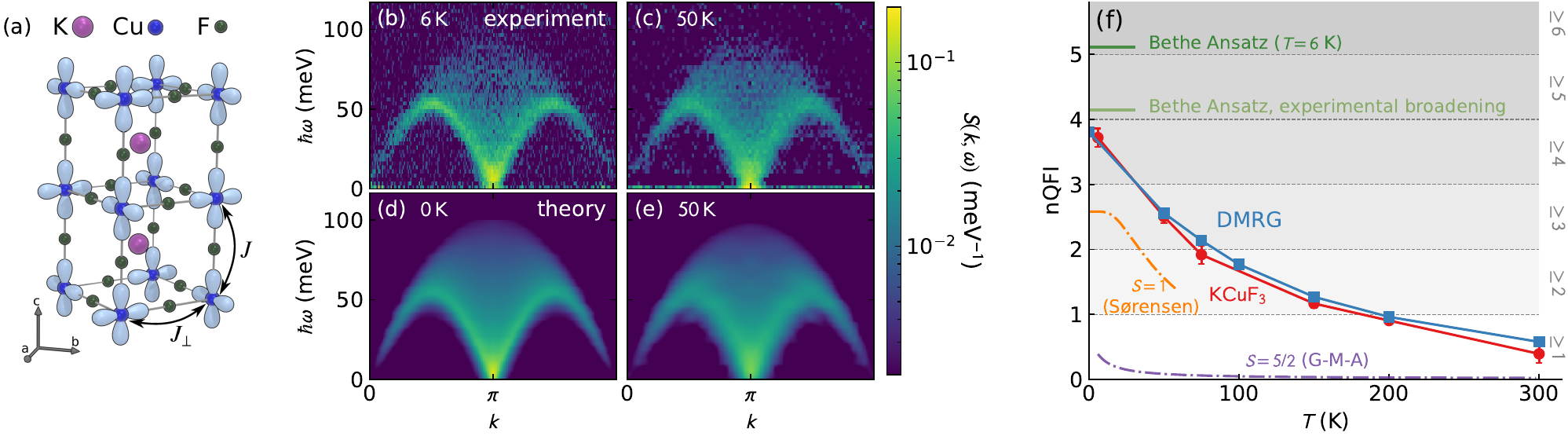}
    \caption{The spin-$1/2$ Heisenberg antiferromagnetic chain compound \ce{KCuF3}.
    (a) The crystal structure features chains of \ce{Cu} ions. Owing to the orbital order, the interchain exchange $J_\perp=\SI{-1.6}{meV}$ is much weaker than the intrachain coupling $J=\SI{34}{meV}$, making the magnetism largely one-dimensional.
    (b)-(e) Measured and DMRG-simulated inelastic neutron scattering spectra at selected temperatures.
    (f) Normalized quantum Fisher information as a function of temperature, calculated at the antiferromagnetic momentum $k=\pi$. Values from experiment (red line) and DMRG (blue line) closely agree throughout the entire temperature range. At least quadpartite entanglement is witnessed at the lowest temperatures, and at least bipartite entanglement is witnessed up to \SI{150}{K}. 
    The algebraic Bethe ansatz predictions with and without experimental broadening are shown in light and dark green, respectively.
    Also shown are estimated values for the $S=1$ chain \scite{PhysRevB.99.045117} (orange line) and for $S=5/2$ (purple line). 
    In the integer-$S$ case, the Haldane spin gap produces a plateau at low temperatures. 
    The QFI at any finite temperature decays as $S$ increases, reflecting a quantum-to-classical crossover.
    Reproduced with permission.\scite{PhysRevB.103.224434, PhysRevB.107.059902} 2021-2023, American Physical Society.}
    \label{fig:kcuf3}
\end{figure*}

\ce{KCuF3}, Fig.~\ref{fig:kcuf3}(a), is among the most well-studied realizations of the isotropic $S=1/2$ Heisenberg antiferromagnetic chain \cite{Hutchings_1979, PhysRevB.21.2001, PhysRevB.44.12361, PhysRevLett.70.4003, PhysRevB.52.13368, PhysRevLett.85.832, Lake2005, doi:10.1063/1.4709772, PhysRevLett.111.137205, PhysRevB.103.224434, Scheie2021, Scheie2022}. It can be obtained in large single crystals suitable to inelastic neutron scattering, and features robust enough intrachain exchange coupling ($J\approx\SI{34}{meV}$) that the scattering continuum remains at room temperature. The system orders magnetically at low temperatures, below $T_N=\SI{39}{K}$, due to weak interchain coupling ($J_\perp\approx\SI{-1.6}{K}$). However, such effects affect only the low-energy scattering, with high-energy scattering reflecting the universality of the Heisenberg chain \scite{Lake2005}. The scattering intensity, and thus the dynamical spin structure factor, was obtained in absolute units as shown in Fig.~\ref{fig:kcuf3}(b),(c). The entanglement properties of \ce{KCuF3} were investigated experimentally in Ref.~\scite{PhysRevB.103.224434, PhysRevB.107.059902}, and found to closely agree with finite-temperature DMRG simulations [Fig.~\ref{fig:kcuf3}(d),(e)]. Concurrence indicates short-range pairwise entanglement. Most interestingy, the QFI was found to witness substantial entanglement depths; see Fig.~\ref{fig:kcuf3}(f). At the lowest measured temperature of \SI{6}{K}, \emph{at least} quadpartite entanglement was witnessed, meaning that the thermal state features entanglement between at least four spins. (We stress that, since the entanglement bounds take the form of inequalities, QFI can only witness a minimal entanglement depth. In other words, QFI can certify the presence of a certain entanglement depth, but never its absence.) This number is comparable with entanglement depths probed in atomic spin chains in optical lattices \scite{Dai_2016, PhysRevLett.131.073401}. \ce{KCuF3} also features at least bipartite entanglement up to at least \SI{150}{K}. 

\begin{figure*}
    \includegraphics[width=\textwidth]{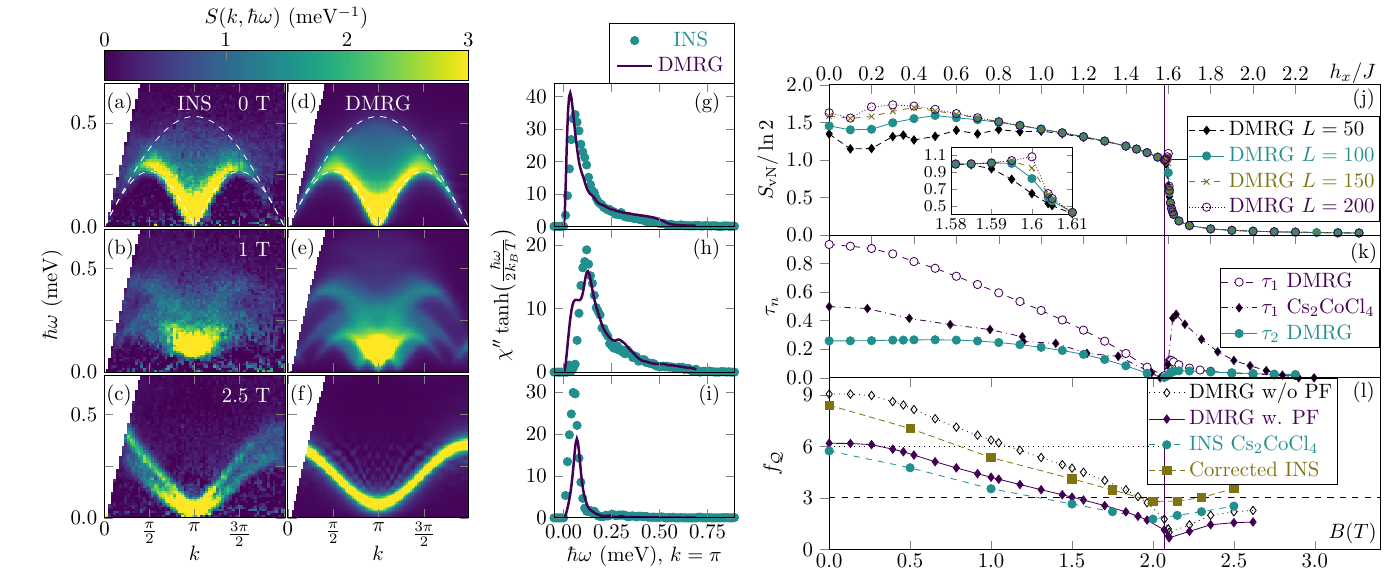}
    \caption{The spin-1/2 transverse-field XXZ chain \ce{Cs2CoCl4}.
    Left: experimental and DMRG-simulated inelastic neutron scattering spectra and QFI integrands. The agreement is excellent at weak and intermediate fields, with deviations caused above $h_c\approx \SI{2.1}{T}$ due to weak interchain couplings not accounted for in the theoretical 1D model.
    Right: entanglement properties. The bottom panel (l) shows the QFI $f_\mathcal{Q}$. For $f_\mathcal{Q}>3$ (dashed line), at least bipartite entanglement is witnessed. For $f_\mathcal{Q}>6$, at least tripartite entanglement is witnessed. The cyan curve represents the experimental data, subject to the experimental polarization factor (PF). It is in good agreement with the purple DMRG line, for which the DMRG data was convoluted with the same polarization factor. In general, polarization factors and resolution effects tend to suppress the QFI. Here, because the DMRG calculation can be done without applying the polarization factor (black diamonds), it is possible to correct the experimental data (yellow squares), and to witness a higher entanglement depth. 
    The top panel (j) shows the theoretically calculated von Neumann entanglement entropy, and the middle panel (k) shows the one- and two-tangle. 
    Adapted with permission.\scite{PhysRevLett.127.037201,PhysRevLett.130.129902} 2021-2023, American Physical Society.}
    \label{fig:cs2cocl4}
\end{figure*}
It should be noted that the spin isotropy in \ce{KCuF3} and similar compounds substantially simplifies the data analysis. For spin-anisotropic compounds, it is necessary to take polarization factor effects into account, either by spin-polarization-resolved experiments or through theoretical modeling. The latter approach was taken in a study on \ce{Cs2CoCl4} \scite{PhysRevLett.127.037201, PhysRevLett.130.129902}, a compound that can be described as a spin-$1/2$ transverse-field XXZ chain \scite{PhysRevB.65.144432, PhysRevLett.111.187202}, with the Hamiltonian
\begin{align}
    H   &=  \sum_{j=1}^L \left[ J \left( S_j^x S_{j+1}^x + S_j^y S_{j+1}^y + \Delta S_j^z S_{j+1}^z  \right) + h_x S_j^x \right],
\end{align}
where $\Delta$ is a parameter controlling the anisotropy, and $h_x$ is a magnetic field. \ce{Cs2CoCl4} is in a regime with $J>0$ and $\left| \Delta \right| <1$, where the model has two quantum critical points:\scite{PhysRevB.65.172409} (i) $h_x=0$, in the same universality class as the isotropic Heisenberg chain and with central charge $c=1$, and (ii) $h_x=h_c\approx 1.6J$, which is in the Ising universality class and has central charge $c=1/2$. For $0<h_x<h_c$, the field induces a new source of fluctuations and a so-called spin-flop magnetic order with an excitation gap. For $h_x>h_c$ the 1D model becomes spin-polarized, entering a product state. In addition there exists a factoring field, $h_f<h_c$, where the system also assumes a product state in the form of a classical spin-flop state. The entanglement of the system can thus be controlled by changing the strength of the magnetic field. Its pairwise entanglement also changes qualitatively at $h_f$, in what is known as an entanglement transition \cite{PhysRevLett.93.167203, Abouie_2010, PhysRevLett.108.240503, PhysRevA.96.052303}. Figure~\ref{fig:cs2cocl4} summarizes the study experimentally characterizing the entanglement in \ce{Cs2CoCl4}. In particular, we note that the QFI witnesses entanglement at low fields, but does not capture the entanglement at $h_c$. In general, at such transitions there is no inelastic spectral weight available for witnessing the entanglement using the dynamical spin susceptibility, indicating the need for additional experimentally accessible entanglement witnesses.

Given sufficient momentum resolution, it is also possible to experimentally extract \emph{spatial} quantum correlation functions that go beyond witnesses of pairwise entanglement and two-site discord. This has been done for \ce{KCuF3}. In Ref.~\scite{Scheie2022}, the dynamical spin structure factor $S(k,\omega)$ measured with neutrons was Fourier transformed into real space and time, yielding the so-called Van Hove correlation function
\begin{equation}
    G(r,t)  = \langle S_i ^z (0) S_{i+r}^z (t) \rangle,
\end{equation}
whose imaginary part can be expressed as a commutator, $\mathrm{Im}\left[ G(r,t)\right] \propto \left[ S_i^z(0), S_{i+r}^z(t)\right]$, and is an indicator of quantum coherence between spins at two different sites. Notably, this quantity reveals a ``light-cone'' limiting the information transfer rate and spread of correlations. This type of speed limit generically occurs in quantum lattice systems with local interactions due to Lieb-Robinson bounds \scite{Lieb1972, PhysRevLett.97.050401,(Anthony)Chen_2023}, and was first experimentally demonstrated in a cold atom system \scite{Cheneau2012}.

In Ref.~\scite{Scheie2023} it was shown how quantum correlation functions---defined as the difference between two clas\-sically equivalent correlation functions---can be extracted from inelastic scattering by taking the spatial Fourier transform over a generalization of the QFI integral. By introducing a quantum filter function, the quantum Fisher information matrix \scite{Liu_2020}, quantum covariance \scite{MalpettiR2016, Frerot2019}, and skew information matrix \scite{Wigner_1963} can be discussed on equal footing. These spatial quantum correlation functions were found to decay with a characteristic length scale, the quantum coherence length \scite{MalpettiR2016} in agreement with numerical and theoretical results \scite{MalpettiR2016, Frerot2019, PhysRevX.12.021022}.

\subsection{Towards quantum spin liquids}
\begin{figure*}
    \includegraphics[width=\textwidth]{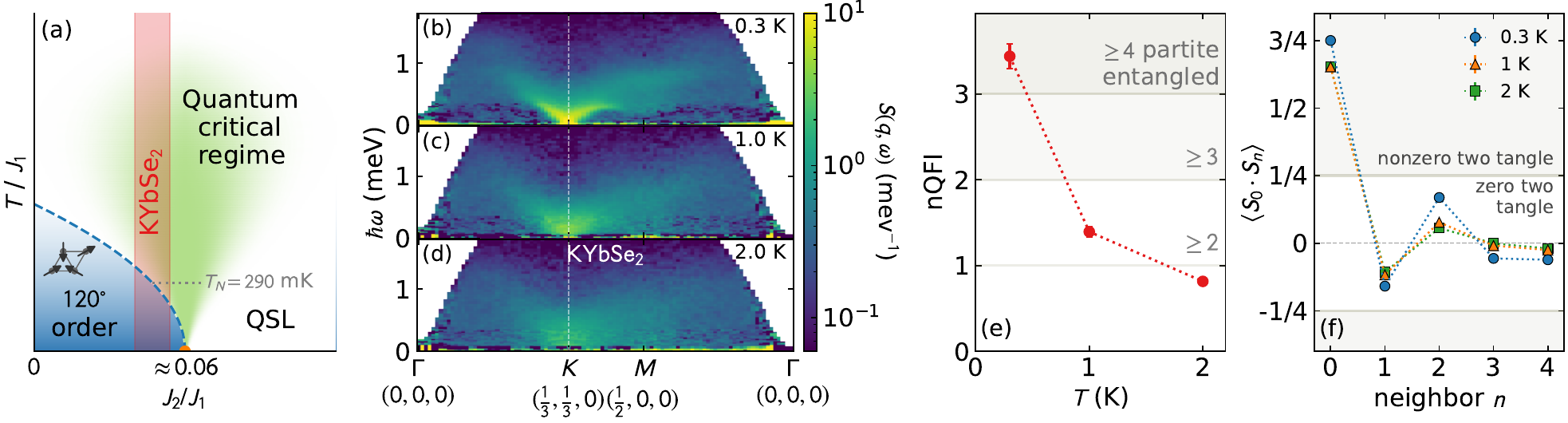}
    \caption{
    The triangular lattice antiferromagnet \ce{KYbSe2}.
    (a) Phase diagram of the $J_1$--$J_2$ model on the triangular lattice. \ce{KYbSe2} exists close to a quantum critical point that marks a transition form a magnetically ordered state to a quantum spin liquid. The proximity to a quantum critical point promotes quantum fluctuations.
    (b),(c),(d) Inelastic neutron scattering spectra from \ce{KYbSe2} at temperatures $T=0.3,\,1.0,\,\text{and }\SI{2.0}{K}$.
    (e) Normalized quantum Fisher information, evaluated at the ordering vector $K$. At \SI{0.3}{K}, at least four-partite entanglement is witnessed, showing that the ground state is strongly entangled. 
    (f) Only the on-site spin-spin correlations exceed the classical bound $1/4$. Further-range spin-spin correlations and the two-tangle do not witness quantumness or entanglement. This behavior is expected due to monogamy.
    Reproduced by permission.\scite{Scheie2023tlafm} 2023, Springer Nature.}
    \label{fig:kybse2:1}
\end{figure*}
A highly promising application for entanglement witnesses currently available in neutron scattering is in the search for quantum spin liquids. These elusive states are topologically ordered, fundamentally quantum states of matter characterized by a lack of magnetic order even at zero temperature and nonlocal entanglement \scite{Savary2017, Knolle2019, Broholm2020}. However, unambiguously identifying their presence in materials is a longstanding challenge, complicated by the fact that disorder effects can mimic the proposed signatures of quantum spin liquids \scite{Khatua2023}.  Ref.~\cite{PhysRevB.103.224434} proposed a diagnosis protocol based on entanglement witnesses to discriminate between genuine quantum spin liquid candidates and other types of disorder. The protocol is to look for materials with (i) substantial $\tau_1$, to avoid states that have weak quantum correlations or are strongly magnetically ordered, (ii) vanishing $\tau_2$, as quantum spin liquids distribute the entanglement between all sites, making pairwise between any two sites weak due to monogamy effects, 
and (iii) finite nQFI. All three conditions being met would strongly indicate long-range entanglement. 
It is important to note that the three witnesses involved in the protocol are all based on local observables, and thus cannot directly probe the nonlocal entanglement inherent to topological order. Nevertheless, derivatives of the QFI from local operators have theoretically been shown to be capable of detecting topological quantum phase transitions \scite{PhysRevLett.119.250401, Gabbrielli2018, PhysRevB.102.224401}. 

The protocol has been applied to the triangular lattice antiferromagnet \ce{KYbSe2} \scite{Scheie2023tlafm}; see Fig.~\ref{fig:kybse2:1}. This material is part of a family of delafossite materials in which magnetic \ce{Yb^3+} ions form two-dimensional triangular lattice networks with antiferromagnetic nearest- and next-nearest neighbor interactions $J_1$ and $J_2$. For $J_2/J_1 \lesssim 0.06$, a noncollinear $120^\circ$ magnetic order with spins pointing in or out of triangles is realized. However, for $J_2/J_1 \gtrsim 0.06$ theory predicts a quantum spin liquid phase. \ce{KYbSe2}, with $J_2/J_1\approx 0.044(5)$ \cite{PhysRevB.109.014425}, is very close to the critical point at $J_2/J_1\approx 0.06$. Experimentally, a one-tangle $\tau_1=0.85(2)$ and vanishing two-tangle is obtained. Based on the inelastic neutron spectra, the QFI at the K point is extracted, indicating at least quadpartite entanglement at the lowest temperatures. Such high QFI is due to the proximity to the quantum critical point. Although \ce{KYbSe2} is on the ``wrong'' side of the critical point, it is within the quantum critical fan emanating from it at finite temperature. The cousin material \ce{NaYbSe2} is expected to have higher $J_2/J_1$ than does \ce{KYbSe2}, and has been argued to fall within the quantum spin liquid phase \scite{PhysRevX.11.021044}. Experimentally probing its entanglement would be helpful for settling this point.

\begin{figure}
    \includegraphics[width=\columnwidth]{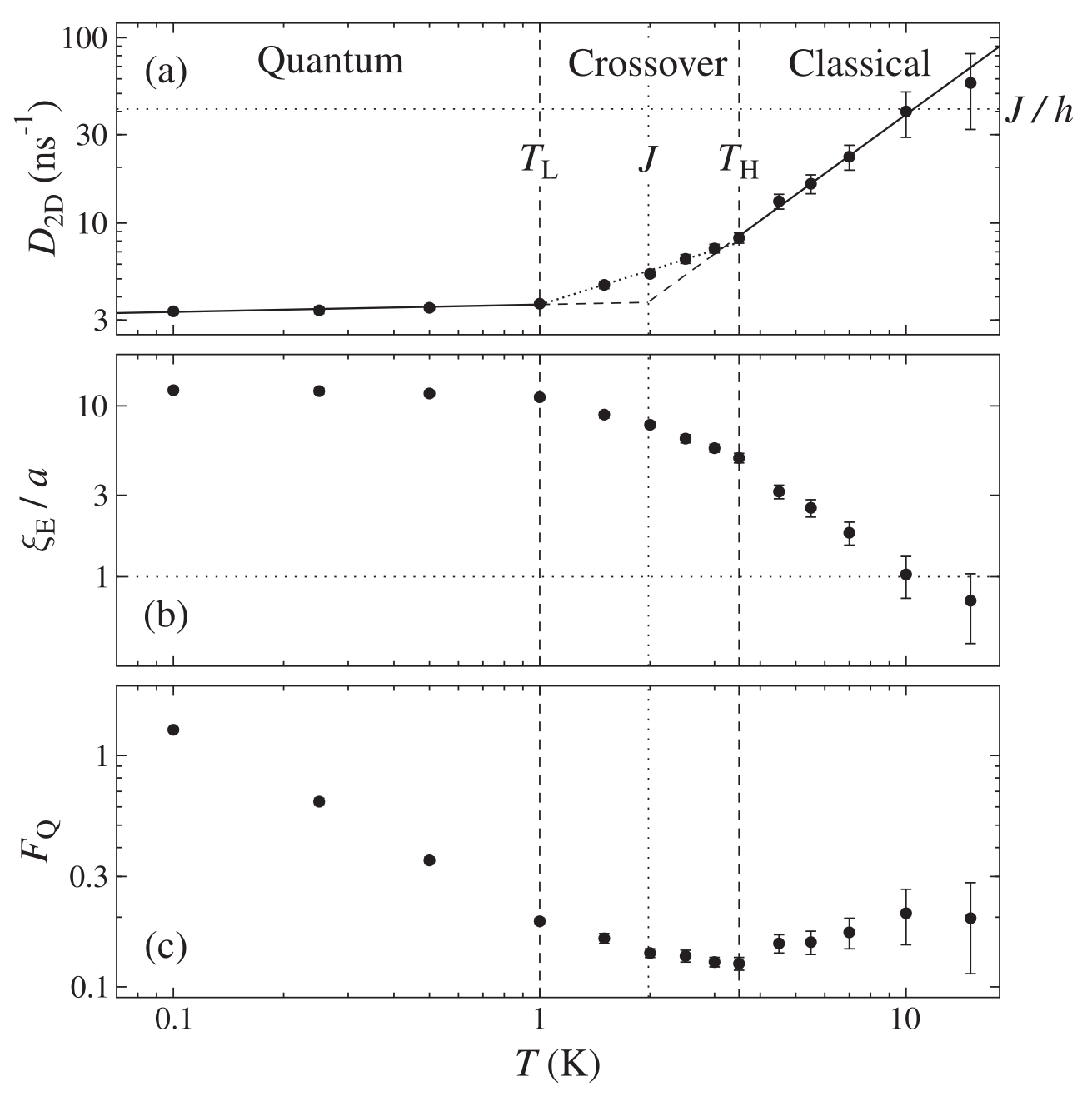}
    \caption{
    Entanglement properties of the triangular lattice antiferromagnet \ce{YbZnGaO4} from muon spin relaxation. 
    Comparison of 2D spin diffusion rate $D_\mathrm{2D}$ (a), entanglement length $\xi_\mathrm{E}=aJ/(h D_\mathrm{2D})$ (b), and QFI (c).
    Reproduced with permission.\scite{PhysRevB.106.L060401} 2022, American Physical Society.}
    \label{fig:ybzngao4}
\end{figure}
\ce{YbZnGaO4} is another candidate quantum spin liquid material featuring a triangular lattice of antiferromagnetically interacting \ce{Yb^3+} ions. It is a close relative of \ce{YbMnGaO4}, which was initially looked at as a very promising spin liquid candidate, but found to be very susceptible to site disorder \scite{PhysRevLett.119.157201}. The situation in \ce{YbZnGaO4} is at present less clear \scite{PhysRevB.104.224433, PhysRevB.106.L060401}. Partial entanglement information on the system has been extracted from muon spin relaxation ($\mu$SR) by Pratt et al. \scite{PhysRevB.106.L060401, Pratt_2023}; see Fig.~\ref{fig:ybzngao4}. It was argued that, since the 2D spin diffusion rate has a clear quantum-to-classical crossover as a function of temperature, its inverse provides a mean-free path that Pratt et al. interpret as an estimate for the entanglement depth. As far as we are aware, this is not a rigorously defined quantum correlation. Here, we instead want to highlight that they also obtained quantum Fisher information from the diffusive spectral density measured by the muon probe in Fig.~\ref{fig:ybzngao4}, showing the potential for measuring QFI with different experimental techniques. We will return to this point when discussing future directions for the field.

\subsection{Other systems}
In addition to the above magnetic systems, we note an inelastic neutron study of the two-leg ladder $S=1/2$ antiferromagnet \ce{C9H18N2CuBr4} \scite{Hong2023v3}. This material has a critical pressure $P_c~$\SI{1.0}{GPa} pressure above which the N\'eel-ordered phase breaks down \cite{Hong_2022}, and may host unconventional states. QFI was used to witness at least bipartite entanglement at a pressure of \SI{1.05}{GPa} up to at least \SI{1.1}{K}.\scite{Hong2023v3}

Experimental QFI results were also recently reported for two heavy fermion compounds: \ce{CeCu_{5.9}Au_{0.1}} \cite{fang_2024} and \ce{Ce3Pd20Si6} \cite{mazza_2024}. In both cases, spin-sector QFI obtained from inelastic neutron scattering indicates multipartite entanglement. 
\ce{Ce3Pd20Si6} was tuned to a quantum phase transition using an applied magnetic field, where a significant entanglement depth was found \cite{mazza_2024}. These studies provide promising results in the application of witnesses to correlated electrons.

\begin{sidewaystable*}[t]
  \setlength\tabcolsep{3 pt}
  \centering
  \caption{Table listing the materials reviewed in Section \ref{sec:applications}. 
  Models of the relevant underlying quantum system for each material are given including
  alternating Heisenberg chain (AHC), Heisenberg antiferromagnetic chain (HAFC), triangular antiferromagnet (TAF), and anisotropic Heisenberg chain with a uniaxial coupling anisotropy (XXZ chain); see text and references for more information. The entanglement and quantum correlation witnesses are listed in Table \ref{tab:witnesses}  and described in Section \ref{sec:theory}. (*Note, these authors also extract entanglement of formation from the concurrence. See Section \ref{sec:concurrence-two-tangle} for more details on these quantitites).
  Experimental probes that have been applied to the materials are also listed; probes are magnetic susceptibility ($\chi_m$), inelastic neutron scattering (INS), muon spin resonance ($\mu$SR), and relevant heat capacity ($C_v$) of the quantum subsystem e.g. magnetic part with phonon contribution subtracted.}  
  \begin{tabular}{ccccc}
    {\bf Materials} & {\bf Model} & \begin{tabular}{c} {\bf Entanglement and} \\ {\bf Correlation Witnesses} \end{tabular} & {\bf Experimental Probes} & {\bf References} \\
    \hline
    \ce{C8H16Cu2O10} & $S=1/2$ dimer & $\chi_\mathrm{EW}$, two-site QD & $\chi_m$ & Athira (2023) \scite{Athira_2023},Yurishchev (2011)\scite{PhysRevB.84.024418,10.1063/1.4862469} \\
    \ce{C9H18N2CuBr4} & $S=1/2$ ladder &  QFI  & INS (high pressure) & Hong (2023) \cite{Hong2023v3} \\
    \ce{CeCu_{5.9}Au_{0.1}} & heavy fermion & QFI & INS & Fang (2024)\cite{fang_2024} \\
    \ce{Ce3Pd20Si6} & \begin{tabular}{c} heavy fermion/\\strange metal \end{tabular} & QFI & INS & Mazza(2024)\cite{mazza_2024} \\
    copper carboxylate & $S=1/2$ dimer & $\chi_\mathrm{EW}$, concurrence & $\chi_m$ & Souza (2009)\scite{PhysRevB.79.054408} \\
    \ce{(Cr7Ni)2} supramolecular dimers & $S=1/2$ coupled rings & concurrence & INS, $\chi_m$ & Candini (2010), Garlatti (2017) \scite{PhysRevLett.104.037203, Garlatti2017, Garlatti_2019}\\
    Cs$_2$CoCl$_4$ & $S=1/2$ XXZ chain & one-tangle, two-tangle, QFI  & INS & Laurell (2021) \cite{PhysRevLett.127.037201} \\
    \ce{[Cu(\mu-C2O4)(4-aminopyridine)2(H2O)]_n} & $S=1/2$ HAFC &  $\chi_\mathrm{EW}$, QFI (qualitative T-dependence) & INS (powder) & Matthew (2020) \cite{PhysRevResearch.2.043329}\\
    \ce{Cu(NO3)2 2.5H2O} & $S=1/2$ AHC &  $\chi_\mathrm{EW}$, concurrence, two-site QD  & INS, $\chi_m$, $C_v$ & \begin{tabular}{c} Brukner (2006) \scite{PhysRevA.73.012110}, Singh (2013) \cite{Singh2013}, \\ Wie{\'{s}}niak (2005) \scite{Wie_niak_2005}, Yurishchev (2011)\cite{PhysRevB.84.024418} \end{tabular} \\
    \ce{Cu(thiazole)2Cl2} & $S=1/2$ HAFC & concurrence & $\chi_m$ &  Chakraborty (2012) \scite{Chakraborty2012} \\
    \ce{Fe2(\mu2-oxo)-(C3H4N2)6(C2O4)2} & $S=5/2$ dimer & $\chi_\mathrm{EW}$ & $\chi_m$ & Reis (2012) \scite{Reis_2012} \\
    KCuF$_3$ & $S=1/2$ HAFC & \begin{tabular}{c} one-tangle, two-tangle, QFI, QV, SI\\ spatial quantum correlation functions \end{tabular} & INS & Scheie (2021) \cite{PhysRevB.103.224434}, 
    Scheie (2023) \cite{Scheie2023}\\
    \ce{KNaMSi4O10} (M=Mn, Fe, or Cu) & $S=5/2,2,1/2$ & $\chi_\mathrm{EW}$ & $\chi_m$ & Pinto (2009) \scite{Soares-Pinto_2009} \\
    KYbSe$_2$ & $S=1/2$ TAF & one-tangle, two-tangle, QFI  & INS & Scheie (2024) \cite{Scheie2023tlafm} \\
    LiHo$_{0.045}$Y$_{0.955}$F$_4$ & 3D dilute Ising model & concurrence  & $\chi_m$, $C_v$ & Ghosh (2003) \cite{Ghosh2003} \\
    \ce{MgMnB2O5} & \begin{tabular}{c} $S=5/2$ dimer \\ (Griffiths phase) \end{tabular} & $\chi_\mathrm{EW}$ & $\chi_m$ & Rappoport (2007) \scite{PhysRevB.75.054422} \\
    \ce{MgTiOBO3} & \begin{tabular}{c} $S=1/2$ dimer \\ (random singlet phase) \end{tabular} & $\chi_\mathrm{EW}$ & $\chi_m$ & Rappoport (2007) \scite{PhysRevB.75.054422} \\
    \ce{NH4CuPO4\cdot H2O} & $S=1/2$ dimer & $\chi_\mathrm{EW}$, concurrence* & $\chi_m$, $C_v$ & Chakraborty (2014)\scite{doi:10.1063/1.4861732} \\
    nitrosyl Fe complexes  & $S=1/2$ dimer & $\chi_\mathrm{EW}$, two-site QD & $\chi_m$, $C_v$ & Aldoshin (2014) \scite{10.1063/1.4862469} \\
    \ce{Na2Cu5Si4O14} & $S=1/2$ cluster chain & $\chi_\mathrm{EW}$, concurrence* & $\chi_m$ & Souza (2008)\scite{PhysRevB.77.104402} \\
    \begin{tabular}{c} spiro-bis (1, 9-disubstituted\\-phenalenyl) boron \end{tabular}  & $S=1/2$ HAFC & concurrence & $\chi_m$ &  Chakraborty (2013)\scite{10.1063/1.4824458} \\
    \ce{Sr14Cu24O41} & $S=1/2$ cluster chain & \begin{tabular}{c} concurrence, two-site QD \end{tabular} & $\chi_m$ & Sahling (2015) \scite{Sahling2015} \\
    \ce{YbZnGaO4} & $S=1/2$ TAF & QFI  & $\mu$SR & Pratt (2023) \scite{PhysRevB.106.L060401, Pratt_2023} \\
    \hline
  \end{tabular}
  \label{tab:Materials}
\end{sidewaystable*}

\section{Broader Perspective and Future Directions}\label{sec:broaderperspective}
In this section, we outline some of the frontiers of the field. 
The past sections show that witnesses already have been applied to a wide assortment of materials, but there is no reason to stop here. Indeed, as pointed out by Brukner et al.\scite{PhysRevA.73.012110}, even reanalysis of data from past experiments is likely to reveal entanglement and other quantum correlations in a much broader range of systems. Until now, many efforts have focused on types of systems for which it is theoretically motivated to look for entanglement. In the future, one can imagine the converse: experimentally witnessing significant quantum correlations in a new material may immediately reveal the need for quantum-mechanical modeling. For this to become a commonplace scenario, it will be necessary to consider a range of experimental techniques and observables, such that we can probe general quantum materials. Since no single witness can certify entanglement in all possible states, or be measured using every technique, it is worthwhile to consider new witnesses, and other ways or contexts for entanglement certification.

\subsection{Electronic systems and other spectroscopies}\label{sec:electronic}
\begin{figure}
    \includegraphics[width=\columnwidth]{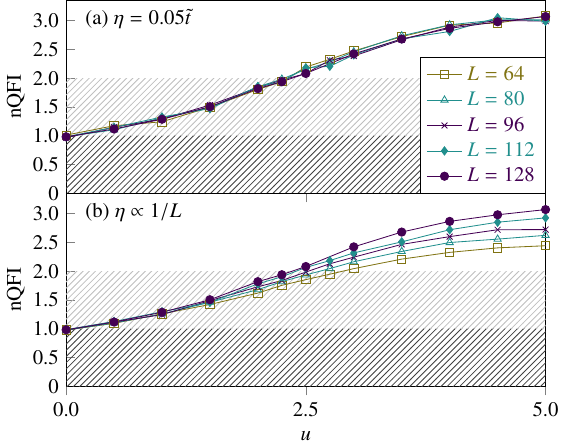}
    \caption{Calculated normalized quantum Fisher information in the half-filled Hubbard chain with hopping energy $\tilde{t}$ and Hubbard repulsion strength $U$. The spin-$1/2$ antiferromagnetic Heisenberg chain is recovered in the strong-coupling limit $u=U/\tilde{t}\rightarrow \infty$. Here, the dynamical spin structure factor $S(k,\omega)$ was computed using DMRG, and used to evaluate the nQFI at the antiferromagnetic momentum $k=\pi$. 
    (a) With fixed realistic energy resolution $\eta$ it is possible to witness at least bipartite entanglement (light shaded region) even at weak interactions, and at least tripartite entanglement (white region) at intermediate interaction strengths. (b) The impact of the energy resolution $\eta$, here chosen to depend on the system size $L$. 
    Adapted with permission.\scite{PhysRevB.106.085110, PhysRevB.107.119901} 
    2022-2023, American Physical Society.}
    \label{fig:hubbard}
\end{figure}

It is striking that \emph{almost} all of the experimental works cited in the 
preceding sections involve materials that have been treated as spin systems. The reason is largely that entanglement witnesses originated in quantum information theory, in the context of qubits, which are two-level systems. Although there is a growing quantum information scientific interest in ``qudits'' with higher $d$-dimensional local Hilbert spaces due to their technological advantages \scite{Cozzolino2019, Erhard_2020}, developing suitable witnesses is mathematically challenging. Thus, the quantum information progress has mainly mapped onto spin-$1/2$ systems, and much less to $S>1/2$ spins and electrons with both charge and spin degrees of freedom.
Nevertheless, electrons are the building blocks of quantum materials, and worth studying more closely. It is also clear that electronic condensed matter systems can be entangled in ways pure spin systems cannot \scite{RevModPhys.80.517}, including in charge and particle channels.

Currently, the most promising witness for electronic systems is the quantum Fisher information. As we discussed earlier, the construction by Hauke et al. \scite{Hauke2016} shows that the QFI remains a witness of multipartite entanglement as long as it is evaluated for a dynamical susceptibility associated with a bounded Hermitian local operator, be it spin, charge density, or otherwise. 
In Ref.~\scite{PhysRevB.106.085110, PhysRevB.107.119901} we demonstrated theoretically that observing multipartite spin-channel entanglement in the Fermi-Hubbard chain is possible using inelastic neutron scattering and realistic energy resolution; see Fig.~\ref{fig:hubbard}. Bipartite entanglement can be witnessed even at very weak interactions $u$, and at least tripartite entanglement can be witnessed at intermediate repulsion. Recently, theoretical QFI results have also been reported for Kondo lattice models relevant to quantum critical strange metals, finding multipartite entanglement near the Kondo destruction quantum critical point \cite{fang_2024, mazza_2024}.

Future directions include measuring the QFI in the charge sector, which is possible using spectroscopies that probe the dynamical charge structure factor, such as non-resonant inelastic x-ray scattering (NRIXS) \cite{schulke2007electron} and momentum-resolved electron energy loss spectroscopy (M-EELS)\cite{SciPostPhys.3.4.026}. However, NRIXS has the drawback of coupling to the entire charge density, including core electrons, and is unlikely to witness entanglement outside of special cases where one can isolate the scattering from specific bands. M-EELS is more promising in this regard, as it probes the physics near the Fermi energy, which is typically where the bands of interest are situated. Resonant inelastic x-ray scattering (RIXS) \cite{RevModPhys.83.705} can also be used for witnessing entanglement in both the spin and charge sectors. However, the RIXS matrix element is quite complicated, which means detailed modeling may be required to rigorously extract dynamical susceptibilities. An approach for this is discussed in Ref. \scite{Ren2024}, and tested on the iridate dimer system \ce{Ba3CeIr2O9}. Although entanglement between the Ir orbitals has yet to be witnessed using this approach, simulations suggest it can be achieved with polarization analysis or by optimizing the choice of incident energy and momentum transfer. 
A recently proposed protocol \scite{Malla2023} goes beyond the Hauke et al. construction, connecting single-particle Green's functions to multipartite entanglement, which could enable entanglement detection using scanning tunneling microscopy and angle-resolved photoemission spectroscopy (ARPES).

\subsection{Novel witnesses and correlation functions}
As the previous section argues, there are clear paths towards broadly applying QFI (and related quantum coher\-ence-based measures) to electronic systems and higher-$S$ spin systems. The witnessing of pairwise entanglement presents further theoretical challenges. 
It involves constructing a two-site reduced density matrix (which is larger than in the $S=1/2$ case), and relating its elements to experimentally accessible quantities, which, with current techniques, largely implies one- and two-site correlation functions of local operators. 
Recent work expressing the two-site density matrix for Hubbard systems in one- and two-particle Green's functions \cite{Roosz2023} may serve as a starting point for work in this direction for electronic systems. Efforts have also been made to generalizing the concurrence to $S>1/2$ systems and, more generally, systems with higher-dimensional local Hilbert spaces \cite{Li2008, Eltschka_2014, Osterloh_2015, Bahmani2020}. However, it is an open question whether simple expressions can be obtained for condensed matter systems of experimental interest.

Another intriguing open question is whether experimental techniques can be developed to probe correlation functions beyond the one-and two point functions we have discussed so far. If possible, it could open up the paths to measuring quantities generalizing the one- ($\tau_1$) and two-tangles ($\tau_2$) discussed in this review into, for example,  three- \cite{PhysRevA.61.052306, PhysRevA.62.062314, PhysRevLett.97.260502, Gour_2010, PhysRevA.94.012323} and $n$-tangles \cite{PhysRevA.63.044301, Li2012}. These were introduced in quantum information to diagnose and understand the patterns of entanglement in systems of $n>2$ qubits, and could potentially also help capture the entanglement structure of clusters of spins within large crystals. 
Alternatively, one can consider relaxing the condition that the correlation functions probe local operators. Entanglement witnessing using cross-correlations of electrical currents has been discussed for devices  \cite{PhysRevLett.118.036804, PhysRevB.104.245425}, and could potentially also be probed optically in materials. Overall, there are many avenues open for further work into extending existing witnesses to new classes of systems, constructing new witnesses, and developing experimental techniques.

\subsection{Beyond equilibrium}\label{sec:beyondeq}
There has been tremendous progress in our understanding of quantum dynamics and nonequilibrium quantum phenomena over the last decades \scite{RevModPhys.83.863, Eisert_2015, RevModPhys.91.021001}. In particular, it is now understood that thermalization processes in closed many-body systems are linked to quantum chaos and the dynamics of the entanglement entropy, which can be understood as a generalization of the entropy familiar from thermodynamics and statistical mechanics \cite{PhysRevLett.106.040401, Eisert_2015, Kaufman2016, RevModPhys.91.021001}. Typical systems, equipped with eigenstates obeying entanglement entropy volume laws, are believed to equilibrate according to the eigenstate thermalization hypothesis \cite{PhysRevA.43.2046, PhysRevE.50.888, Rigol2008, D_Alessio_2016, Kaufman2016, PhysRevLett.124.040605}. However, there appear to exist special quantum systems that can escape thermalization, including integrable systems, many-body localized (MBL) systems \scite{RevModPhys.91.021001}, quantum many-body scars \cite{Chandran2023}, and certain classes of periodically driven systems \scite{Moessner_2017}. On the experimental side, there has been significant advances in the control of cold atom \scite{RevModPhys.80.885} and condensed matter systems \scite{doi:10.1146/annurev-matsci-070813-113258, Basov_2017}, as well as in time-resolved spectroscopic techniques. Stepping away from thermal equilibrium has allowed ultrafast control of material properties \scite{RevModPhys.93.041002}, access to otherwise hidden or metastable phases, and the realization of novel, fundamentally nonequilibrium phases, such as time crystals \scite{Zhang_2017, Choi_2017} and new topological phases \scite{Moessner_2017}. 

\begin{figure*}
	\includegraphics[width=\textwidth]{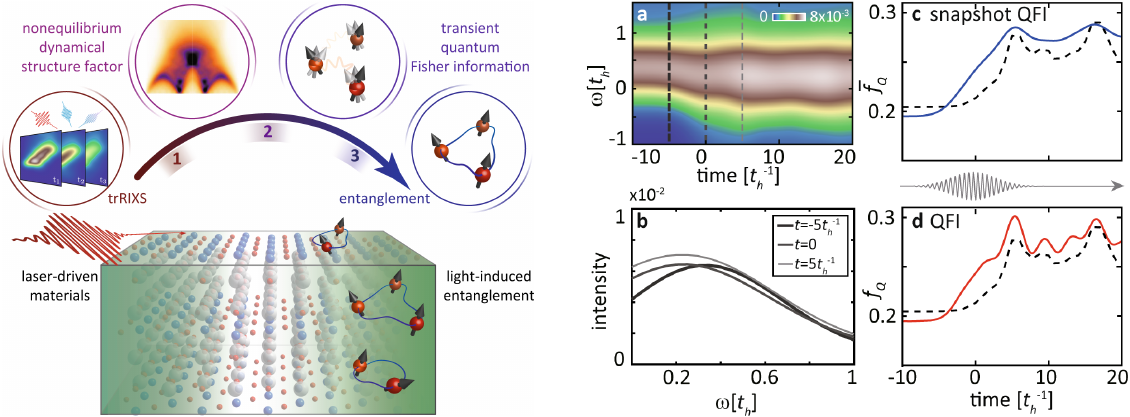}
    \caption{\label{fig:trRIXS}
    Left: a proposed method to probe light-driven entanglement in quantum materials. The system is driven out of equilibrium by a pump laser, and time-resolved resonant inelastic x-ray scattering (trRIXS) is used to probe the collective excitations. From the trRIXS response function, the nonequilibrium dynamical structure factor is recovered, and then entanglement is witnessed using a transient quantum Fisher information.    
    This approach is inherently different from the thermal equilibrium QFI discussed elsewhere in this review, and the reader is referred to \scite{Hales2023} for detailed derivations. 
    Right: (a) Evolution of the nonequilibrium dynamical spin structure factor $S(q=\pi/6,\omega,t)$ for a one-dimensional extended Hubbard model relevant to cuprate chain systems. (b) Spectral distribution. (c) Time-dependence of a ``snapshot'' QFI (blue solid line), calculated from $S(q,\omega,t)$ as if it was an equilibrium spectrum, and of the exact QFI (black dashed line). (d) Time-dependence of a self-consistently corrected QFI (red solid line), and of the exact QFI (black dashed line). This self-consistent calculation includes effects due to higher-order time derivatives, and better captures the exact result.     
    Panels from figures 1 and 4 of \scite{Hales2023}, reproduced under the CC BY 4.0 license\scite{ccby4}. Copyright 2023, J. Hales et al., published by Springer Nature.}
\end{figure*}

All these aspects suggest that probing entanglement in quantum materials as a function of time is a worthwhile direction. While it is not possible to probe the entanglement entropy directly in condensed matter, unlike cold atom systems \scite{Islam2015, Kaufman2016}, the quantum Fisher information again appears promising. However, the construction of Hauke et al. \scite{Hauke2016} does not immediately generalize to the out-of-equilibrium case. 
It is clear that QFI calculated from eigenstates can detect entanglement dynamics in the Ising chain after a quantum quench (i.e. a sudden change of an applied magnetic field) \scite{Pappalardi_2017}, in systems with quantum many-body scars \scite{PhysRevLett.129.020601, PhysRevB.107.035123} and correlated fermionic systems \cite{PhysRevLett.130.106902}. In special cases \scite{Pappalardi_2017}, it turns out to be possible to define a generalized fluctuation-dissipation theorem and rewrite the QFI as an integral over a generalized susceptibility, but this does not extend to generic systems. A better approach is to instead carefully relate QFI to time-dependent response functions that can be probed experimentally by time-resolved spectroscopy. There have been recent proposals to use time-resolved resonant xray scattering (trRIXS) \scite{Hales2023, Suresh2022}, see Fig.~\ref{fig:trRIXS}, but, to the best of our knowledge, experimental tests await.

\section{Technical developments and challenges for scattering experiments}\label{sec:techdevelopments}
This more specialized section is primarily intended for readers interested in current experimental technique development. Sec.~\ref{ssec:entbeam} reviews progress made on realizing scattering probes that are themselves entangled. Sec.~\ref{ssec:expchallenges} discusses experimental requirements for resolution and polarization capabilities, as well as future instrumentation.

\subsection{Entangled beams}\label{ssec:entbeam}

So far we have considered measurement of entanglement in materials using conventional probes. However, entangled beams could offer an alternative path to probing quantum correlations in materials. 

While beams of entangled neutron pairs are possible in principle, they are currently not available due to issues of sources and moderators. Conventional reactor and spallation sources produce unentangled fluxes of neutrons so isotope decay is needed to produce pairs through double ($n=2$) ($^5$H, $^{10}$He, and $^{21}$B) emission processes \cite{Kondev_2021}. However, isotope sources produce orders of magnitude less flux than that needed for scattering experiments. Besides, the moderation process involves collisions which cause loss of entanglement. An alternative approach is to self-entangle neutrons. 

Modal entanglement involves entangling disjoint Hil\-bert space properties of the neutron's spin, position, momentum {\it etc}. Engineering of such self-entangled neutron states has been demonstrated with high efficiency and precision \cite{Sponar_2012,Shen2020}: Entanglement of two \cite{Hasegawa2003} and three \cite{Hasegawa_2010, Sponar_2012} degrees of freedom by utilizing neutron polarimetry and radio frequency and static magnetic fields have been achieved. 

\begin{figure*}
    \includegraphics[width=\textwidth]{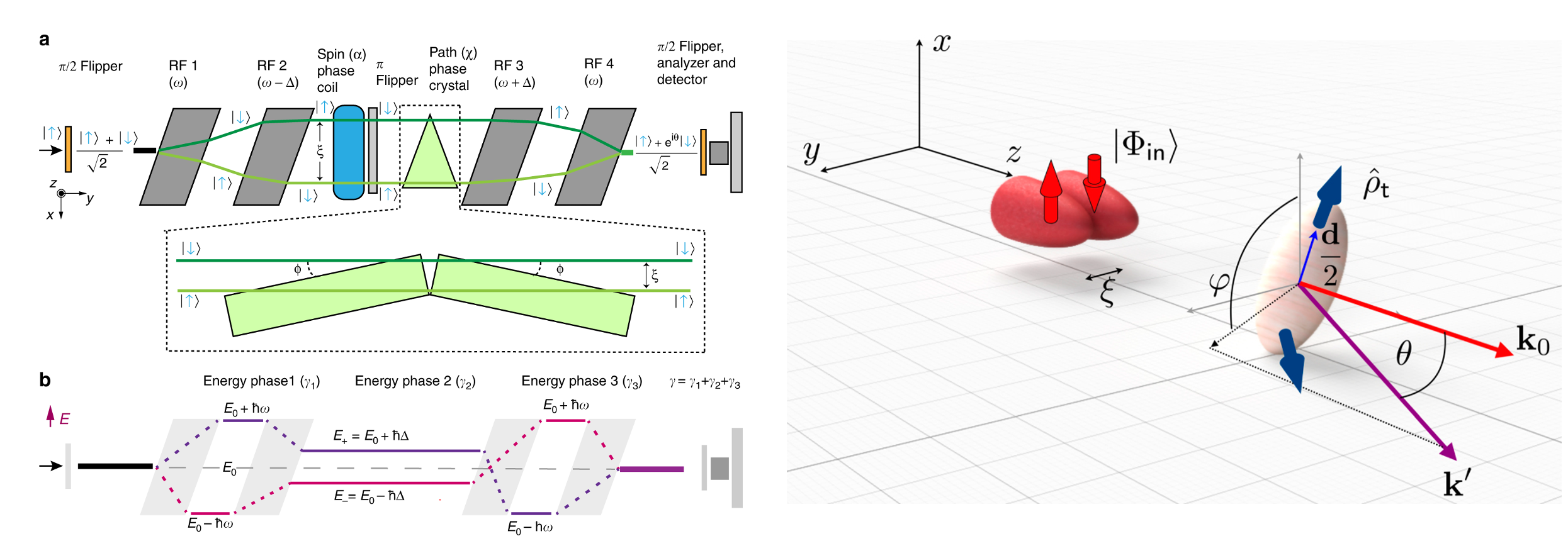}
    \caption{\label{fig:entangled_beams} Left: 
    a Schematic plan view of the main spin manipulation components of the Larmor instrument used to generate the mode-entangled Greenberger-Horne-Zeilenger states in \scite{Shen2020} showing the evolution of the neutron path and spin states along the beam line. A superposition of up and down spin states at the beginning of the instrument are manipulated using RF flippers and magnetic fields. They are refracted along different paths and separated by the entanglement length, $\xi$, in the space between the second and third RF flippers. b A plot of the total neutron energy for each neutron spin state along the beam line. Each RF flipper reverses a neutron spin state at the same time as it exchanges a quantum of RF energy with that state. A difference in the energy phase between the two spin states develops in the space between each pair of RF flippers because the two states have different total energies. Panels from Fig. 1 of \scite{Shen2020}, reproduced under the CC BY 4.0 license\scite{ccby4}. Copyright 2020, J. Shen \textit{et al.}, published by Springer Nature.
    Right: 
    Magnetic scattering of an entangled probe (entanglement length $\xi$) from a dimer of size $\mathbf{d}$ \scite{Irfan_2021} allows the entanglement in the dimer to be quantified when $\xi$ and $d$ are similar in size. 
    Fig. 3 of \scite{Irfan_2021}, reproduced under the CC BY 4.0 license\scite{ccby4}. Copyright 2021, A. A. M. Irfan \textit{et al.}, published by IOP Publishing.
    }
\end{figure*}

 Flexible beams that integrate into neutron scattering instrumentation  have demonstrated entangled properties \cite{Shen2020,Kuhn_2021} proposed to be suitable ``for investigations of microscopic magnetic correlations in systems with strongly entangled phases, such as those believed to emerge in unconventional superconductors'' \cite{Shen2020}. These use spin echo type techniques to manipulate wavepackets \cite{Gahler_1998,Groitl_2016,mckay_2023} with components such as magnetic Wollaston prisms and resonance-field radio-frequency flippers that can fit on diffractometers and spectrometers; see Fig. \ref{fig:entangled_beams}. 
 
Robust beams over different apparatuses and neutron pathways \cite{Kuhn_2021} entangling spin, trajectory, and energy \cite{Shen2020}, as well as orbital angular momentum \cite{Le_2023}, have been generated, showing promising progress towards application.

On the theory side, extensions of standard scattering theory \cite{Lu_2020} to include mode entanglement \cite{Irfan_2021} results in a generalization of van Hove scattering theory \cite{vanHove_1954}. The magnetic response, although still expressed in terms of two-point correlation functions, is modified reflecting the entanglement of the beam and that within the scattering target. For example, tuning the beam's entanglement length allows the interrogation of spatial scales by analyzing interference patterns in the differential cross-section \cite{Irfan_2021}. 

For the simplest case of a spin dimer target, Fig. \ref{fig:entangled_beams}, a Young-like interference pattern observed if the target state is un-entangled becomes quantum erased when the target state becomes maximally entangled. This suggests that features of entanglement in materials may be revealed and interpreted through qualitative signatures in the scattering patterns. More work on scattering from different cases is however needed to determine how useful this could be.

While progress has been made in theory and experiment, open questions remain. Experiments on quantum magnetic materials are required to demonstrate the effectiveness of these techniques. Also, exploitation of properties such as the entanglement length of the probe are novel yet how well these match with the entanglement scales in target materials needs more clarification. Much work remains to be done to explore the potential of this novel experimental approach.

To-date, less work has been undertaken towards entangled photon beams suitable for studying properties of materials. A key bottleneck here is preparing sufficiently intense beams of entangled x-rays. 
Recently, the use of Hong-Ou-Mandel interferometers to create fully entangled N00N ($N=2$) biphoton states at high intensity synchrotron sources has been proposed \cite{Durbin_2022}. Implementation is expected to be achievable with currently available beamsplitters and interferometers on the latest generation of synchrotrons. Beams with a high fraction of biphoton pairs can then be separated using diffraction from the single photon (unentangled) background. 
Intense emission of entangled x-ray photon pairs may also be achievable using undulators at free electron lasers \cite{Zhang_2023}. This would bring significant advantages in intensity and time structure which could open many new directions of study. 

\subsection{Experimental capabilities and requirements}\label{ssec:expchallenges}

Entanglement witnesses, quantum correlators, and entangled beams bring up new measurement challenges and instrumentation needs. For spin systems probed with neutron spectroscopy, fully polarized neutron scattering (FPNS) \cite{Boothroyd_2020,Lovesey1986} where the ${x,y,z}$ spin components of the incident and scattered neutrons can be selected, will ideally be needed. FPNS on diffractometers can determine moment directions and sizes in more complex magnetic structures than is possible with unpolarized beams. FPNS is also useful to extract sum rules from elastic and inelastic scattering as an alternative to sometimes hard to implement absolute normalizations (especially on reactor based instrumentation). 

Current instruments such as HYSPEC \cite{Zaliznyak_2017} at the Spallation Neutron Source at Oak Ridge National Laboratory and D7 \cite{Gabrys1997} at the Institut Laue-Langevin (ILL) use supermirror analyzers, giving them limited capabilities for measuring $S^{\alpha\beta}(\textbf{Q})$ ($\alpha,\beta=x,y,z$) in terms of polarization components and reciprocal space mapping respectively. Efficient mapping for diffuse FPNS is an important future goal for instrumentation given the essential information locked in the scattering from the different spin components. 

The demands on inelastic instrumentation are greater than for diffraction. The extraction of witnesses require wavevector and energy ranges sufficient to be used for transformation into combinations of coordinate spaces {\it i.e.} $\textbf{R,}\textbf{Q}, t, \omega$ which need to be accurately transformed. Mixtures of resolution conditions can be tolerated {\it e.g.} for the extraction of $G(\textbf{r},t)$, if transformations are taken with care \cite{Scheie2022}. This can be achieved with many current direct geometry time-of-flight spectrometers using $\textbf{Q}, \omega$ mapping executed with a combination of incident energies to gain coverage across 
multiple Brillouin zones. 

The extraction of spin components [$S^{\alpha\beta}(\textbf{Q},\omega)$] is hard to accomplish as nearly all instruments are either unpolarized or can extract only one component at a time. 
One approach is to use fully polarized triple-axis spectroscopy \cite{Shirane_Shapiro_Tranquada_2002, Boothroyd_2020} to identify the spin components involved in the signal. This can be aided by application of magnetic fields to Zeeman split in energy excitations with different spin quantum numbers.
Computational modeling can also be used as an aid to separating components as was utilized for QFI anaysis of Cs$_2$CoCl$_4$ \cite{PhysRevLett.127.037201}. 

Simplifications of EWs to more easily measured quantities such as the use of $S(\textbf{Q})$ for QFI rather than an energy integral requiring $S(\textbf{Q},\omega)$ has been proposed \cite{PhysRevB.107.054422}. Adoption of such strategies could make a significant practical impact and this needs to be pursued further.  

Given the importance of inelastic FPNS it is notable that next-generation neutron spectrometers, such as CHE\-SS \cite{STS2019} at the Second Target Station at Oak Ridge National Laboratory, are expected to offer orders of magnitude increases in performance and will implement full polarization analysis. Such capabilities could also be implemented on reactor based CAMEA-type \cite{Camea2016} instruments. These would be revolutionary capabilities for extraction of quantum witnesses and correlators in materials.  

For QFI much of the integral weight is contributed by low-energy scattering. This requires high energy resolution and at the same time sufficient wave-vector resolution. This can be challenging for neutron spectroscopy and detailed experimental studies will be needed to find optimal 
scattering configurations including the use of ultra-high-resolution spin-echo techniques \cite{Bayrakci2013, Groitl_2016}. Spin echo can also measure correlations in $\textbf{R}$ and $t$ \cite{Groitl_2016} so can avoid Fourier transforms.

Polarization is also crucial for entangled beam studies, Section \ref{ssec:entbeam}. 
Specialized high flux polarized beamlines will be needed for efficient counting. Current 
polarized triple axis 
spectrometers suffer from having a single detector and need to be scanned point-by-point, making measurements orders of magnitude slower than conventional unentangled experiments. Spin manipulation components, such as 
magnetic Wollaston prisms \cite{PYNN2009}, that operate over wide scattering angles would bring the significant efficiency gains needed if this were to become a mainstream technique.

Finally, as noted in sections \ref{sec:electronic} and \ref{sec:beyondeq} photons hold great promise for application to EWs. However, as there has been little work so far, we believe further discussion of experimental capabilities and requirements is premature until technical needs become clearer. It can be expected though that both nonresonant and resonant xray techniques will be of interest.

The measurement of entangled charge-charge correlations in materials \cite{PhysRevB.106.085110} is of obvious interest and non-resonant inelastic x-ray scattering could be useful here. Meanwhile RIXS, which has a more complex scattering matrix element which depends on the orbital transition involved, can provide access to two-point and four-point spin correlation functions as well as entanglement of spin, charge, and orbital degrees of freedom. Although the energy resolution is typically in the tens of millivolts, measurement times are fast enabling out-of equilibrium studies \cite{Hales2023} as well as the dynamics in thin films, not accessible to neutrons due to flux constraints, to be probed. Concerted development efforts along with experiments on candidate materials will be needed to fully utilize the potential of xrays.

\section{Conclusion}\label{sec:conclusion}
We have reviewed applications of entanglement and quantum correlation witnesses to condensed matter, including important derivations, past experimental successes, and future directions and challenges. The field is clearly at an inflection point: new, model-independent witnesses have opened up the study of complicated and poorly understood materials. Although it is often possible to guess, based on (fallible) heuristics, whether a given system hosts an entangled state, witness measurements allow \emph{quantitative} and therefore definitive statements to be made. Going forward, such quantitative information can help inform theoretical modeling and reasoning about states in materials, enabling new understanding.

Finally, as is evident from Table \ref{tab:Materials}, within condensed matter, entanglement witnesses have been applied so far almost exclusively to quantum magnets, where they have diversified from dimerized materials to more complex quantum critical and spin liquid states. 
However, as detailed in this review there is significant scope for wider application to quantum critical systems, heavy fermions, liquid helium, exotic superconductors, and correlated electron systems generally. 
Indeed, recent INS results on the strange metal \ce{Ce3Pd20Si6} show that QFI can effectively witness the entanglement in quantum critical metallic systems \cite{fang_2024, mazza_2024}. This opens the way for protocols, similar to that used for quantum spin liquids \cite{Scheie2023tlafm} to be developed. 
The EWs in Table \ref{tab:Materials} utilized measurements based on susceptibility, heat capacity, and neutron scattering. As this review has outlined, there is significant scope for, and activity towards, expanding the range of quantities and experimental techniques that can be effectively used to probe entanglement in quantum matter, both in- and out-of-equilibrium. For these reasons, the future of the field will undoubtedly bring exciting new developments.

\appendix
\section{Sketch of linear response theory}\label{sec:qfi:linearresponse}
This appendix provides a brief primer on linear response theory, including different expressions of $\chi^{\prime\prime}$. The purpose is to set a consistent notation, as mixing different notations can lead to mistaken conclusions about the entanglement depth. We follow the convention used in Appendix B of Lovesey's book \cite{Lovesey1986}, to which readers are referred for additional details. For the convenience of experimentally minded readers we will keep factors of $\hbar$ and $k_B$ explicit, but suppress potential momentum dependence.

In many experimental techniques one applies a (relatively weak) perturbation to a system and then observes the effect in the response of a measured quantity. This may be described by considering an isolated system that is initially (at time $t=-\infty$) at thermal equilibrium with temperature $T$, and described by a time-independent Hamiltonian $\mathcal{H}_0$. A time-dependent external perturbation $\mathcal{H}_1$ is allowed to act on the system, giving the total Hamiltonian at time $t$
\begin{align}
	\mathcal{H}(t)	&=	\mathcal{H}_0 - \mathcal{H}_1 \equiv \mathcal{H}_0 - \hat{B} h(t),	
\end{align}
where the time dependence of $\mathcal{H}_1$ is captured by the real-valued function $h(t)$ and $\hat{B}$ is a Hermitian operator. 
The response of the system is reflected in a change of a variable $A$ that is not itself explicitly time-dependent and corresponds to a Hermitian operator $\hat{A}$. 
For a linear response we have
\begin{align}
	\overline{A(t)}	&=	\langle \hat{A}\rangle_0 + \int_{-\infty}^t \mathrm{d}t' \phi_{AB}\left( t-t'\right) h\left( t'\right).	\label{eq:responsefunction:implicitdefinition}
\end{align}
where the equilibrium average $\langle \hat{A}\rangle_0	=	\mathrm{Tr} \left[ \rho_0 \hat{A}\right]$ and $\rho_0$ is the density matrix at $t=-\infty$. Eq.~\eqref{eq:responsefunction:implicitdefinition} implicitly defines the real-valued response function $\phi_{AB}(t)$ that incorporates history effects. The perturbation can generally be Fourier decomposed into a set of frequencies $\left\{ \omega \right\}$, each of which has the time dependence $e^{i\omega t}$. We assume that the perturbation $\mathcal{H}_1$ describes an adiabatic process, such that the system is in equilibrium at each time $t$, albeit with a state that is time-dependent. This is achieved by turning on the perturbation very slowly, by making the replacement $e^{i\omega t}\rightarrow e^{i\omega t+\epsilon t}$, where $\epsilon>0$ is a small number. At the end of the calculation, we will let $\epsilon\rightarrow 0^+$. Taking the time dependence to be $h(t)=h\exp\left( \epsilon t\right) \cos \left( \omega t\right)$, where $h$ is real, we can write
\begin{align}
	\overline{A(t)}	&=	\langle \hat{A}\rangle_0 - h \mathrm{Re} \left\{ \exp \left( i\omega t\right) \chi_{AB} \left[ \omega\right] \right\}
\end{align}
where the generalized susceptibility $\chi\left[ \omega\right]$ is defined as
\begin{align}
	\chi_{AB}\left[\omega\right]	&=	-\lim_{\epsilon\rightarrow 0^+} \int_0^\infty \mathrm{d}t \phi_{AB}(t) \exp \left( -i\omega t - \epsilon t\right)	\label{eq:generalizedsusceptibility:def}\\
	&=	\chi'_{AB} \left[ \omega\right] + i \chi''_{AB} \left[ \omega\right].	\label{eq:generalizedsusceptibility:decomposition}
\end{align}
Following Lovesey we use the notation $\left[ \omega\right]$ to indicate a one-sided Fourier transform, in which the integral is taken only over positive real axis. 
$\chi'_{AB}\left[ \omega\right]$ and $\chi''_{AB}\left[ \omega\right]$ denote the real and imaginary parts of the generalized susceptibility, respectively.  Since $\phi_{AB}(t) \in \mathbb{R}$ for real external perturbations, we must have $\chi'\left[ \omega\right]	= \chi'\left[ -\omega\right]$, $\chi''\left[ \omega\right]	= -\chi''\left[ -\omega\right]$.

We next derive the fluctuation-dissipation theorem. By assumption of a weak perturbation, the average value of $\mathcal{H}_1$ is very small compared to $\langle \mathcal{H}_0\rangle$. The Heisenberg equation gives the time evolution of the density matrix $\rho$, $i\hbar \dot{\rho}(t)	= \left[ \mathcal{H},\rho\right]$ 
with initial condition $\rho(-\infty)=\rho_0$, Letting $\rho(t)	=	\rho_0	+ \Delta \rho(t)$ and ignoring second-order effects,
\begin{align}
	i\hbar \dot{\rho(t)}	&=	- \left[ \mathcal{H}_1,\rho_0\right] - \left[ \Delta\rho(t), \mathcal{H}_0\right].
\end{align}
Expressing $\rho(t)$ in the interaction picture\cite{Sakurai1994},
\begin{align}
	\rho_I(t)	&=	e^{it\mathcal{H}_0/\hbar} \rho(t)	e^{-it\mathcal{H}_0/\hbar}\\
    \Rightarrow i\hbar \dot{\rho_I}(t)  &= e^{it\mathcal{H}_0/\hbar} \left[ \rho_0, \mathcal{H}_1 \right]	e^{-it\mathcal{H}_0/\hbar}.	\label{eq:deltarhodot}
\end{align}
Returning to the Schr\"odinger picture and integrating yields
\begin{align}
	\Delta \rho (t)	&=\frac{1}{i\hbar}	\int_{-\infty}^t dt' h(t') \left[ \rho_0, \hat{B}(t'-t)\right].	\label{eq:deltarho}
\end{align}
Since the process is adiabatic, $\overline{A(t)}	=	\mathrm{Tr} \left[ \left( \rho_0 + \Delta \rho(t) \right) \hat{A}\right]$. Using the cyclic property of the trace the time dependence can be moved from $\hat{B}$ to $\hat{A}$,
\begin{align}
	\overline{A(t)} - \langle \hat{A} \rangle_0	&=	\frac{1}{i\hbar} \int_{-\infty}^t dt' h(t') \mathrm{Tr} \left\{ \left[\rho_0, \hat{B}(0)\right] \hat{A}(t-t') \right\},
\end{align}
By comparing with Eq.~\eqref{eq:responsefunction:implicitdefinition} and using the cyclic property of the trace we find
\begin{align}
	\phi_{AB}(t-t')	&=	\frac{1}{i\hbar} \mathrm{Tr} \left\{ \left[\rho_0, \hat{B}(0)\right] \hat{A}(t-t') \right\} \\
	&=	\frac{i}{\hbar}	\left\langle \left[ \hat{A}(t), \hat{B}(t')\right]\right\rangle_0,	\label{eq:responsefunction:commutator}
\end{align}
establishing the relation between spontaneous fluctuations and the linear response.

In the Van Hove formulation of scattering experiments, the observed cross section is directly related to a dynamical structure factor
\begin{align}
	S(\omega)	&=	\frac{1}{2\pi\hbar}\int_{-\infty}^\infty dt \exp \left( -i\omega t\right) \left\langle \hat{B}(0) \hat{B}^\dagger (t)\right\rangle,\label{eq:DSF}
\end{align}
which, unlike $\chi\left[ \omega\right]$, is a purely real function. We want to relate it to the response function $\phi(t)	\equiv \phi_{B^\dagger B}(t)	=	\frac{i}{\hbar} \left\langle \left[ \hat{B}^\dagger (t), \hat{B} \right] \right\rangle.$ To achieve this, we consider the Fourier transform of $\phi(t)$:
\begin{align}
	\phi (\omega)	&=	\frac{1}{2\pi} \int_{-\infty}^\infty dt \exp \left( -i\omega t \right) \phi(t)	\label{eq:response:ft:def}\\
	&=	\frac{i}{2\pi\hbar} \int_{-\infty}^\infty dt \exp \left( -i\omega t \right) \left\langle  \hat{B}^\dagger(t) \hat{B}(0) - \hat{B}(0) \hat{B}^\dagger(t) \right\rangle.
\end{align}
The last term is proportional to Eq.~\eqref{eq:DSF}, and we can write
\begin{align}
	\phi(\omega)	&=	-iS(\omega ) + \frac{i}{2\pi\hbar} \int_{-\infty}^\infty dt \exp \left( -i\omega t \right) \left\langle  \hat{B}^\dagger(t) \hat{B}(0)  \right\rangle.	\label{eq:response:ft}
\end{align}
Applying the identity $\langle \hat{B}^\dagger (t) \hat{B}(0)\rangle	=\left\langle \hat{B}(0) \hat{B}^\dagger \left( t+ i\beta\hbar\right)\right\rangle$ to the second term and assuming analyticity in an appropriate region of the complex plane, we can perform a complex frequency/time shift of the Fourier transform to get
\begin{align}
	& \int_{-\infty}^\infty dt \exp \left( -i\omega t\right) \left\langle \hat{B}^\dagger(t) \hat{B}(0) \right\rangle	\nonumber\\
	&=	\int_{-\infty}^\infty dt \exp \left( -i\omega t\right) \left\langle \hat{B}(0) \hat{B}^\dagger \left( t + i\hbar\beta\right) \right\rangle	\nonumber\\
	&= e^{-\beta\hbar \omega} 2\pi \hbar S(\omega).	
\end{align}
Eq.~\eqref{eq:response:ft} can now be written
\begin{align}
	\phi(\omega)	&=	i \left( e^{-\beta\hbar \omega} - 1 \right) S(\omega),	\label{eq:fluctdiss1}
\end{align}
or, alternatively, using Eq.~\eqref{eq:response:ft:def},
\begin{align}
	S(\omega)	&=	\left( 1-e^{-\beta\hbar\omega} \right)^{-1} \frac{i}{2\pi} \int_{-\infty}^\infty dt e^{-i\omega t} \phi(t).	\label{eq:fluctdiss2}
\end{align}
If $\phi(t)$ is odd in $t$, the expression reduces to
\begin{align}
	S(\omega)	&=	\left( 1-e^{-\beta\hbar\omega} \right)^{-1} \frac{1}{\pi} \int_{0}^\infty dt \sin \left( \omega t\right) \phi(t).
\end{align}
Recalling Eq.~\eqref{eq:generalizedsusceptibility:def}, we obtain a form of the fluctuation-dissipation theorem familiar in scattering,
\begin{align}
	S(\omega)	&=\left( 1-e^{-\beta\hbar\omega} \right)^{-1} \frac{1}{\pi}	\chi''\left[ \omega\right].	\label{eq:fluctdiss:DSFchi}
\end{align}

We next derive the K\"all\'en-Lehmann spectral representation of the dynamical susceptibility. Using equations \eqref{eq:generalizedsusceptibility:def} and \eqref{eq:responsefunction:commutator} we have, for a Hermitian operator $\mathcal{O}$ (suppressing the $\hat{~}$ from now on),
\begin{equation}
	\chi_{\mathcal{O}\mathcal{O}}\left[ \omega\right]	=		-\int_0^\infty dt \frac{i}{\hbar} \left\langle \left[ \mathcal{O}(t), \mathcal{O}(0)\right]\right\rangle_0 e^{-i\omega t}	
 \end{equation}
 \begin{align}
    &=	-\frac{i}{Z \hbar}	\int_0^\infty dt e^{-i\omega t} \sum_\lambda e^{-iE_\lambda \beta }\left\{ \left\langle \lambda \middle| e^{iH_0 t/\hbar}\mathcal{O} e^{-iH_0 t/\hbar} \mathcal{O}\middle| \lambda\right\rangle  \right. \nonumber\\
			&\left. - \left\langle \lambda \middle| \mathcal{O} e^{iH_0 t/\hbar}\mathcal{O} e^{-iH_0t/\hbar} \middle| \lambda \right\rangle_0 \right\}
\end{align}
where $|\lambda\rangle$ is an energy eigenstate with eigenvalue $E_\lambda$, and where $Z$ is the partition function. Now, introduce the resolution of identity,
\begin{align}
	&\chi_{\mathcal{O}\mathcal{O}}\left[ \omega\right]	=	-\frac{i}{Z \hbar}	\int_0^\infty dt e^{-i\omega t} \sum_{\lambda,\lambda'} e^{-iE_\lambda \beta }	\\
	&\left\{ \left\langle \lambda \middle| e^{iH_0 t/\hbar}\mathcal{O} \middle| \lambda' \middle\rangle \middle\langle \lambda' \middle| e^{-iH_0 t/\hbar} \mathcal{O}\middle| \lambda\right\rangle  \right.\nonumber\\
    &\left. - \nonumber 
 \left\langle \lambda \middle| \mathcal{O} e^{iH_0 t/\hbar} \middle| \lambda' \middle\rangle\middle\langle \lambda'\middle|  \mathcal{O} e^{-iH_0t/\hbar} \middle| \lambda \right\rangle_0 \right\} \nonumber .
\end{align}
Recalling that $e^{-iH_0t/\hbar}|\lambda\rangle = e^{-iE_\lambda t/\hbar}|\lambda\rangle$ and $\langle \lambda | e^{iH_0t/\hbar} = \langle \lambda | e^{iE_\lambda t/\hbar}$, one obtains
\begin{align}
	\chi_{\mathcal{O}\mathcal{O}}\left[ \omega\right]	&=	-\frac{i}{Z \hbar}	\int_0^\infty dt e^{-i\omega t} \sum_{\lambda,\lambda'} e^{-iE_\lambda \beta }	\left| \middle\langle \lambda \middle| \mathcal{O} \middle| \lambda' \middle\rangle \right|^2	\nonumber\\
			&\times \left\{ e^{i \left( E_\lambda - E_{\lambda'}\right)t/\hbar}  - e^{i \left( E_{\lambda'} - E_\lambda \right)t/\hbar} \right\}	\\
			&=	-\frac{i}{Z \hbar}	\sum_{\lambda,\lambda'} e^{-iE_\lambda \beta }	\left| \middle\langle \lambda \middle| \mathcal{O} \middle| \lambda' \middle\rangle \right|^2	\\
			&\times\pi \left\{ \delta\left( \omega + \frac{E_{\lambda'}}{\hbar} - \frac{E_\lambda}{\hbar} \right)- \delta\left( \omega - \frac{E_{\lambda'}}{\hbar} + \frac{E_\lambda}{\hbar} \right) \right\}\nonumber
\end{align} 
where $\delta(x)$ is the Dirac delta function. 
Using the definition that $p_\lambda = e^{-E_\lambda/k_B T}/Z$,
\begin{align}
	\chi_{\mathcal{O}\mathcal{O}}\left[ \omega\right]	&=		\frac{i}{\hbar}	\sum_{\lambda,\lambda'} \left| \middle\langle \lambda \middle| \mathcal{O} \middle| \lambda' \middle\rangle \right|^2 \pi \\
		&	\times \left[ p_\lambda \delta\left( \omega - \frac{E_{\lambda'}}{\hbar} +  \frac{E_\lambda}{\hbar} \right)- p_\lambda \delta\left( \omega +\frac{E_{\lambda'}}{\hbar} - \frac{E_\lambda}{\hbar} \right)\right].\nonumber
\end{align}
Relabeling $\lambda \leftrightarrow \lambda'$ in the second term, and absorbing the $1/\hbar$ into the argument of the $\delta$ function, we obtain
\begin{align}
	\chi''\left[ \omega\right]	&=	\sum_{\lambda,\lambda'} \left( p_\lambda - p_{\lambda'}\right) \left| \left\langle \lambda \middle| \mathcal{O}\middle| \lambda' \right\rangle \right|^2 \pi \delta \left( \hbar \omega - E_{\lambda'} + E_\lambda \right).	\label{eq:chiprimeprime:lehmann}
\end{align}

\subsection{Notes on notational differences}
To minimize confusion, we want to highlight some differences between the convention adopted here, and that adopted in the seminal work of Hauke et al. \cite{Hauke2016}. Their definition of the dynamical structure factor does not include the factor $1/\pi$ present in Eq.~\eqref{eq:DSF}, which is convention for neutron scattering. Furthermore, they work with a frequency-symmetrized structure factor $\tilde{S}(\omega)=S(\omega)+S(-\omega)$, which modifies the fluctuation-dissipation theorem from Eq.~\eqref{eq:fluctdiss:DSFchi} to $\chi''[\omega]=\frac{1}{\hbar} \tanh \left( \hbar\omega\beta/2\right) \tilde{S}(\omega)$. Finally, we treat all susceptibilities and structure factors as intensive quantities, i.e. including a system size normalization factor as is conventional in the magnetism literature, whereas this is not assumed in Ref.~\cite{Hauke2016}.

\medskip
\textbf{Acknowledgements} \par

The work of P.L. and E.D. was supported by the U.S. Department of Energy, Office of Science, Basic Energy Sciences, Materials Sciences and Engineering Division. The work by A.S. and D.A.T. is supported by the Quantum Science Center (QSC), a National Quantum Information Science Research Center of the U.S. Department of Energy (DOE). D.A.T. would like to thank J. Quintanilla, R. Pynn, and G. Ortiz for enlightening discussions.

\medskip
\textbf{Conflict of Interest}
The authors declare no conflicts of interest.

\medskip
\textbf{Data Availability Statement}
Data sharing is not applicable to this article as no new data were created or analyzed in this review. Several of the cited works have associated public data archives.

\medskip
\bibliographystyle{MSP}


\begin{mcbibliography}{100}
	\providecommand{\url}[1]{\texttt{#1}}
	\providecommand{\urlprefix}{URL }
	
	\bibitem{Guehne2009}
	O.~G\"uhne, G.~T\'oth,
	\newblock \emph{Phys. Rep.} \textbf{2009}, \emph{474} 1 \relax
	\relax
	\bibitem{BESreport2016}
	Basic research needs workshop on quantum materials for energy relevant
	technology,
	\newblock Technical report, US Department of Energy, Office of Science, United
	States, \textbf{2016},
	\newblock \urlprefix\url{https://doi.org/10.2172/1616509}\relax
	\relax
	\bibitem{NSFreport2018}
	Midscale instrumentation for quantum materials,
	\newblock Technical report, National Science Foundation, Division of Materials
	Research, United States, \textbf{2018},
	\newblock \urlprefix\url{https://www.nsf.gov/mps/dmr/MIQM_report_v15.pdf}\relax
	\relax
	\bibitem{Friis_2018}
	N.~Friis, G.~Vitagliano, M.~Malik, M.~Huber,
	\newblock \emph{Nat. Rev. Phys.} \textbf{2018}, \emph{1} 72\relax
	\relax
	\bibitem{Yu2022a}
	X.-D. Yu, J.~Shang, O.~Gühne,
	\newblock \emph{Adv. Quantum Technol.} \textbf{2022}, \emph{5} 2100126\relax
	\relax
	\bibitem{Wang_2022}
	K.~Wang, Z.~Song, X.~Zhao, Z.~Wang, X.~Wang,
	\newblock \emph{npj Quantum Inf.} \textbf{2022}, \emph{8}\relax
	\relax
	\bibitem{RevModPhys.80.517}
	L.~Amico, R.~Fazio, A.~Osterloh, V.~Vedral,
	\newblock \emph{Rev. Mod. Phys.} \textbf{2008}, \emph{80} 517\relax
	\relax
	\bibitem{RevModPhys.81.865}
	R.~Horodecki, P.~Horodecki, M.~Horodecki, K.~Horodecki,
	\newblock \emph{Rev. Mod. Phys.} \textbf{2009}, \emph{81} 865\relax
	\relax
	\bibitem{Laflorencie2016}
	N.~Laflorencie,
	\newblock \emph{Phys. Rep.} \textbf{2016}, \emph{646} 1 \relax
	\relax
	\bibitem{Chiara2018}
	G.~D. Chiara, A.~Sanpera,
	\newblock \emph{Rep. Prog. Phys.} \textbf{2018}, \emph{81} 074002\relax
	\relax
	\bibitem{zeng2019quantum}
	B.~Zeng, X.~Chen, D.-L. Zhou, X.-G. Wen,
	\newblock \emph{Quantum Information Meets Quantum Matter: From Quantum
		Entanglement to Topological Phases of Many-Body Systems},
	\newblock Quantum Science and Technology. Springer New York,
	\textbf{2019}\relax
	\relax
	\bibitem{bayat2022entanglement}
	A.~Bayat, S.~Bose, H.~Johannesson, editors,
	\newblock \emph{Entanglement in Spin Chains: From Theory to Quantum Technology
		Applications},
	\newblock Quantum Science and Technology. Springer International Publishing,
	\textbf{2022}\relax
	\relax
	\bibitem{Tokura_2017}
	Y.~Tokura, M.~Kawasaki, N.~Nagaosa,
	\newblock \emph{Nat. Phys.} \textbf{2017}, \emph{13} 1056\relax
	\relax
	\bibitem{keimer2017physics}
	B.~Keimer, J.~Moore,
	\newblock \emph{Nat. Phys.} \textbf{2017}, \emph{13} 1045\relax
	\relax
	\bibitem{Cava2021}
	R.~Cava, N.~de~Leon, W.~Xie,
	\newblock \emph{Chem. Rev.} \textbf{2021}, \emph{121} 2777\relax
	\relax
	\bibitem{Iyengar2023}
	S.~A. Iyengar, A.~B. Puthirath, V.~Swaminathan,
	\newblock \emph{Adv. Mater.} \textbf{2023}, \emph{35} 2107839\relax
	\relax
	\bibitem{Frerot2023}
	I.~Frérot, M.~Fadel, M.~Lewenstein,
	\newblock \emph{Rep. Prog. Phys.} \textbf{2023}, \emph{86} 114001\relax
	\relax
	\bibitem{Cruz2023}
	C.~Cruz,
	\newblock \emph{Physica B: Condens. Matter} \textbf{2023}, \emph{653}
	414485\relax
	\relax
	\bibitem{PhysRevLett.28.938}
	S.~J. Freedman, J.~F. Clauser,
	\newblock \emph{Phys. Rev. Lett.} \textbf{1972}, \emph{28} 938\relax
	\relax
	\bibitem{PhysRevLett.23.880}
	J.~F. Clauser, M.~A. Horne, A.~Shimony, R.~A. Holt,
	\newblock \emph{Phys. Rev. Lett.} \textbf{1969}, \emph{23} 880\relax
	\relax
	\bibitem{PhysRevLett.90.227902}
	G.~Vidal, J.~I. Latorre, E.~Rico, A.~Kitaev,
	\newblock \emph{Phys. Rev. Lett.} \textbf{2003}, \emph{90} 227902\relax
	\relax
	\bibitem{PhysRevD.96.126007}
	M.~A. Rajabpour,
	\newblock \emph{Phys. Rev. D} \textbf{2017}, \emph{96} 126007\relax
	\relax
	\bibitem{RevModPhys.89.041004}
	X.-G. Wen,
	\newblock \emph{Rev. Mod. Phys.} \textbf{2017}, \emph{89} 041004\relax
	\relax
	\bibitem{PhysRevLett.96.110404}
	A.~Kitaev, J.~Preskill,
	\newblock \emph{Phys. Rev. Lett.} \textbf{2006}, \emph{96} 110404\relax
	\relax
	\bibitem{PhysRevLett.96.110405}
	M.~Levin, X.-G. Wen,
	\newblock \emph{Phys. Rev. Lett.} \textbf{2006}, \emph{96} 110405\relax
	\relax
	\bibitem{RevModPhys.89.025003}
	Y.~Zhou, K.~Kanoda, T.-K. Ng,
	\newblock \emph{Rev. Mod. Phys.} \textbf{2017}, \emph{89} 025003\relax
	\relax
	\bibitem{Savary2017}
	L.~Savary, L.~Balents,
	\newblock \emph{Rep. Prog. Phys.} \textbf{2017}, \emph{80} 016502\relax
	\relax
	\bibitem{Broholm2020}
	C.~Broholm, R.~J. Cava, S.~A. Kivelson, D.~G. Nocera, M.~R. Norman, T.~Senthil,
	\newblock \emph{Science} \textbf{2020}, \emph{367} 263\relax
	\relax
	\bibitem{Kitaev2003}
	A.~Y. Kitaev,
	\newblock \emph{Ann. Phys. (N.Y.)} \textbf{2003}, \emph{303} 2\relax
	\relax
	\bibitem{RevModPhys.80.1083}
	C.~Nayak, S.~H. Simon, A.~Stern, M.~Freedman, S.~Das~Sarma,
	\newblock \emph{Rev. Mod. Phys.} \textbf{2008}, \emph{80} 1083\relax
	\relax
	\bibitem{Krasnok2024}
	A.~Krasnok, P.~Dhakal, A.~Fedorov, P.~Frigola, M.~Kelly, S.~Kutsaev,
	\newblock \emph{Appl. Phys. Rev.} \textbf{2024}, \emph{11} 011302\relax
	\relax
	\bibitem{Doherty2013}
	M.~W. Doherty, N.~B. Manson, P.~Delaney, F.~Jelezko, J.~Wrachtrup, L.~C.
	Hollenberg,
	\newblock \emph{Phys. Rep.} \textbf{2013}, \emph{528} 1\relax
	\relax
	\bibitem{Pezzagna2021}
	S.~Pezzagna, J.~Meijer,
	\newblock \emph{Appl. Phys. Rev.} \textbf{2021}, \emph{8} 011308\relax
	\relax
	\bibitem{RevModPhys.82.2313}
	M.~Saffman, T.~G. Walker, K.~M\o{}lmer,
	\newblock \emph{Rev. Mod. Phys.} \textbf{2010}, \emph{82} 2313\relax
	\relax
	\bibitem{Schaefer2020}
	F.~Schäfer, T.~Fukuhara, S.~Sugawa, Y.~Takasu, Y.~Takahashi,
	\newblock \emph{Nat. Rev. Phys.} \textbf{2020}, \emph{2} 411\relax
	\relax
	\bibitem{Nakamura_2020}
	J.~Nakamura, S.~Liang, G.~C. Gardner, M.~J. Manfra,
	\newblock \emph{Nat. Phys.} \textbf{2020}, \emph{16} 931\relax
	\relax
	\bibitem{Kundu_2023}
	H.~K. Kundu, S.~Biswas, N.~Ofek, V.~Umansky, M.~Heiblum,
	\newblock \emph{Nat. Phys.} \textbf{2023}, \emph{19} 515\relax
	\relax
	\bibitem{PhysRevX.13.041012}
	J.~Nakamura, S.~Liang, G.~C. Gardner, M.~J. Manfra,
	\newblock \emph{Phys. Rev. X} \textbf{2023}, \emph{13} 041012\relax
	\relax
	\bibitem{Hauke2016}
	P.~Hauke, M.~Heyl, L.~Tagliacozzo, P.~Zoller,
	\newblock \emph{Nat. Phys.} \textbf{2016}, \emph{12} 778\relax
	\relax
	\bibitem{PhysRev.47.777}
	A.~Einstein, B.~Podolsky, N.~Rosen,
	\newblock \emph{Phys. Rev.} \textbf{1935}, \emph{47} 777\relax
	\relax
	\bibitem{PhysRev.48.696}
	N.~Bohr,
	\newblock \emph{Phys. Rev.} \textbf{1935}, \emph{48} 696\relax
	\relax
	\bibitem{Schroedinger1935}
	E.~Schr\"odinger,
	\newblock \emph{Naturwissenschaften} \textbf{1935}, \emph{23} 807\relax
	\relax
	\bibitem{PhysicsPhysiqueFizika.1.195}
	J.~S. Bell,
	\newblock \emph{Physics Physique Fizika} \textbf{1964}, \emph{1} 195\relax
	\relax
	\bibitem{Bell_2004}
	J.~S. Bell,
	\newblock \emph{Speakable and Unspeakable in Quantum Mechanics: Collected
		Papers on Quantum Philosophy},
	\newblock Cambridge University Press, second edition, \textbf{2004}\relax
	\relax
	\bibitem{RevModPhys.86.419}
	N.~Brunner, D.~Cavalcanti, S.~Pironio, V.~Scarani, S.~Wehner,
	\newblock \emph{Rev. Mod. Phys.} \textbf{2014}, \emph{86} 419\relax
	\relax
	\bibitem{Cirelson1980}
	B.~Cirel'son,
	\newblock \emph{Lett. Math. Phys.} \textbf{1980}, \emph{4} 93\relax
	\relax
	\bibitem{Aspect_1999}
	A.~Aspect,
	\newblock \emph{Nature} \textbf{1999}, \emph{398} 189\relax
	\relax
	\bibitem{RevModPhys.71.S288}
	A.~Zeilinger,
	\newblock \emph{Rev. Mod. Phys.} \textbf{1999}, \emph{71} S288\relax
	\relax
	\bibitem{Hensen_2015}
	B.~Hensen, H.~Bernien, A.~E. Dr\'eau, A.~Reiserer, N.~Kalb, M.~S. Blok,
	J.~Ruitenberg, R.~F.~L. Vermeulen, R.~N. Schouten, C.~Abell\'an, W.~Amaya,
	V.~Pruneri, M.~W. Mitchell, M.~Markham, D.~J. Twitchen, D.~Elkouss,
	S.~Wehner, T.~H. Taminiau, R.~Hanson,
	\newblock \emph{Nature} \textbf{2015}, \emph{526} 682\relax
	\relax
	\bibitem{PhysRevLett.115.250401}
	M.~Giustina, M.~A.~M. Versteegh, S.~Wengerowsky, J.~Handsteiner, A.~Hochrainer,
	K.~Phelan, F.~Steinlechner, J.~Kofler, J.-A. Larsson, C.~Abell\'an, W.~Amaya,
	V.~Pruneri, M.~W. Mitchell, J.~Beyer, T.~Gerrits, A.~E. Lita, L.~K. Shalm,
	S.~W. Nam, T.~Scheidl, R.~Ursin, B.~Wittmann, A.~Zeilinger,
	\newblock \emph{Phys. Rev. Lett.} \textbf{2015}, \emph{115} 250401\relax
	\relax
	\bibitem{PhysRevLett.115.250402}
	L.~K. Shalm, E.~Meyer-Scott, B.~G. Christensen, P.~Bierhorst, M.~A. Wayne,
	M.~J. Stevens, T.~Gerrits, S.~Glancy, D.~R. Hamel, M.~S. Allman, K.~J.
	Coakley, S.~D. Dyer, C.~Hodge, A.~E. Lita, V.~B. Verma, C.~Lambrocco,
	E.~Tortorici, A.~L. Migdall, Y.~Zhang, D.~R. Kumor, W.~H. Farr, F.~Marsili,
	M.~D. Shaw, J.~A. Stern, C.~Abell\'an, W.~Amaya, V.~Pruneri, T.~Jennewein,
	M.~W. Mitchell, P.~G. Kwiat, J.~C. Bienfang, R.~P. Mirin, E.~Knill, S.~W.
	Nam,
	\newblock \emph{Phys. Rev. Lett.} \textbf{2015}, \emph{115} 250402\relax
	\relax
	\bibitem{PhysRevLett.119.010402}
	W.~Rosenfeld, D.~Burchardt, R.~Garthoff, K.~Redeker, N.~Ortegel, M.~Rau,
	H.~Weinfurter,
	\newblock \emph{Phys. Rev. Lett.} \textbf{2017}, \emph{119} 010402\relax
	\relax
	\bibitem{PhysRevLett.121.080404}
	M.-H. Li, C.~Wu, Y.~Zhang, W.-Z. Liu, B.~Bai, Y.~Liu, W.~Zhang, Q.~Zhao, H.~Li,
	Z.~Wang, L.~You, W.~J. Munro, J.~Yin, J.~Zhang, C.-Z. Peng, X.~Ma, Q.~Zhang,
	J.~Fan, J.-W. Pan,
	\newblock \emph{Phys. Rev. Lett.} \textbf{2018}, \emph{121} 080404\relax
	\relax
	\bibitem{Storz_2023}
	S.~Storz, J.~Sch\"ar, A.~Kulikov, P.~Magnard, P.~Kurpiers, J.~L\"utolf,
	T.~Walter, A.~Copetudo, K.~Reuer, A.~Akin, J.-C. Besse, M.~Gabureac, G.~J.
	Norris, A.~Rosario, F.~Martin, J.~Martinez, W.~Amaya, M.~W. Mitchell,
	C.~Abellan, J.-D. Bancal, N.~Sangouard, B.~Royer, A.~Blais, A.~Wallraff,
	\newblock \emph{Nature} \textbf{2023}, \emph{617} 265\relax
	\relax
	\bibitem{Horodecki1997}
	P.~Horodecki,
	\newblock \emph{Phys. Lett. A} \textbf{1997}, \emph{232} 333 \relax
	\relax
	\bibitem{Terhal2000}
	B.~M. Terhal,
	\newblock \emph{Phys. Lett. A} \textbf{2000}, \emph{271} 319\relax
	\relax
	\bibitem{PhysRevA.62.052310}
	M.~Lewenstein, B.~Kraus, J.~I. Cirac, P.~Horodecki,
	\newblock \emph{Phys. Rev. A} \textbf{2000}, \emph{62} 052310\relax
	\relax
	\bibitem{PhysRevA.72.012321}
	P.~Hyllus, O.~G\"uhne, D.~Bru\ss{}, M.~Lewenstein,
	\newblock \emph{Phys. Rev. A} \textbf{2005}, \emph{72} 012321\relax
	\relax
	\bibitem{Chruscinski2014}
	D.~Chruściński, G.~Sarbicki,
	\newblock \emph{J. Phys. A: Math. Theor.} \textbf{2014}, \emph{47} 483001\relax
	\relax
	\bibitem{PhysRevA.69.022304}
	L.~Amico, A.~Osterloh, F.~Plastina, R.~Fazio, G.~Massimo~Palma,
	\newblock \emph{Phys. Rev. A} \textbf{2004}, \emph{69} 022304\relax
	\relax
	\bibitem{PhysRevLett.93.167203}
	T.~Roscilde, P.~Verrucchi, A.~Fubini, S.~Haas, V.~Tognetti,
	\newblock \emph{Phys. Rev. Lett.} \textbf{2004}, \emph{93} 167203\relax
	\relax
	\bibitem{PhysRevA.72.032309}
	L.-A. Wu, S.~Bandyopadhyay, M.~S. Sarandy, D.~A. Lidar,
	\newblock \emph{Phys. Rev. A} \textbf{2005}, \emph{72} 032309\relax
	\relax
	\bibitem{PhysRevA.73.012110}
	{\v{C}}.~Brukner, V.~Vedral, A.~Zeilinger,
	\newblock \emph{Phys. Rev. A} \textbf{2006}, \emph{73} 012110\relax
	\relax
	\bibitem{PhysRevA.74.022322}
	L.~Amico, F.~Baroni, A.~Fubini, D.~Patan\`e, V.~Tognetti, P.~Verrucchi,
	\newblock \emph{Phys. Rev. A} \textbf{2006}, \emph{74} 022322\relax
	\relax
	\bibitem{Wie_niak_2005}
	M.~Wie{\'{s}}niak, V.~Vedral, {\v{C}}.~Brukner,
	\newblock \emph{New J. Phys.} \textbf{2005}, \emph{7} 258\relax
	\relax
	\bibitem{PhysRevB.78.064108}
	M.~Wie\ifmmode~\acute{s}\else \'{s}\fi{}niak, V.~Vedral, i.~c.~v. Brukner,
	\newblock \emph{Phys. Rev. B} \textbf{2008}, \emph{78} 064108\relax
	\relax
	\bibitem{Singh2013}
	H.~Singh, T.~Chakraborty, D.~Das, H.~S. Jeevan, Y.~Tokiwa, P.~Gegenwart,
	C.~Mitra,
	\newblock \emph{New J. Phys.} \textbf{2013}, \emph{15} 113001\relax
	\relax
	\bibitem{PhysRevLett.106.020401}
	M.~Cramer, M.~B. Plenio, H.~Wunderlich,
	\newblock \emph{Phys. Rev. Lett.} \textbf{2011}, \emph{106} 020401\relax
	\relax
	\bibitem{Kwek2011}
	L.~C. Kwek,
	\newblock \emph{Laser Phys.} \textbf{2011}, \emph{21} 1511\relax
	\relax
	\bibitem{PhysRevB.107.054422}
	V.~Menon, N.~E. Sherman, M.~Dupont, A.~O. Scheie, D.~A. Tennant, J.~E. Moore,
	\newblock \emph{Phys. Rev. B} \textbf{2023}, \emph{107} 054422\relax
	\relax
	\bibitem{PhysRevA.61.052306}
	V.~Coffman, J.~Kundu, W.~K. Wootters,
	\newblock \emph{Phys. Rev. A} \textbf{2000}, \emph{61} 052306\relax
	\relax
	\bibitem{PhysRevLett.88.017901}
	H.~Ollivier, W.~H. Zurek,
	\newblock \emph{Phys. Rev. Lett.} \textbf{2001}, \emph{88} 017901\relax
	\relax
	\bibitem{PhysRevA.80.022108}
	M.~S. Sarandy,
	\newblock \emph{Phys. Rev. A} \textbf{2009}, \emph{80} 022108\relax
	\relax
	\bibitem{Ghosh2003}
	S.~Ghosh, T.~F. Rosenbaum, G.~Aeppli, S.~N. Coppersmith,
	\newblock \emph{Nature} \textbf{2003}, \emph{425} 48\relax
	\relax
	\bibitem{Vedral2003Nature}
	V.~Vedral,
	\newblock \emph{Nature} \textbf{2003}, \emph{425} 28\relax
	\relax
	\bibitem{PhysRevLett.70.4003}
	D.~A. Tennant, T.~G. Perring, R.~A. Cowley, S.~E. Nagler,
	\newblock \emph{Phys. Rev. Lett.} \textbf{1993}, \emph{70} 4003\relax
	\relax
	\bibitem{Christensen2007}
	N.~B. Christensen, H.~M. R{\o}nnow, D.~F. McMorrow, A.~Harrison, T.~G. Perring,
	M.~Enderle, R.~Coldea, L.~P. Regnault, G.~Aeppli,
	\newblock \emph{Proc. Natl. Acad. Sci. U.S.A.} \textbf{2007}, \emph{104}
	15264\relax
	\relax
	\bibitem{Mourigal2013}
	M.~Mourigal, M.~Enderle, A.~Kl\"opperpieper, J.-S. Caux, A.~Stunault, H.~M.
	R\o{}nnow,
	\newblock \emph{Nat. Phys.} \textbf{2013}, \emph{9} 435\relax
	\relax
	\bibitem{Piazza2015}
	B.~D. Piazza, M.~Mourigal, N.~B. Christensen, G.~J. Nilsen,
	P.~Tregenna-Piggott, T.~G. Perring, M.~Enderle, D.~F. McMorrow, D.~A. Ivanov,
	H.~M. R{\o}nnow,
	\newblock \emph{Nature Physics} \textbf{2015}, \emph{11} 62\relax
	\relax
	\bibitem{Tennant2019}
	D.~A. Tennant,
	\newblock \emph{J. Phys. Soc. Jpn.} \textbf{2019}, \emph{88} 081009\relax
	\relax
	\bibitem{Vasiliev2018}
	A.~Vasiliev, O.~Volkova, E.~Zvereva, M.~Markina,
	\newblock \emph{npj Quantum Mater.} \textbf{2018}, \emph{3} 18\relax
	\relax
	\bibitem{Sachdev_2011}
	S.~Sachdev,
	\newblock \emph{Quantum Phase Transitions},
	\newblock Cambridge University Press, second edition, \textbf{2011}\relax
	\relax
	\bibitem{Yuan2022}
	H.~Yuan, Y.~Cao, A.~Kamra, R.~A. Duine, P.~Yan,
	\newblock \emph{Phys. Rep.} \textbf{2022}, \emph{965} 1\relax
	\relax
	\bibitem{Paschen_2020}
	S.~Paschen, Q.~Si,
	\newblock \emph{Nat. Rev. Phys.} \textbf{2020}, \emph{3} 9\relax
	\relax
	\bibitem{volovik}
	G.~E. Volovik,
	\newblock \emph{{The Universe in a Helium Droplet}},
	\newblock Oxford University Press, \textbf{2009}\relax
	\relax
	\bibitem{Herdman2017}
	C.~M. Herdman, P.-N. Roy, R.~G. Melko, A.~Del~Maestro,
	\newblock \emph{Nat. Phys.} \textbf{2017}, \emph{13} 556\relax
	\relax
	\bibitem{delmaestro_2011}
	A.~Del~Maestro, M.~Boninsegni, I.~Affleck,
	\newblock \emph{Phys. Rev. Lett.} \textbf{2011}, \emph{106} 105303\relax
	\relax
	\bibitem{DelMaestro2022}
	A.~Del~Maestro, N.~S. Nichols, T.~R. Prisk, G.~Warren, P.~E. Sokol,
	\newblock \emph{Nat. Commun.} \textbf{2022}, \emph{13} 3168\relax
	\relax
	\bibitem{Dmowski2017}
	W.~Dmowski, S.~O. Diallo, K.~Lokshin, G.~Ehlers, G.~Ferr\'e, J.~Boronat,
	T.~Egami,
	\newblock \emph{Nat. Commun.} \textbf{2017}, \emph{8} 15294\relax
	\relax
	\bibitem{kardar2007statistical}
	M.~Kardar,
	\newblock \emph{Statistical physics of particles},
	\newblock Cambridge University Press, \textbf{2007}\relax
	\relax
	\bibitem{Witten_2020}
	E.~Witten,
	\newblock \emph{Riv. Nuovo. Cim.} \textbf{2020}, \emph{43} 187\relax
	\relax
	\bibitem{PhysRevB.103.224434}
	A.~Scheie, P.~Laurell, A.~M. Samarakoon, B.~Lake, S.~E. Nagler, G.~E. Granroth,
	S.~Okamoto, G.~Alvarez, D.~A. Tennant,
	\newblock \emph{Phys. Rev. B} \textbf{2021}, \emph{103} 224434\relax
	\relax
	\bibitem{Gurvits2003}
	L.~Gurvits,
	\newblock In \emph{Proceedings of the Thirty-Fifth Annual ACM Symposium on
		Theory of Computing}, STOC '03. Association for Computing Machinery, New
	York, NY, USA,
	\newblock ISBN 1581136749, \textbf{2003} 10–19,
	\newblock \urlprefix\url{https://doi.org/10.1145/780542.780545}\relax
	\relax
	\bibitem{PhysRevA.54.3824}
	C.~H. Bennett, D.~P. DiVincenzo, J.~A. Smolin, W.~K. Wootters,
	\newblock \emph{Phys. Rev. A} \textbf{1996}, \emph{54} 3824\relax
	\relax
	\bibitem{PhysRevLett.80.2245}
	W.~K. Wootters,
	\newblock \emph{Phys. Rev. Lett.} \textbf{1998}, \emph{80} 2245\relax
	\relax
	\bibitem{Hyllus2012}
	P.~Hyllus, W.~Laskowski, R.~Krischek, C.~Schwemmer, W.~Wieczorek,
	H.~Weinfurter, L.~Pezz\'e, A.~Smerzi,
	\newblock \emph{Phys. Rev. A} \textbf{2012}, \emph{85} 022321\relax
	\relax
	\bibitem{PhysRevA.68.032103}
	H.~F. Hofmann, S.~Takeuchi,
	\newblock \emph{Phys. Rev. A} \textbf{2003}, \emph{68} 032103\relax
	\relax
	\bibitem{PhysRevA.79.042334}
	G.~T\'oth, C.~Knapp, O.~G\"uhne, H.~J. Briegel,
	\newblock \emph{Phys. Rev. A} \textbf{2009}, \emph{79} 042334\relax
	\relax
	\bibitem{Fubini2006}
	A.~Fubini, T.~Roscilde, V.~Tognetti, M.~Tusa, P.~Verrucchi,
	\newblock \emph{Eur. Phys. J. D} \textbf{2006}, \emph{38} 563\relax
	\relax
	\bibitem{PhysRevA.67.022110}
	T.-C. Wei, K.~Nemoto, P.~M. Goldbart, P.~G. Kwiat, W.~J. Munro, F.~Verstraete,
	\newblock \emph{Phys. Rev. A} \textbf{2003}, \emph{67} 022110\relax
	\relax
	\bibitem{PhysRevLett.96.220503}
	T.~J. Osborne, F.~Verstraete,
	\newblock \emph{Phys. Rev. Lett.} \textbf{2006}, \emph{96} 220503\relax
	\relax
	\bibitem{PhysRevB.102.064409}
	P.~Thakur, P.~Durganandini,
	\newblock \emph{Phys. Rev. B} \textbf{2020}, \emph{102} 064409\relax
	\relax
	\bibitem{PhysRevA.68.060301}
	O.~F. Sylju\aa{}sen,
	\newblock \emph{Phys. Rev. A} \textbf{2003}, \emph{68} 060301(R)\relax
	\relax
	\bibitem{PhysRevLett.78.5022}
	S.~A. Hill, W.~K. Wootters,
	\newblock \emph{Phys. Rev. Lett.} \textbf{1997}, \emph{78} 5022\relax
	\relax
	\bibitem{Sakurai1994}
	J.~J. Sakurai,
	\newblock \emph{Modern Quantum Mechanics},
	\newblock Addison-Wesley, revised edition, \textbf{1994}\relax
	\relax
	\bibitem{PhysRevB.77.104402}
	A.~M. Souza, M.~S. Reis, D.~O. Soares-Pinto, I.~S. Oliveira, R.~S. Sarthour,
	\newblock \emph{Phys. Rev. B} \textbf{2008}, \emph{77} 104402\relax
	\relax
	\bibitem{PhysRevA.63.052302}
	K.~M. O'Connor, W.~K. Wootters,
	\newblock \emph{Phys. Rev. A} \textbf{2001}, \emph{63} 052302\relax
	\relax
	\bibitem{Wang2002}
	X.~Wang, P.~Zanardi,
	\newblock \emph{Phys. Lett. A} \textbf{2002}, \emph{301} 1 \relax
	\relax
	\bibitem{Terhal_2004}
	B.~M. Terhal,
	\newblock \emph{{IBM} J. Res. Dev.} \textbf{2004}, \emph{48} 71\relax
	\relax
	\bibitem{Scheie2023}
	A.~Scheie, P.~Laurell, E.~Dagotto, D.~A. Tennant, T.~Roscilde,
	\newblock Reconstructing the spatial structure of quantum correlations,
	\textbf{2023},
	\newblock \urlprefix\url{https://doi.org/10.48550/arXiv.2306.11723},
	\newblock ArXiv:2306.11723\relax
	\relax
	\bibitem{Syljuaasen_2004}
	O.~F. Sylju\aa{}sen,
	\newblock \emph{Phys. Lett. A} \textbf{2004}, \emph{322} 25\relax
	\relax
	\bibitem{PhysRevA.69.062314}
	B.-Q. Jin, V.~E. Korepin,
	\newblock \emph{Phys. Rev. A} \textbf{2004}, \emph{69} 062314\relax
	\relax
	\bibitem{PhysRevLett.94.147208}
	T.~Roscilde, P.~Verrucchi, A.~Fubini, S.~Haas, V.~Tognetti,
	\newblock \emph{Phys. Rev. Lett.} \textbf{2005}, \emph{94} 147208\relax
	\relax
	\bibitem{Baroni_2007}
	F.~Baroni, A.~Fubini, V.~Tognetti, P.~Verrucchi,
	\newblock \emph{J. Phys. A: Math. Theor.} \textbf{2007}, \emph{40} 9845\relax
	\relax
	\bibitem{PhysRevLett.98.247201}
	G.~Baskaran, S.~Mandal, R.~Shankar,
	\newblock \emph{Phys. Rev. Lett.} \textbf{2007}, \emph{98} 247201\relax
	\relax
	\bibitem{LHenderson_2001}
	L.~Henderson, V.~Vedral,
	\newblock \emph{J. Phys. A} \textbf{2001}, \emph{34} 6899\relax
	\relax
	\bibitem{RevModPhys.84.1655}
	K.~Modi, A.~Brodutch, H.~Cable, T.~Paterek, V.~Vedral,
	\newblock \emph{Rev. Mod. Phys.} \textbf{2012}, \emph{84} 1655\relax
	\relax
	\bibitem{PhysRevA.81.052318}
	A.~Ferraro, L.~Aolita, D.~Cavalcanti, F.~M. Cucchietti, A.~Ac\'{\i}n,
	\newblock \emph{Phys. Rev. A} \textbf{2010}, \emph{81} 052318\relax
	\relax
	\bibitem{PhysRevA.77.042303}
	S.~Luo,
	\newblock \emph{Phys. Rev. A} \textbf{2008}, \emph{77} 042303\relax
	\relax
	\bibitem{PhysRevB.78.224413}
	R.~Dillenschneider,
	\newblock \emph{Phys. Rev. B} \textbf{2008}, \emph{78} 224413\relax
	\relax
	\bibitem{PhysRevA.81.042105}
	M.~Ali, A.~R.~P. Rau, G.~Alber,
	\newblock \emph{Phys. Rev. A} \textbf{2010}, \emph{81} 042105\relax
	\relax
	\bibitem{PhysRevA.82.069902}
	M.~Ali, A.~R.~P. Rau, G.~Alber,
	\newblock \emph{Phys. Rev. A} \textbf{2010}, \emph{82} 069902\relax
	\relax
	\bibitem{PhysRevA.84.042124}
	N.~Li, S.~Luo,
	\newblock \emph{Phys. Rev. A} \textbf{2011}, \emph{84} 042124\relax
	\relax
	\bibitem{10.1063/1.4862469}
	S.~M. Aldoshin, E.~B. Fel'dman, M.~A. Yurishchev,
	\newblock \emph{Low Temp, Phys.} \textbf{2014}, \emph{40} 3\relax
	\relax
	\bibitem{helstrom1976quantum}
	C.~Helstrom,
	\newblock \emph{Quantum Detection and Estimation Theory},
	\newblock Mathematics in Science and Engineering : a series of monographs and
	textbooks. Academic Press, \textbf{1976}\relax
	\relax
	\bibitem{Holevo2011}
	A.~Holevo,
	\newblock \emph{Probabilistic and Statistical Aspects of Quantum Theory},
	\newblock Scuola Normale Superiore Pisa, second {English} edition,
	\textbf{2011}\relax
	\relax
	\bibitem{PhysRevLett.72.3439}
	S.~L. Braunstein, C.~M. Caves,
	\newblock \emph{Phys. Rev. Lett.} \textbf{1994}, \emph{72} 3439\relax
	\relax
	\bibitem{Pezze2014}
	L.~Pezz\`e, A.~Smerzi,
	\newblock In G.~Tino, M.~Kasevich, editors, \emph{{Atom Interferometry,
			Proceedings of the International School of Physics `{Enrico Fermi}', Course
			188, {Varenna}}}, 691--741. IOS Press, Amsterdam, \textbf{2014}\relax
	\relax
	\bibitem{Toth_2014}
	G.~T{\'{o}}th, I.~Apellaniz,
	\newblock \emph{J. Phys. A: Math. Theor.} \textbf{2014}, \emph{47} 424006\relax
	\relax
	\bibitem{Lambert2023}
	J.~Lambert, E.~S. S\o{}rensen,
	\newblock \emph{New. J. Phys.} \textbf{2023}, \emph{25} 081201\relax
	\relax
	\bibitem{10.1063/1.4818323}
	G.~Xu, Z.~Xu, J.~M. Tranquada,
	\newblock \emph{Rev. Sci. Instrum.} \textbf{2013}, \emph{84} 083906\relax
	\relax
	\bibitem{PhysRevLett.102.100401}
	L.~Pezz\'e, A.~Smerzi,
	\newblock \emph{Phys. Rev. Lett.} \textbf{2009}, \emph{102} 100401\relax
	\relax
	\bibitem{Toth2012}
	G.~T\'oth,
	\newblock \emph{Phys. Rev. A} \textbf{2012}, \emph{85} 022322\relax
	\relax
	\bibitem{Cohen1968}
	M.~Cohen,
	\newblock \emph{IEEE Trans. Inf. Th.} \textbf{1968}, \emph{14} 591\relax
	\relax
	\bibitem{PhysRevLett.127.037201}
	P.~Laurell, A.~Scheie, C.~J. Mukherjee, M.~M. Koza, M.~Enderle, Z.~Tylczynski,
	S.~Okamoto, R.~Coldea, D.~A. Tennant, G.~Alvarez,
	\newblock \emph{Phys. Rev. Lett.} \textbf{2021}, \emph{127} 037201\relax
	\relax
	\bibitem{Petz1996}
	D.~Petz,
	\newblock \emph{Linear Algebra Its Appl.} \textbf{1996}, \emph{244} 81\relax
	\relax
	\bibitem{Gibilisco2009}
	P.~Gibilisco, D.~Imparato, T.~Isola,
	\newblock \emph{Proc. Amer. Math. Soc.} \textbf{2009}, \emph{137} 317\relax
	\relax
	\bibitem{Frerot2017}
	I.~Fr\'erot,
	\newblock {Ph.D.} thesis, École normale supérieure de Lyon, \textbf{2017},
	\newblock \urlprefix\url{https://tel.archives-ouvertes.fr/tel-01679743}\relax
	\relax
	\bibitem{FrerotR2016}
	I.~Fr\'erot, T.~Roscilde,
	\newblock \emph{Phys. Rev. B} \textbf{2016}, \emph{94} 075121\relax
	\relax
	\bibitem{Frerot2019}
	I.~Fr\'erot, T.~Roscilde,
	\newblock \emph{Nat. Commun.} \textbf{2019}, \emph{10} 577\relax
	\relax
	\bibitem{Wigner_1963}
	E.~P. Wigner, M.~M. Yanase,
	\newblock \emph{Proc. Natl. Acad. Sci. U.S.A.} \textbf{1963}, \emph{49}
	910\relax
	\relax
	\bibitem{RevModPhys.50.221}
	A.~Wehrl,
	\newblock \emph{Rev. Mod. Phys.} \textbf{1978}, \emph{50} 221\relax
	\relax
	\bibitem{luo2004wigner}
	S.~Luo,
	\newblock \emph{Proc. Amer. Math. Soc.} \textbf{2004}, \emph{132} 885\relax
	\relax
	\bibitem{PhysRevA.104.042416}
	A.~E. Sifain, F.~Fassioli, G.~D. Scholes,
	\newblock \emph{Phys. Rev. A} \textbf{2021}, \emph{104} 042416\relax
	\relax
	\bibitem{Nielsen2010}
	M.~A. Nielsen, I.~L. Chuang,
	\newblock \emph{Quantum Computation and Quantum Information},
	\newblock Cambridge University Press, Cambridge, UK, \textbf{2010}\relax
	\relax
	\bibitem{Igloi2005}
	F.~Igl\'o{}i, C.~Monthus,
	\newblock \emph{Phys. Rep.} \textbf{2005}, \emph{412} 277\relax
	\relax
	\bibitem{Igloi2018}
	F.~Igl\'o{}i, C.~Monthus,
	\newblock \emph{Eur. Phys. J. B} \textbf{2018}, \emph{91} 290\relax
	\relax
	\bibitem{PhysRevX.8.041040}
	L.~Liu, H.~Shao, Y.-C. Lin, W.~Guo, A.~W. Sandvik,
	\newblock \emph{Phys. Rev. X} \textbf{2018}, \emph{8} 041040\relax
	\relax
	\bibitem{PhysRevLett.123.087201}
	K.~Uematsu, H.~Kawamura,
	\newblock \emph{Phys. Rev. Lett.} \textbf{2019}, \emph{123} 087201\relax
	\relax
	\bibitem{PhysRevLett.79.745}
	A.~W. Garrett, S.~E. Nagler, D.~A. Tennant, B.~C. Sales, T.~Barnes,
	\newblock \emph{Phys. Rev. Lett.} \textbf{1997}, \emph{79} 745\relax
	\relax
	\bibitem{PhysRevB.59.11384}
	T.~Barnes, J.~Riera, D.~A. Tennant,
	\newblock \emph{Phys. Rev. B} \textbf{1999}, \emph{59} 11384\relax
	\relax
	\bibitem{PhysRevB.60.1197}
	R.~Calvo, M.~C.~G. Passeggi, N.~O. Moreno, G.~E. Barberis, A.~B. Chaves,
	B.~C.~M. Torres, L.~Lezama, T.~Rojo,
	\newblock \emph{Phys. Rev. B} \textbf{1999}, \emph{60} 1197\relax
	\relax
	\bibitem{PhysRevLett.99.087204}
	M.~B. Stone, W.~Tian, M.~D. Lumsden, G.~E. Granroth, D.~Mandrus, J.-H. Chung,
	N.~Harrison, S.~E. Nagler,
	\newblock \emph{Phys. Rev. Lett.} \textbf{2007}, \emph{99} 087204\relax
	\relax
	\bibitem{PhysRev.132.1057}
	L.~Berger, S.~A. Friedberg, J.~T. Schriempf,
	\newblock \emph{Phys. Rev.} \textbf{1963}, \emph{132} 1057\relax
	\relax
	\bibitem{PhysRevB.27.248}
	J.~C. Bonner, S.~A. Friedberg, H.~Kobayashi, D.~L. Meier, H.~W.~J. Bl\"ote,
	\newblock \emph{Phys. Rev. B} \textbf{1983}, \emph{27} 248\relax
	\relax
	\bibitem{PhysRevB.67.054414}
	D.~A. Tennant, C.~Broholm, D.~H. Reich, S.~E. Nagler, G.~E. Granroth,
	T.~Barnes, K.~Damle, G.~Xu, Y.~Chen, B.~C. Sales,
	\newblock \emph{Phys. Rev. B} \textbf{2003}, \emph{67} 054414\relax
	\relax
	\bibitem{PhysRevLett.84.4465}
	G.~Xu, C.~Broholm, D.~H. Reich, M.~A. Adams,
	\newblock \emph{Phys. Rev. Lett.} \textbf{2000}, \emph{84} 4465\relax
	\relax
	\bibitem{PhysRevB.85.014402}
	D.~A. Tennant, B.~Lake, A.~J.~A. James, F.~H.~L. Essler, S.~Notbohm, H.-J.
	Mikeska, J.~Fielden, P.~K\"ogerler, P.~C. Canfield, M.~T.~F. Telling,
	\newblock \emph{Phys. Rev. B} \textbf{2012}, \emph{85} 014402\relax
	\relax
	\bibitem{PhysRevB.90.094419}
	M.~B. Stone, Y.~Chen, D.~H. Reich, C.~Broholm, G.~Xu, J.~R.~D. Copley, J.~C.
	Cook,
	\newblock \emph{Phys. Rev. B} \textbf{2014}, \emph{90} 094419\relax
	\relax
	\bibitem{Das_2013}
	D.~Das, H.~Singh, T.~Chakraborty, R.~K. Gopal, C.~Mitra,
	\newblock \emph{New J. Phys.} \textbf{2013}, \emph{15} 013047\relax
	\relax
	\bibitem{PhysRevB.84.024418}
	M.~A. Yurishchev,
	\newblock \emph{Phys. Rev. B} \textbf{2011}, \emph{84} 024418\relax
	\relax
	\bibitem{Singh_2015}
	H.~Singh, T.~Chakraborty, P.~K. Panigrahi, C.~Mitra,
	\newblock \emph{Quantum Inf. Process.} \textbf{2015}, \emph{14} 951\relax
	\relax
	\bibitem{PhysRevB.79.054408}
	A.~M. Souza, D.~O. Soares-Pinto, R.~S. Sarthour, I.~S. Oliveira, M.~S. Reis,
	P.~Brand\~ao, A.~M. dos Santos,
	\newblock \emph{Phys. Rev. B} \textbf{2009}, \emph{79} 054408\relax
	\relax
	\bibitem{Soares-Pinto_2009}
	D.~O. Soares-Pinto, A.~M. Souza, R.~S. Sarthour, I.~S. Oliveira, M.~S. Reis,
	P.~Brandão, J.~Rocha, A.~M. dos Santos,
	\newblock \emph{Europhys. Lett.} \textbf{2009}, \emph{87} 40008\relax
	\relax
	\bibitem{Reis_2012}
	M.~S. Reis, S.~Soriano, A.~M. dos Santos, B.~C. Sales, D.~O. Soares-Pinto,
	P.~Brandão,
	\newblock \emph{Europhys. Lett.} \textbf{2012}, \emph{100} 50001\relax
	\relax
	\bibitem{doi:10.1063/1.4861732}
	T.~Chakraborty, H.~Singh, C.~Mitra,
	\newblock \emph{J. Appl. Phys.} \textbf{2014}, \emph{115} 034909\relax
	\relax
	\bibitem{Athira_2023}
	S.~Athira, S.~L.~L. Silva, P.~Nag, S.~Lakshmi, S.~K. C, D.~P. Panda, S.~Das,
	S.~Rajput, A.~P. Alex, A.~Sundaresan, S.~R. Vennapusa, T.~Maitra,
	D.~Jaiswal-Nagar,
	\newblock \emph{New J. Phys.} \textbf{2023}, \emph{25} 103002\relax
	\relax
	\bibitem{PhysRevB.75.054422}
	T.~G. Rappoport, L.~Ghivelder, J.~C. Fernandes, R.~B. Guimar\~aes, M.~A.
	Continentino,
	\newblock \emph{Phys. Rev. B} \textbf{2007}, \emph{75} 054422\relax
	\relax
	\bibitem{Sahling2015}
	S.~Sahling, G.~Remenyi, C.~Paulsen, P.~Monceau, V.~Saligrama, C.~Marin,
	A.~Revcolevschi, L.~P. Regnault, S.~Raymond, J.~E. Lorenzo,
	\newblock \emph{Nat. Phys.} \textbf{2015}, \emph{11} 255\relax
	\relax
	\bibitem{PhysRevB.100.235103}
	W.~Y. Kon, T.~Krisnanda, P.~Sengupta, T.~Paterek,
	\newblock \emph{Phys. Rev. B} \textbf{2019}, \emph{100} 235103\relax
	\relax
	\bibitem{PhysRevLett.104.037203}
	A.~Candini, G.~Lorusso, F.~Troiani, A.~Ghirri, S.~Carretta, P.~Santini,
	G.~Amoretti, C.~Muryn, F.~Tuna, G.~Timco, E.~J.~L. McInnes, R.~E.~P.
	Winpenny, W.~Wernsdorfer, M.~Affronte,
	\newblock \emph{Phys. Rev. Lett.} \textbf{2010}, \emph{104} 037203\relax
	\relax
	\bibitem{Garlatti2017}
	E.~Garlatti, T.~Guidi, S.~Ansbro, P.~Santini, G.~Amoretti, J.~Ollivier,
	H.~Mutka, G.~Timco, I.~J. Vitorica-Yrezabal, G.~F.~S. Whitehead, R.~E.~P.
	Winpenny, S.~Carretta,
	\newblock \emph{Nat. Commun.} \textbf{2017}, \emph{8} 14543\relax
	\relax
	\bibitem{Garlatti_2019}
	E.~Garlatti, A.~Chiesa, T.~Guidi, G.~Amoretti, P.~Santini, S.~Carretta,
	\newblock \emph{Eur. J. Inorg. Chem.} \textbf{2019}, \emph{2019} 1106\relax
	\relax
	\bibitem{PhysRevLett.94.207208}
	F.~Troiani, A.~Ghirri, M.~Affronte, S.~Carretta, P.~Santini, G.~Amoretti,
	S.~Piligkos, G.~Timco, R.~E.~P. Winpenny,
	\newblock \emph{Phys. Rev. Lett.} \textbf{2005}, \emph{94} 207208\relax
	\relax
	\bibitem{10.1063/5.0053378}
	S.~Carretta, D.~Zueco, A.~Chiesa, A.~G\'omez-Le\'on, F.~Luis,
	\newblock \emph{Appl. Phys. Lett.} \textbf{2021}, \emph{118} 240501\relax
	\relax
	\bibitem{Moreno_Pineda_2021}
	E.~Moreno-Pineda, W.~Wernsdorfer,
	\newblock \emph{Nat. Rev. Phys.} \textbf{2021}, \emph{3} 645\relax
	\relax
	\bibitem{PhysRevLett.107.230502}
	P.~Santini, S.~Carretta, F.~Troiani, G.~Amoretti,
	\newblock \emph{Phys. Rev. Lett.} \textbf{2011}, \emph{107} 230502\relax
	\relax
	\bibitem{Heisenberg1928}
	W.~Heisenberg,
	\newblock \emph{Z. Physik} \textbf{1928}, \emph{49} 619\relax
	\relax
	\bibitem{PhysRevLett.91.207901}
	S.~Bose,
	\newblock \emph{Phys. Rev. Lett.} \textbf{2003}, \emph{91} 207901\relax
	\relax
	\bibitem{Bose_2007}
	S.~Bose,
	\newblock \emph{Contemp. Phys.} \textbf{2007}, \emph{48} 13\relax
	\relax
	\bibitem{Haldane1983}
	F.~Haldane,
	\newblock \emph{Phys. Lett. A} \textbf{1983}, \emph{93} 464 \relax
	\relax
	\bibitem{PhysRevLett.50.1153}
	F.~D.~M. Haldane,
	\newblock \emph{Phys. Rev. Lett.} \textbf{1983}, \emph{50} 1153\relax
	\relax
	\bibitem{Affleck1989a}
	I.~Affleck,
	\newblock \emph{J. Phys. Condens. Matter} \textbf{1989}, \emph{1} 3047\relax
	\relax
	\bibitem{Auerbach1994}
	A.~Auerbach,
	\newblock \emph{Interacting Electrons and Quantum Magnetism},
	\newblock Springer-Verlag, New York, \textbf{1994}\relax
	\relax
	\bibitem{Faddeev1981}
	L.~Faddeev, L.~Takhtajan,
	\newblock \emph{Phys. Lett. A} \textbf{1981}, \emph{85} 375\relax
	\relax
	\bibitem{PhysRevB.99.045117}
	J.~Lambert, E.~S. S\o{}rensen,
	\newblock \emph{Phys. Rev. B} \textbf{2019}, \emph{99} 045117\relax
	\relax
	\bibitem{PhysRevB.108.144414}
	F.~Dell'Anna, S.~Pradhan, C.~D.~E. Boschi, E.~Ercolessi,
	\newblock \emph{Phys. Rev. B} \textbf{2023}, \emph{108} 144414\relax
	\relax
	\bibitem{PhysRevB.107.059902}
	A.~Scheie, P.~Laurell, A.~M. Samarakoon, B.~Lake, S.~E. Nagler, G.~E. Granroth,
	S.~Okamoto, G.~Alvarez, D.~A. Tennant,
	\newblock \emph{Phys. Rev. B} \textbf{2023}, \emph{107} 059902\relax
	\relax
	\bibitem{Bethe1931}
	H.~Bethe,
	\newblock \emph{Z. Physik} \textbf{1931}, \emph{71} 205\relax
	\relax
	\bibitem{Hulthen1938}
	L.~Hulth{\'e}n,
	\newblock Ph.D. thesis, Stockholm College, \textbf{1938},
	\newblock
	\urlprefix\url{http://urn.kb.se/resolve?urn=urn:nbn:se:su:diva-72311},
	\newblock Ark. Mat. Astron. Fys.\relax
	\relax
	\bibitem{PhysRevLett.92.096402}
	V.~E. Korepin,
	\newblock \emph{Phys. Rev. Lett.} \textbf{2004}, \emph{92} 096402\relax
	\relax
	\bibitem{Calabrese2004}
	P.~Calabrese, J.~Cardy,
	\newblock \emph{J. Stat. Mech.} \textbf{2004}, \emph{2004} P06002\relax
	\relax
	\bibitem{PhysRevLett.87.017901}
	M.~C. Arnesen, S.~Bose, V.~Vedral,
	\newblock \emph{Phys. Rev. Lett.} \textbf{2001}, \emph{87} 017901\relax
	\relax
	\bibitem{PhysRevB.52.13368}
	D.~A. Tennant, R.~A. Cowley, S.~E. Nagler, A.~M. Tsvelik,
	\newblock \emph{Phys. Rev. B} \textbf{1995}, \emph{52} 13368\relax
	\relax
	\bibitem{Lake2005}
	B.~Lake, D.~A. Tennant, C.~D. Frost, S.~E. Nagler,
	\newblock \emph{Nat. Mater.} \textbf{2005}, \emph{4} 329\relax
	\relax
	\bibitem{PhysRevLett.111.137205}
	B.~Lake, D.~A. Tennant, J.-S. Caux, T.~Barthel, U.~Schollw\"ock, S.~E. Nagler,
	C.~D. Frost,
	\newblock \emph{Phys. Rev. Lett.} \textbf{2013}, \emph{111} 137205\relax
	\relax
	\bibitem{Blanc2018}
	N.~Blanc, J.~Trinh, L.~Dong, X.~Bai, A.~A. Aczel, M.~Mourigal, L.~Balents,
	T.~Siegrist, A.~P. Ramirez,
	\newblock \emph{Nat. Phys.} \textbf{2018}, \emph{14} 273\relax
	\relax
	\bibitem{Wu2019}
	L.~S. Wu, S.~E. Nikitin, Z.~Wang, W.~Zhu, C.~D. Batista, A.~M. Tsvelik, A.~M.
	Samarakoon, D.~A. Tennant, M.~Brando, L.~Vasylechko, M.~Frontzek, A.~T.
	Savici, G.~Sala, G.~Ehlers, A.~D. Christianson, M.~D. Lumsden, A.~Podlesnyak,
	\newblock \emph{Nat. Commun.} \textbf{2019}, \emph{10} 698\relax
	\relax
	\bibitem{Gao2023}
	S.~Gao, L.-F. Lin, P.~Laurell, Q.~Chen, Q.~Huang, C.~d. Cruz, K.~V. Vemuru,
	M.~D. Lumsden, S.~E. Nagler, G.~Alvarez, E.~Dagotto, H.~Zhou, A.~D.
	Christianson, M.~B. Stone,
	\newblock \emph{Phys. Rev. B} \textbf{2024}, \emph{109} L020402\relax
	\relax
	\bibitem{Chakraborty2012}
	T.~Chakraborty, H.~Singh, D.~Das, T.~K. Sen, C.~Mitra,
	\newblock \emph{Phys. Lett. A} \textbf{2012}, \emph{376} 2967\relax
	\relax
	\bibitem{10.1063/1.4824458}
	T.~Chakraborty, T.~K. Sen, H.~Singh, D.~Das, S.~K. Mandal, C.~Mitra,
	\newblock \emph{J. Appl. Phys.} \textbf{2013}, \emph{114} 144904\relax
	\relax
	\bibitem{PhysRevResearch.2.043329}
	G.~Mathew, S.~L.~L. Silva, A.~Jain, A.~Mohan, D.~T. Adroja, V.~G. Sakai, C.~V.
	Tomy, A.~Banerjee, R.~Goreti, A.~V.N., R.~Singh, D.~Jaiswal-Nagar,
	\newblock \emph{Phys. Rev. Research} \textbf{2020}, \emph{2} 043329\relax
	\relax
	\bibitem{ccby4}
	Creative commons attribution 4.0 international license,
	\newblock \urlprefix\url{https://creativecommons.org/licenses/by/4.0/}\relax
	\relax
	\bibitem{Hutchings_1979}
	M.~T. Hutchings, J.~M. Milne, H.~Ikeda,
	\newblock \emph{J. Phys. C: Solid State Phys.} \textbf{1979}, \emph{12}
	L739\relax
	\relax
	\bibitem{PhysRevB.21.2001}
	S.~K. Satija, J.~D. Axe, G.~Shirane, H.~Yoshizawa, K.~Hirakawa,
	\newblock \emph{Phys. Rev. B} \textbf{1980}, \emph{21} 2001\relax
	\relax
	\bibitem{PhysRevB.44.12361}
	S.~E. Nagler, D.~A. Tennant, R.~A. Cowley, T.~G. Perring, S.~K. Satija,
	\newblock \emph{Phys. Rev. B} \textbf{1991}, \emph{44} 12361\relax
	\relax
	\bibitem{PhysRevLett.85.832}
	B.~Lake, D.~A. Tennant, S.~E. Nagler,
	\newblock \emph{Phys. Rev. Lett.} \textbf{2000}, \emph{85} 832\relax
	\relax
	\bibitem{doi:10.1063/1.4709772}
	V.~Gnezdilov, J.~Deisenhofer, P.~Lemmens, D.~Wulferding, O.~Afanasiev,
	P.~Ghigna, A.~Loidl, A.~Yeremenko,
	\newblock \emph{Low Temp. Phys.} \textbf{2012}, \emph{38} 419\relax
	\relax
	\bibitem{Scheie2021}
	A.~Scheie, N.~E. Sherman, M.~Dupont, S.~E. Nagler, M.~B. Stone, G.~E. Granroth,
	J.~E. Moore, D.~A. Tennant,
	\newblock \emph{Nat. Phys.} \textbf{2021}, \emph{17} 726\relax
	\relax
	\bibitem{Scheie2022}
	A.~Scheie, P.~Laurell, B.~Lake, S.~E. Nagler, M.~B. Stone, J.-S. Caux, D.~A.
	Tennant,
	\newblock \emph{Nat. Commun.} \textbf{2022}, \emph{13} 5796\relax
	\relax
	\bibitem{Dai_2016}
	H.-N. Dai, B.~Yang, A.~Reingruber, X.-F. Xu, X.~Jiang, Y.-A. Chen, Z.-S. Yuan,
	J.-W. Pan,
	\newblock \emph{Nat. Phys.} \textbf{2016}, \emph{12} 783\relax
	\relax
	\bibitem{PhysRevLett.131.073401}
	W.-Y. Zhang, M.-G. He, H.~Sun, Y.-G. Zheng, Y.~Liu, A.~Luo, H.-Y. Wang, Z.-H.
	Zhu, P.-Y. Qiu, Y.-C. Shen, X.-K. Wang, W.~Lin, S.-T. Yu, B.-C. Li, B.~Xiao,
	M.-D. Li, Y.-M. Yang, X.~Jiang, H.-N. Dai, Y.~Zhou, X.~Ma, Z.-S. Yuan, J.-W.
	Pan,
	\newblock \emph{Phys. Rev. Lett.} \textbf{2023}, \emph{131} 073401\relax
	\relax
	\bibitem{PhysRevLett.130.129902}
	P.~Laurell, A.~Scheie, C.~J. Mukherjee, M.~M. Koza, M.~Enderle, Z.~Tylczynski,
	S.~Okamoto, R.~Coldea, D.~A. Tennant, G.~Alvarez,
	\newblock \emph{Phys. Rev. Lett.} \textbf{2023}, \emph{130} 129902\relax
	\relax
	\bibitem{PhysRevB.65.144432}
	M.~Kenzelmann, R.~Coldea, D.~A. Tennant, D.~Visser, M.~Hofmann, P.~Smeibidl,
	Z.~Tylczynski,
	\newblock \emph{Phys. Rev. B} \textbf{2002}, \emph{65} 144432\relax
	\relax
	\bibitem{PhysRevLett.111.187202}
	O.~Breunig, M.~Garst, E.~Sela, B.~Buldmann, P.~Becker, L.~Bohat\'y,
	R.~M\"uller, T.~Lorenz,
	\newblock \emph{Phys. Rev. Lett.} \textbf{2013}, \emph{111} 187202\relax
	\relax
	\bibitem{PhysRevB.65.172409}
	D.~V. Dmitriev, V.~Y. Krivnov, A.~A. Ovchinnikov,
	\newblock \emph{Phys. Rev. B} \textbf{2002}, \emph{65} 172409\relax
	\relax
	\bibitem{Abouie_2010}
	J.~Abouie, A.~Langari, M.~Siahatgar,
	\newblock \emph{J. Phys. Condens. Matter} \textbf{2010}, \emph{22} 216008\relax
	\relax
	\bibitem{PhysRevLett.108.240503}
	L.~Amico, D.~Rossini, A.~Hamma, V.~E. Korepin,
	\newblock \emph{Phys. Rev. Lett.} \textbf{2012}, \emph{108} 240503\relax
	\relax
	\bibitem{PhysRevA.96.052303}
	S.~Mahdavifar, S.~Mahdavifar, R.~Jafari,
	\newblock \emph{Phys. Rev. A} \textbf{2017}, \emph{96} 052303\relax
	\relax
	\bibitem{Lieb1972}
	E.~H. Lieb, D.~W. Robinson,
	\newblock \emph{Commun. Math. Phys.} \textbf{1972}, \emph{28} 251\relax
	\relax
	\bibitem{PhysRevLett.97.050401}
	S.~Bravyi, M.~B. Hastings, F.~Verstraete,
	\newblock \emph{Phys. Rev. Lett.} \textbf{2006}, \emph{97} 050401\relax
	\relax
	\bibitem{(Anthony)Chen_2023}
	C.-F.~A. Chen, A.~Lucas, C.~Yin,
	\newblock \emph{Rep. Prog. Phys.} \textbf{2023}, \emph{86} 116001\relax
	\relax
	\bibitem{Cheneau2012}
	M.~Cheneau, P.~Barmettler, D.~Poletti, M.~Endres, P.~Schauß, T.~Fukuhara,
	C.~Gross, I.~Bloch, C.~Kollath, S.~Kuhr,
	\newblock \emph{Nature} \textbf{2012}, \emph{481} 484\relax
	\relax
	\bibitem{Liu_2020}
	J.~Liu, H.~Yuan, X.-M. Lu, X.~Wang,
	\newblock \emph{J. Phys. A: Math. Theor.} \textbf{2019}, \emph{53} 023001\relax
	\relax
	\bibitem{MalpettiR2016}
	D.~Malpetti, T.~Roscilde,
	\newblock \emph{Phys. Rev. Lett.} \textbf{2016}, \emph{117} 130401\relax
	\relax
	\bibitem{PhysRevX.12.021022}
	T.~Kuwahara, K.~Saito,
	\newblock \emph{Phys. Rev. X} \textbf{2022}, \emph{12} 021022\relax
	\relax
	\bibitem{Scheie2023tlafm}
	A.~O. Scheie, E.~A. Ghioldi, J.~Xing, J.~A.~M. Paddison, N.~E. Sherman,
	M.~Dupont, L.~D. Sanjeewa, S.~Lee, A.~J. Woods, D.~Abernathy, D.~M.
	Pajerowski, T.~J. Williams, S.-S. Zhang, L.~O. Manuel, A.~E. Trumper, C.~D.
	Pemmaraju, A.~S. Sefat, D.~S. Parker, T.~P. Devereaux, R.~Movshovich, J.~E.
	Moore, C.~D. Batista, D.~A. Tennant,
	\newblock \emph{Nat. Phys.} \textbf{2024}, \emph{20} 74\relax
	\relax
	\bibitem{Knolle2019}
	J.~Knolle, R.~Moessner,
	\newblock \emph{Annu. Rev. Condens. Matter Phys.} \textbf{2019}, \emph{10}
	451\relax
	\relax
	\bibitem{Khatua2023}
	J.~Khatua, B.~Sana, A.~Zorko, M.~Gomilšek, K.~Sethupathi, M.~R. Rao,
	M.~Baenitz, B.~Schmidt, P.~Khuntia,
	\newblock \emph{Phys. Rep.} \textbf{2023}, \emph{1041} 1\relax
	\relax
	\bibitem{PhysRevLett.119.250401}
	L.~Pezz\`e, M.~Gabbrielli, L.~Lepori, A.~Smerzi,
	\newblock \emph{Phys. Rev. Lett.} \textbf{2017}, \emph{119} 250401\relax
	\relax
	\bibitem{Gabbrielli2018}
	M.~Gabbrielli, A.~Smerzi, L.~Pezz\`e,
	\newblock \emph{Sci. Rep.} \textbf{2018}, \emph{8} 15663\relax
	\relax
	\bibitem{PhysRevB.102.224401}
	J.~Lambert, E.~S. S\o{}rensen,
	\newblock \emph{Phys. Rev. B} \textbf{2020}, \emph{102} 224401\relax
	\relax
	\bibitem{PhysRevB.109.014425}
	A.~O. Scheie, Y.~Kamiya, H.~Zhang, S.~Lee, A.~J. Woods, M.~O. Ajeesh, M.~G.
	Gonzalez, B.~Bernu, J.~W. Villanova, J.~Xing, Q.~Huang, Q.~Zhang, J.~Ma,
	E.~S. Choi, D.~M. Pajerowski, H.~Zhou, A.~S. Sefat, S.~Okamoto, T.~Berlijn,
	L.~Messio, R.~Movshovich, C.~D. Batista, D.~A. Tennant,
	\newblock \emph{Phys. Rev. B} \textbf{2024}, \emph{109} 014425\relax
	\relax
	\bibitem{PhysRevX.11.021044}
	P.-L. Dai, G.~Zhang, Y.~Xie, C.~Duan, Y.~Gao, Z.~Zhu, E.~Feng, Z.~Tao, C.-L.
	Huang, H.~Cao, A.~Podlesnyak, G.~E. Granroth, M.~S. Everett, J.~C. Neuefeind,
	D.~Voneshen, S.~Wang, G.~Tan, E.~Morosan, X.~Wang, H.-Q. Lin, L.~Shu,
	G.~Chen, Y.~Guo, X.~Lu, P.~Dai,
	\newblock \emph{Phys. Rev. X} \textbf{2021}, \emph{11} 021044\relax
	\relax
	\bibitem{PhysRevB.106.L060401}
	F.~L. Pratt, F.~Lang, W.~Steinhardt, S.~Haravifard, S.~J. Blundell,
	\newblock \emph{Phys. Rev. B} \textbf{2022}, \emph{106} L060401\relax
	\relax
	\bibitem{PhysRevLett.119.157201}
	Z.~Zhu, P.~A. Maksimov, S.~R. White, A.~L. Chernyshev,
	\newblock \emph{Phys. Rev. Lett.} \textbf{2017}, \emph{119} 157201\relax
	\relax
	\bibitem{PhysRevB.104.224433}
	Z.~Ma, Z.-Y. Dong, J.~Wang, S.~Zheng, K.~Ran, S.~Bao, Z.~Cai, Y.~Shangguan,
	W.~Wang, M.~Boehm, P.~Steffens, L.-P. Regnault, X.~Wang, Y.~Su, S.-L. Yu,
	J.-M. Liu, J.-X. Li, J.~Wen,
	\newblock \emph{Phys. Rev. B} \textbf{2021}, \emph{104} 224433\relax
	\relax
	\bibitem{Pratt_2023}
	F.~L. Pratt, F.~Lang, S.~J. Blundell, W.~Steinhardt, S.~Haravifard,
	S.~Mañas-Valero, E.~Coronado, B.~M. Huddart, T.~Lancaster,
	\newblock \emph{J. Phys. Conf. Ser.} \textbf{2023}, \emph{2462} 012038\relax
	\relax
	\bibitem{Hong2023v3}
	T.~Hong, I.~Makhfudz, X.~Ke, A.~F. May, A.~A. Podlesnyak, D.~Pajerowski,
	B.~Winn, M.~Deumal, Y.~Takano, M.~M. Turnbull,
	\newblock Coexistence of symmetry-protected topological order and {N\'eel}
	order in the spin-1/2 ladder antiferromagnet
	{C}$_9${H}$_{18}${N}$_2${CuBr}$_4$, \textbf{2023},
	\newblock \urlprefix\url{https://arxiv.org/abs/2306.06021v3},
	\newblock ArXiv:2306.06021\relax
	\relax
	\bibitem{Hong_2022}
	T.~Hong, T.~Ying, Q.~Huang, S.~E. Dissanayake, Y.~Qiu, M.~M. Turnbull, A.~A.
	Podlesnyak, Y.~Wu, H.~Cao, Y.~Liu, I.~Umehara, J.~Gouchi, Y.~Uwatoko,
	M.~Matsuda, D.~A. Tennant, G.-W. Chern, K.~P. Schmidt, S.~Wessel,
	\newblock \emph{Nat. Commun.} \textbf{2022}, \emph{13} 3073\relax
	\relax
	\bibitem{fang_2024}
	Y.~Fang, M.~Mahankali, Y.~Wang, L.~Chen, H.~Hu, S.~Paschen, Q.~Si,
	\newblock Amplified entanglement witnessed in a quantum critical metal,
	\textbf{2024},
	\newblock \urlprefix\url{https://arxiv.org/abs/2402.18552},
	\newblock ArXiv:2402.18552\relax
	\relax
	\bibitem{mazza_2024}
	F.~Mazza, S.~Biswas, X.~Yan, A.~Prokofiev, P.~Steffens, Q.~Si, F.~F. Assaad,
	S.~Paschen,
	\newblock Quantum {Fisher} information in a strange metal, \textbf{2024},
	\newblock \urlprefix\url{https://arxiv.org/abs/2403.12779},
	\newblock ArXiv:2403.12779\relax
	\relax
	\bibitem{PhysRevB.106.085110}
	P.~Laurell, A.~Scheie, D.~A. Tennant, S.~Okamoto, G.~Alvarez, E.~Dagotto,
	\newblock \emph{Phys. Rev. B} \textbf{2022}, \emph{106} 085110\relax
	\relax
	\bibitem{PhysRevB.107.119901}
	P.~Laurell, A.~Scheie, D.~A. Tennant, S.~Okamoto, G.~Alvarez, E.~Dagotto,
	\newblock \emph{Phys. Rev. B} \textbf{2023}, \emph{107} 119901\relax
	\relax
	\bibitem{Cozzolino2019}
	D.~Cozzolino, B.~Da~Lio, D.~Bacco, L.~K. Oxenløwe,
	\newblock \emph{Adv. Quantum Technol.} \textbf{2019}, \emph{2} 1900038\relax
	\relax
	\bibitem{Erhard_2020}
	M.~Erhard, M.~Krenn, A.~Zeilinger,
	\newblock \emph{Nat. Rev. Phys.} \textbf{2020}, \emph{2} 365\relax
	\relax
	\bibitem{schulke2007electron}
	W.~Sch{\"u}lke,
	\newblock \emph{Electron Dynamics by Inelastic X-ray Scattering}, volume~7 of
	\emph{Oxford Series on Synchrotron Radiation},
	\newblock Oxford University Press, \textbf{2007}\relax
	\relax
	\bibitem{SciPostPhys.3.4.026}
	S.~Vig, A.~Kogar, M.~Mitrano, A.~A. Husain, V.~Mishra, M.~S. Rak, L.~Venema,
	P.~D. Johnson, G.~D. Gu, E.~Fradkin, M.~R. Norman, P.~Abbamonte,
	\newblock \emph{SciPost Phys.} \textbf{2017}, \emph{3} 026\relax
	\relax
	\bibitem{RevModPhys.83.705}
	L.~J.~P. Ament, M.~van Veenendaal, T.~P. Devereaux, J.~P. Hill, J.~van~den
	Brink,
	\newblock \emph{Rev. Mod. Phys.} \textbf{2011}, \emph{83} 705\relax
	\relax
	\bibitem{Ren2024}
	T.~Ren, Y.~Shen, S.~F.~R. TenHuisen, J.~Sears, W.~He, M.~H. Upton, D.~Casa,
	P.~Becker, M.~Mitrano, M.~P.~M. Dean, R.~M. Konik,
	\newblock Witnessing quantum entanglement using resonant inelastic x-ray
	scattering, \textbf{2024},
	\newblock \urlprefix\url{https://doi.org/10.48550/arXiv.2404.05850},
	\newblock ArXiv:2404.05850\relax
	\relax
	\bibitem{Malla2023}
	R.~K. Malla, A.~Weichselbaum, T.-C. Wei, R.~M. Konik,
	\newblock Detecting multipartite entanglement patterns using single particle
	{Green's} functions, \textbf{2023},
	\newblock \urlprefix\url{https://doi.org/10.48550/arXiv.2310.05870},
	\newblock ArXiv:2310.05870\relax
	\relax
	\bibitem{Roosz2023}
	G.~Ro\'o{}sz, A.~Kauch, F.~Bippus, D.~Wieser, K.~Held,
	\newblock Two-site reduced density matrix from one- and two-particle {Green's}
	functions, \textbf{2024},
	\newblock \urlprefix\url{https://doi.org/10.48550/arXiv.2312.14275},
	\newblock ArXiv:2312.14275\relax
	\relax
	\bibitem{Li2008}
	Y.-Q. Li, G.-Q. Zhu,
	\newblock \emph{Front. Phys. China} \textbf{2008}, \emph{3} 250\relax
	\relax
	\bibitem{Eltschka_2014}
	C.~Eltschka, J.~Siewert,
	\newblock \emph{J. Phys. A: Math. Theor.} \textbf{2014}, \emph{47} 424005\relax
	\relax
	\bibitem{Osterloh_2015}
	A.~Osterloh,
	\newblock \emph{J. Phys. A: Math. Theor.} \textbf{2015}, \emph{48} 065303\relax
	\relax
	\bibitem{Bahmani2020}
	H.~Bahmani, G.~Najarbashi, A.~Tavana,
	\newblock \emph{Phys. Scr.} \textbf{2020}, \emph{95} 055701\relax
	\relax
	\bibitem{PhysRevA.62.062314}
	W.~D\"ur, G.~Vidal, J.~I. Cirac,
	\newblock \emph{Phys. Rev. A} \textbf{2000}, \emph{62} 062314\relax
	\relax
	\bibitem{PhysRevLett.97.260502}
	R.~Lohmayer, A.~Osterloh, J.~Siewert, A.~Uhlmann,
	\newblock \emph{Phys. Rev. Lett.} \textbf{2006}, \emph{97} 260502\relax
	\relax
	\bibitem{Gour_2010}
	G.~Gour, N.~R. Wallach,
	\newblock \emph{J. Math. Phys.} \textbf{2010}, \emph{51} 112201\relax
	\relax
	\bibitem{PhysRevA.94.012323}
	A.~Osterloh,
	\newblock \emph{Phys. Rev. A} \textbf{2016}, \emph{94} 012323\relax
	\relax
	\bibitem{PhysRevA.63.044301}
	A.~Wong, N.~Christensen,
	\newblock \emph{Phys. Rev. A} \textbf{2001}, \emph{63} 044301\relax
	\relax
	\bibitem{Li2012}
	D.~Li,
	\newblock \emph{Quantum Inf. Process.} \textbf{2012}, \emph{11} 481\relax
	\relax
	\bibitem{PhysRevLett.118.036804}
	F.~Brange, O.~Malkoc, P.~Samuelsson,
	\newblock \emph{Phys. Rev. Lett.} \textbf{2017}, \emph{118} 036804\relax
	\relax
	\bibitem{PhysRevB.104.245425}
	M.~Tam, C.~Flindt, F.~Brange,
	\newblock \emph{Phys. Rev. B} \textbf{2021}, \emph{104} 245425\relax
	\relax
	\bibitem{RevModPhys.83.863}
	A.~Polkovnikov, K.~Sengupta, A.~Silva, M.~Vengalattore,
	\newblock \emph{Rev. Mod. Phys.} \textbf{2011}, \emph{83} 863\relax
	\relax
	\bibitem{Eisert_2015}
	J.~Eisert, M.~Friesdorf, C.~Gogolin,
	\newblock \emph{Nat. Phys.} \textbf{2015}, \emph{11} 124\relax
	\relax
	\bibitem{RevModPhys.91.021001}
	D.~A. Abanin, E.~Altman, I.~Bloch, M.~Serbyn,
	\newblock \emph{Rev. Mod. Phys.} \textbf{2019}, \emph{91} 021001\relax
	\relax
	\bibitem{PhysRevLett.106.040401}
	C.~Gogolin, M.~P. M\"uller, J.~Eisert,
	\newblock \emph{Phys. Rev. Lett.} \textbf{2011}, \emph{106} 040401\relax
	\relax
	\bibitem{Kaufman2016}
	A.~M. Kaufman, M.~E. Tai, A.~Lukin, M.~Rispoli, R.~Schittko, P.~M. Preiss,
	M.~Greiner,
	\newblock \emph{Science} \textbf{2016}, \emph{353} 794\relax
	\relax
	\bibitem{PhysRevA.43.2046}
	J.~M. Deutsch,
	\newblock \emph{Phys. Rev. A} \textbf{1991}, \emph{43} 2046\relax
	\relax
	\bibitem{PhysRevE.50.888}
	M.~Srednicki,
	\newblock \emph{Phys. Rev. E} \textbf{1994}, \emph{50} 888\relax
	\relax
	\bibitem{Rigol2008}
	M.~Rigol, V.~Dunko, M.~Olshanii,
	\newblock \emph{Nature} \textbf{2008}, \emph{452} 854\relax
	\relax
	\bibitem{D_Alessio_2016}
	L.~D’Alessio, Y.~Kafri, A.~Polkovnikov, M.~Rigol,
	\newblock \emph{Adv. Phys.} \textbf{2016}, \emph{65} 239\relax
	\relax
	\bibitem{PhysRevLett.124.040605}
	M.~Brenes, S.~Pappalardi, J.~Goold, A.~Silva,
	\newblock \emph{Phys. Rev. Lett.} \textbf{2020}, \emph{124} 040605\relax
	\relax
	\bibitem{Chandran2023}
	A.~Chandran, T.~Iadecola, V.~Khemani, R.~Moessner,
	\newblock \emph{Annu. Rev. Condens. Matter Phys.} \textbf{2023}, \emph{14}
	443\relax
	\relax
	\bibitem{Moessner_2017}
	R.~Moessner, S.~L. Sondhi,
	\newblock \emph{Nat. Phys.} \textbf{2017}, \emph{13} 424\relax
	\relax
	\bibitem{RevModPhys.80.885}
	I.~Bloch, J.~Dalibard, W.~Zwerger,
	\newblock \emph{Rev. Mod. Phys.} \textbf{2008}, \emph{80} 885\relax
	\relax
	\bibitem{doi:10.1146/annurev-matsci-070813-113258}
	J.~Zhang, R.~Averitt,
	\newblock \emph{Annu. Rev. Mater. Res.} \textbf{2014}, \emph{44} 19\relax
	\relax
	\bibitem{Basov_2017}
	D.~N. Basov, R.~D. Averitt, D.~Hsieh,
	\newblock \emph{Nat. Mater.} \textbf{2017}, \emph{16} 1077\relax
	\relax
	\bibitem{RevModPhys.93.041002}
	A.~de~la Torre, D.~M. Kennes, M.~Claassen, S.~Gerber, J.~W. McIver, M.~A.
	Sentef,
	\newblock \emph{Rev. Mod. Phys.} \textbf{2021}, \emph{93} 041002\relax
	\relax
	\bibitem{Zhang_2017}
	J.~Zhang, P.~W. Hess, A.~Kyprianidis, P.~Becker, A.~Lee, J.~Smith, G.~Pagano,
	I.-D. Potirniche, A.~C. Potter, A.~Vishwanath, N.~Y. Yao, C.~Monroe,
	\newblock \emph{Nature} \textbf{2017}, \emph{543} 217\relax
	\relax
	\bibitem{Choi_2017}
	S.~Choi, J.~Choi, R.~Landig, G.~Kucsko, H.~Zhou, J.~Isoya, F.~Jelezko,
	S.~Onoda, H.~Sumiya, V.~Khemani, C.~von Keyserlingk, N.~Y. Yao, E.~Demler,
	M.~D. Lukin,
	\newblock \emph{Nature} \textbf{2017}, \emph{543} 221\relax
	\relax
	\bibitem{Islam2015}
	R.~Islam, R.~Ma, P.~M. Preiss, M.~E. Tai, A.~Lukin, M.~Rispoli, M.~Greiner,
	\newblock \emph{Nature} \textbf{2015}, \emph{528} 77\relax
	\relax
	\bibitem{Pappalardi_2017}
	S.~Pappalardi, A.~Russomanno, A.~Silva, R.~Fazio,
	\newblock \emph{J. Stat. Mech.: Theor. Exp.} \textbf{2017}, \emph{2017}
	053104\relax
	\relax
	\bibitem{PhysRevLett.129.020601}
	J.-Y. Desaules, F.~Pietracaprina, Z.~Papi\ifmmode~\acute{c}\else \'{c}\fi{},
	J.~Goold, S.~Pappalardi,
	\newblock \emph{Phys. Rev. Lett.} \textbf{2022}, \emph{129} 020601\relax
	\relax
	\bibitem{PhysRevB.107.035123}
	S.~Dooley, S.~Pappalardi, J.~Goold,
	\newblock \emph{Phys. Rev. B} \textbf{2023}, \emph{107} 035123\relax
	\relax
	\bibitem{PhysRevLett.130.106902}
	D.~R. Baykusheva, M.~H. Kalthoff, D.~Hofmann, M.~Claassen, D.~M. Kennes, M.~A.
	Sentef, M.~Mitrano,
	\newblock \emph{Phys. Rev. Lett.} \textbf{2023}, \emph{130} 106902\relax
	\relax
	\bibitem{Hales2023}
	J.~Hales, U.~Bajpai, T.~Liu, D.~R. Baykusheva, M.~Li, M.~Mitrano, Y.~Wang,
	\newblock \emph{Nat. Commun.} \textbf{2023}, \emph{14} 3512\relax
	\relax
	\bibitem{Suresh2022}
	A.~Suresh, R.~D. Soares, P.~Mondal, J.~P.~S. Pires, J.~M. V.~P. Lopes,
	A.~Ferreira, A.~E. Feiguin, P.~Plech\'a\ifmmode~\check{c}\else \v{c}\fi{},
	B.~K. Nikoli\ifmmode~\acute{c}\else \'{c}\fi{},
	\newblock \emph{Phys. Rev. A} \textbf{2024}, \emph{109} 022414\relax
	\relax
	\bibitem{Kondev_2021}
	F.~Kondev, M.~Wang, W.~Huang, S.~Naimi, G.~Audi,
	\newblock \emph{Chin. Phys. C} \textbf{2021}, \emph{45} 030001\relax
	\relax
	\bibitem{Sponar_2012}
	S.~Sponar, J.~Klepp, K.~Durstberger-Rennhofer, C.~Schmitzer, H.~Bartosik,
	H.~Geppert, M.~Both, G.~Badurek, Y.~Hasegawa,
	\newblock \emph{New J. Phys.} \textbf{2012}, \emph{14} 053032\relax
	\relax
	\bibitem{Shen2020}
	J.~Shen, S.~J. Kuhn, R.~M. Dalgliesh, V.~O. de~Haan, N.~Geerits, A.~A.~M.
	Irfan, F.~Li, S.~Lu, S.~R. Parnell, J.~Plomp, A.~A. van Well, A.~Washington,
	D.~V. Baxter, G.~Ortiz, W.~M. Snow, R.~Pynn,
	\newblock \emph{Nat. Commun.} \textbf{2020}, \emph{11} 930\relax
	\relax
	\bibitem{Hasegawa2003}
	Y.~Hasegawa, R.~Loidl, G.~Badurek, M.~Baron, H.~Rauch,
	\newblock \emph{Nature} \textbf{2003}, \emph{425} 45\relax
	\relax
	\bibitem{Hasegawa_2010}
	Y.~Hasegawa, R.~Loidl, G.~Badurek, K.~Durstberger-Rennhofer, S.~Sponar,
	H.~Rauch,
	\newblock \emph{Phys. Rev. A} \textbf{2010}, \emph{81} 032121\relax
	\relax
	\bibitem{Irfan_2021}
	A.~A.~M. Irfan, P.~Blackstone, R.~Pynn, G.~Ortiz,
	\newblock \emph{New J. Phys.} \textbf{2021}, \emph{23} 083022, {Q}uantum
	entangled-probe scattering theory. DOI: 10.1088/1367-2630/ac12e0\relax
	\relax
	\bibitem{Kuhn_2021}
	S.~J. Kuhn, S.~McKay, J.~Shen, N.~Geerits, R.~M. Dalgliesh, E.~Dees, A.~A.~M.
	Irfan, F.~Li, S.~Lu, V.~Vangelista, D.~V. Baxter, G.~Ortiz, S.~R. Parnell,
	W.~M. Snow, R.~Pynn,
	\newblock \emph{Phys. Rev. Research} \textbf{2021}, \emph{3} 023227\relax
	\relax
	\bibitem{Gahler_1998}
	R.~G\"ahler, J.~Felber, F.~Mezei, R.~Golub,
	\newblock \emph{Phys. Rev. A} \textbf{1998}, \emph{58} 280\relax
	\relax
	\bibitem{Groitl_2016}
	F.~Groitl, T.~Keller, K.~Rolfs, D.~A. Tennant, K.~Habicht,
	\newblock \emph{Phys. Rev. B} \textbf{2016}, \emph{93} 134404\relax
	\relax
	\bibitem{mckay_2023}
	S.~McKay, A.~A.~M. Irfan, Q.~L. Thien, N.~Geerits, S.~R. Parnell, R.~M.
	Dalgliesh, N.~V. Lavrik, I.~I. Kravchenko, G.~Ortiz, R.~Pynn,
	\newblock Neutron spin echo is a "quantum tale of two paths'', \textbf{2023},
	\newblock \urlprefix\url{https://doi.org/10.48550/arXiv.2309.03987},
	\newblock ArXiv:2309.03987\relax
	\relax
	\bibitem{Le_2023}
	Q.~Le~Thien, S.~McKay, R.~Pynn, G.~Ortiz,
	\newblock \emph{Phys. Rev. B} \textbf{2023}, \emph{107} 134403\relax
	\relax
	\bibitem{Lu_2020}
	S.~Lu, A.~A.~M. Irfan, J.~Shen, S.~J. Kuhn, W.~M. Snow, D.~V. Baxter, R.~Pynn,
	G.~Ortiz,
	\newblock \emph{Phys. Rev. A} \textbf{2020}, \emph{101} 042318\relax
	\relax
	\bibitem{vanHove_1954}
	L.~Van~Hove,
	\newblock \emph{Phys. Rev.} \textbf{1954}, \emph{95} 249\relax
	\relax
	\bibitem{Durbin_2022}
	S.~M. Durbin,
	\newblock \emph{J. Appl. Phys.} \textbf{2022}, \emph{131} 224401\relax
	\relax
	\bibitem{Zhang_2023}
	L.~Zhang, Z.~Li, D.~Liu, C.~Wu, H.~Xu, Z.~Li,
	\newblock \emph{Phys. Rev. Lett.} \textbf{2023}, \emph{131} 073601\relax
	\relax
	\bibitem{Boothroyd_2020}
	A.~T. Boothroyd,
	\newblock \emph{Principles of Neutron Scattering from Condensed Matter},
	\newblock Oxford University Press, \textbf{2020}\relax
	\relax
	\bibitem{Lovesey1986}
	S.~W. Lovesey,
	\newblock \emph{Theory of Neutron Scattering from Condensed Matter: Volume 1
		(The International Series of Monographs on Physics)},
	\newblock Oxford University Press, \textbf{1986}\relax
	\relax
	\bibitem{Zaliznyak_2017}
	I.~A. Zaliznyak, A.~T. Savici, V.~O. Garlea, B.~Winn, U.~Filges, J.~Schneeloch,
	J.~M. Tranquada, G.~Gu, A.~Wang, C.~Petrovic,
	\newblock \emph{J. Phys.: Conf. Ser.} \textbf{2017}, \emph{862} 012030\relax
	\relax
	\bibitem{Gabrys1997}
	B.~Gabrys, K.~H. Andersen,
	\newblock \emph{Neutron News} \textbf{1997}, \emph{8} 15\relax
	\relax
	\bibitem{Shirane_Shapiro_Tranquada_2002}
	G.~Shirane, S.~M. Shapiro, J.~M. Tranquada,
	\newblock \emph{Neutron Scattering with a Triple-Axis Spectrometer: Basic
		Techniques},
	\newblock Cambridge University Press, \textbf{2002}\relax
	\relax
	\bibitem{STS2019}
	P.~Adams, J.~F. Ankner, L.~Anovitz, A.~Banerjee, E.~Begoli, H.~Bilheux, J.~J.
	Billings, R.~Boehler, S.~Calder, B.~C. Chakoumakos, T.~R. Charlton, W.-R.
	Chen, Y.~Cheng, L.~Coates, M.~J. Cuneo, L.~L. Daemen, C.~R. Dela~Cruz, C.~Do,
	A.~Moreira~dos Santos, N.~Dudney, G.~Ehlers, M.~R. Fitzsimmons, F.~X.
	Gallmeier, A.~Geist~II, V.~Graves, B.~Haberl, L.~He, W.~T. Heller, K.~Herwig,
	J.~P. Hodges, C.~Hoffmann, K.~Hong, A.~Huq, A.~Johs, U.~C. Kalluri,
	J.~Katsaras, A.~I. Kolesnikov, A.~Y. Kovalevskyi, J.~o. J.~L. Labbe, J.~Liu,
	E.~Mamontov, M.~E. Manley, M.~McDonnell, M.~A. McGuire, D.~A.~A. Myles,
	S.~Nagler, J.~C. Neuefeind, D.~P. Olds, K.~Page, A.~Payzant, L.~Petridis,
	S.~V. Pingali, S.~Qian, A.~Ramirez~Cuesta, J.~B. Roberto, L.~Robertson,
	P.~Rosenblad, T.~Saito, G.~Sala, G.~S. Smith, C.~Stanley, A.~D. Stoica,
	A.~Tennant, C.~Tulk, T.-M. Usher-Ditzian, S.~Vazhkudai, J.~Warren, H.-W.
	Wang, Z.~Wang, D.~J. Wesolowski, T.~J. Williams,
	\newblock {First Experiments: New Science Opportunities at the Spallation
		Neutron Source Second Target Station},
	\newblock Technical Report ORNL/SPR-2019/1407, Oak Ridge National Laboratory,
	\textbf{2019},
	\newblock \urlprefix\url{https://www.osti.gov/biblio/1784183}\relax
	\relax
	\bibitem{Camea2016}
	F.~Groitl, D.~Graf, J.~O. Birk, M.~Mark, M.~Bartkowiak, U.~Filges,
	C.~Niedermayer, C.~Ruegg, H.~M. Ronnow,
	\newblock \emph{Rev. Sci. Instrum.} \textbf{2016}, \emph{87} 035109\relax
	\relax
	\bibitem{Bayrakci2013}
	S.~P. Bayrakci, D.~A. Tennant, P.~Leininger, T.~Keller, M.~C.~R. Gibson, S.~D.
	Wilson, R.~J. Birgeneau, B.~Keimer,
	\newblock \emph{Phys. Rev. Lett.} \textbf{2013}, \emph{111} 017204\relax
	\relax
	\bibitem{PYNN2009}
	R.~Pynn, M.~Fitzsimmons, W.~Lee, P.~Stonaha, V.~Shah, A.~Washington, B.~Kirby,
	C.~Majkrzak, B.~Maranville,
	\newblock \emph{Physica B: Condens. Matter} \textbf{2009}, \emph{404}
	2582\relax
	\relax
\end{mcbibliography}

\end{document}